# Chasing Contests

Zhuo Chen* and Yun Liu†

January 6, 2025

**Abstract**

This paper proposes a dynamic research contest, namely *chasing contest*, in which two asymmetric contestants exert costly effort to accomplish two breakthroughs. The contestants are asymmetric in that one of them is present-biased and has already achieved one breakthrough (the *leader*), whereas the other is time-consistent and needs to achieve two breakthroughs to win (the *chaser*). The principal can choose between two disclosure policies: immediately announcing the chaser's first breakthrough (*public chasing contest*) or announcing only the final result (*hidden chasing contest*). We characterize the unique *$x$-start and $y$-stop* equilibrium under both disclosure policies, in which the leader starts working from an instant $x$ to the end while the chaser stops exerting effort by the instant $y$. In addition, the chaser will never stop earlier in the hidden chasing contest, whereas a late deadline extends the leader's effort in the public contest.

## 1 Introduction

Contests have been widely recognized as an effective mechanism for promoting innovation vitality. However, participants are not always homogeneous in many real-world research contests. On one hand, newcomers have to compete with incumbents who are already leading in progress, which discourages the former from making effort; on the other

We thank Bo Chen, Liang Dai, Simona Fabrizi, Xin Feng, Jaimie Lien, Chang Liu, Shuo Liu, Jingfeng Lu, Steffen Lippert, Paul Schweinzer, Xi Weng, Zenan Wu, Jie Zheng, and various seminar and conference participants for many valuable comments and suggestions. The authors acknowledge financial support from the Ministry of Education Project of Youth Fund of Humanities and Social Science (No. 23YJC790090), and the Shandong Provincial Natural Science Foundation (No. ZR2023MG009). All remaining errors are our own.

*Center for Economic Research, Shandong University, Jinan 250100, China. Email: zhuochen@sdu.edu.cn.

†Corresponding author. Center for Economic Research, Shandong University, Jinan 250100, China. Email: yunliucer@sdu.edu.cn.

hand, those incumbents might cut back on intensive and costly exertions given their established competition advantages.[1] In these situations, the asymmetric dynamic contest manifests as some participants chasing the others, and we refer to this type of research contest as *chasing contests*.

This paper formalizes the chasing contest from a behavioral perspective, and addresses the question: instead of imposing discrimination instruments to level the playing field,[2] how to incentivize two asymmetric contestants, a stronger *present-biased leader* and a weaker *time-consistent chaser*, through proper information disclosure policies.[3]

Such behavioral concern of the leader's bias for present is not a mere fable as *"The Tortoise and the Hare"* in the Aesopica. In real life, competition pressure is often utilized to motivate a procrastinating agent. For instance, a supervisor may have limited means to motivate a PhD student with some preliminary progress on a long-term research project. He can formulate a contest by hiring a new research assistant, if incentivizing the current procrastinating student is too costly (e.g., a more complicated contracting scheme or higher pecuniary payments). Although the new research assistant needs to start the project from scratch, the competition pressure from the newcomer could accelerate the research project progress without imposing additional rewards to the leader. Alternatively, we can metaphorically interpret the procrastinated incumbent as the management team of a *short-termistic* firm (Thaler and Shefrin, 1981; Fudenberg and Levine, 2006), who prioritizes short-term revenues from their established product lines over long-term R&D investment on disruptive technologies.[4] A prominent example is Nokia's struggle to adapt to the wave of smartphone innovation (Doz and Wilson, 2017). The company's rising bureaucratic tendencies and fear of destabilizing its core product lines resulted in slow reactions to the technological shift towards mobile phones with advanced operating systems. As newer entrants such as Apple and Android manufacturers brought disruptive innovation to the industry, Nokia gradually lost its dominance in the market, which eventually led to the sale of its mobile phone busi-

[1]In psychology, the phenomenon where agents temporarily deplete their motivation (and self-control capacity) after a period of intense effort is referred to as "ego depletion" (Baumeister and Vohs, 2007)

[2]Some popular discrimination instruments in asymmetric contests include caps on efforts, differential taxation of the prize, excluding strong agents, and affirmative actions such as giving head starts or handicaps to some agents (i.e., biasing the efforts of one or more agents) to affect the probability of success. See Mealem and Nitzan (2016) and Chowdhury et al. (2023) for two comprehensive surveys of discrimination in asymmetric contests.

[3]Time inconsistency (or equivalently, *dynamic inconsistency*) refers to the situation in which an agent's preferences over present and future choices change over time. Thus, present bias represents a particular type of time inconsistency, such that at each instant the agent overvalues her immediate rewards but puts less worth in long-term consequences.

[4]Supported by both theoretical and empirical studies, Extensive management literature suggests that as organizational structures grow more rigid and bureaucratic, companies tend to become more conservative, risk-averse, and focused on short-term accounting profits (see, Levinthal and March (1993), Miller (1994), and Sheppard and Chowdhury (2005), among others). We also briefly illustrate the mathematical analogy between a procrastinated incumbent (a human) and a short-termistic firm (an organization) in online Appendix B.1.

ness to Microsoft in 2014.[5]

By formalizing the chasing contest, we consider two contestants exert costly efforts to accomplish two breakthroughs in a continuous-time dynamic contest game. The contest ends either when one contestant achieves two successes, or no one completes both breakthroughs until a fixed deadline $T < \infty$. A pre-determined prize is rewarded to the winner, with ties broken randomly. We assume that the *leader* has already achieved a breakthrough, which presumably comes from his previous accumulated achievements (e.g., a senior researcher can utilize his unpublished research outputs instead of initiating a project from scratch), but is subject to present bias in the form of the Harris and Laibson (2013)'s *instantaneous-gratification* model.[6] This behavioral abnormality clearly increases the winning probability of the other contestant, who performs as a time-consistent *chaser* (i.e., she needs to complete two breakthroughs to win the contest but with a conventional time discounting).

Both contestants can observe the arrival of their own breakthroughs. In terms of their rival's progress, we assume they only know the winner who has completed both breakthroughs, but unless notified, the leader is unaware of the chaser's progress on her first breakthrough. The principal also commits a disclosure policy at the beginning of the contest, which either immediately discloses the chaser's first breakthrough to the leader (the *public chasing contest*), or only announces the winner who has completed both breakthroughs when the contest ends (the *hidden chasing contest*).

## 1.1 Main Findings

In the public chasing contest, we characterize a peculiar *$x$-start and $y$-stop* structure of the equilibrium strategies in Proposition 1 and 2: the leader starts working from a time $x < T$ until the contest terminates; on the other hand, the chaser will work to the end of the contest after she has achieved her first breakthrough, while stops exerting efforts if her first breakthrough has not been realized by a time $y < T$. In particular, the leader always starts earlier once the chaser has achieved her first breakthrough; however, when the leader exhibits a stronger bias for present, the presence of the chaser becomes less effective to incentivize the present-biased leader to start earlier. On the other hand, the chaser will opt for a later stopping time in the chasing stage as her effort cost diminishes.

[5]Similar business cases had also occurred in many other markets, such as the digital camera industry (Kodak versus Canon), the search engines (Yahoo versus Google), and the automotive industry (Tesla versus traditional automakers), among others.

[6]Harris and Laibson (2013) generalizes the present bias with discrete quasi-hyperbolic discounting (Barro, 1999; Luttmer and Mariotti, 2003) in continuous time. This approach not only serves as a reasonable approximation to evaluate a present-biased agent whose present rapidly collapses into the future, but brings desirable modeling features (such as smooth policy functions, equilibrium uniqueness, and a single value function for welfare evaluation).

Proposition 3 and 4 portray the *x-start and y-stop* equilibrium in the hidden chasing contest. Despite the absence of closed-form expressions for the leader's equilibrium strategies, our characterizations reveal that the leader's optimal starting time in the hidden chasing contest ($x^N$) always lies in between his two optimal starting times in the public chasing contest conditional on whether the chaser has made her first breakthrough ($x_1^F$) or not ($x_0^F$) (see Figure 4). Intuitively, since *no news is good news* for the leader in the public chasing contest, he can safely wait until $x_0^F$ if the chaser has not made her first breakthrough yet. However, if there is no immediate disclosure of his rival's progress, the present-biased leader becomes more skeptical that the chaser could have made a breakthrough when time is close to $x^N$, he will start to exert effort at $x^N$ rather than waiting until $x_0^F$ as in the public chasing contest; also, the leader will not start at $x_1^F$ when there is no real-time alarm about the arrival of the chaser's first breakthrough.

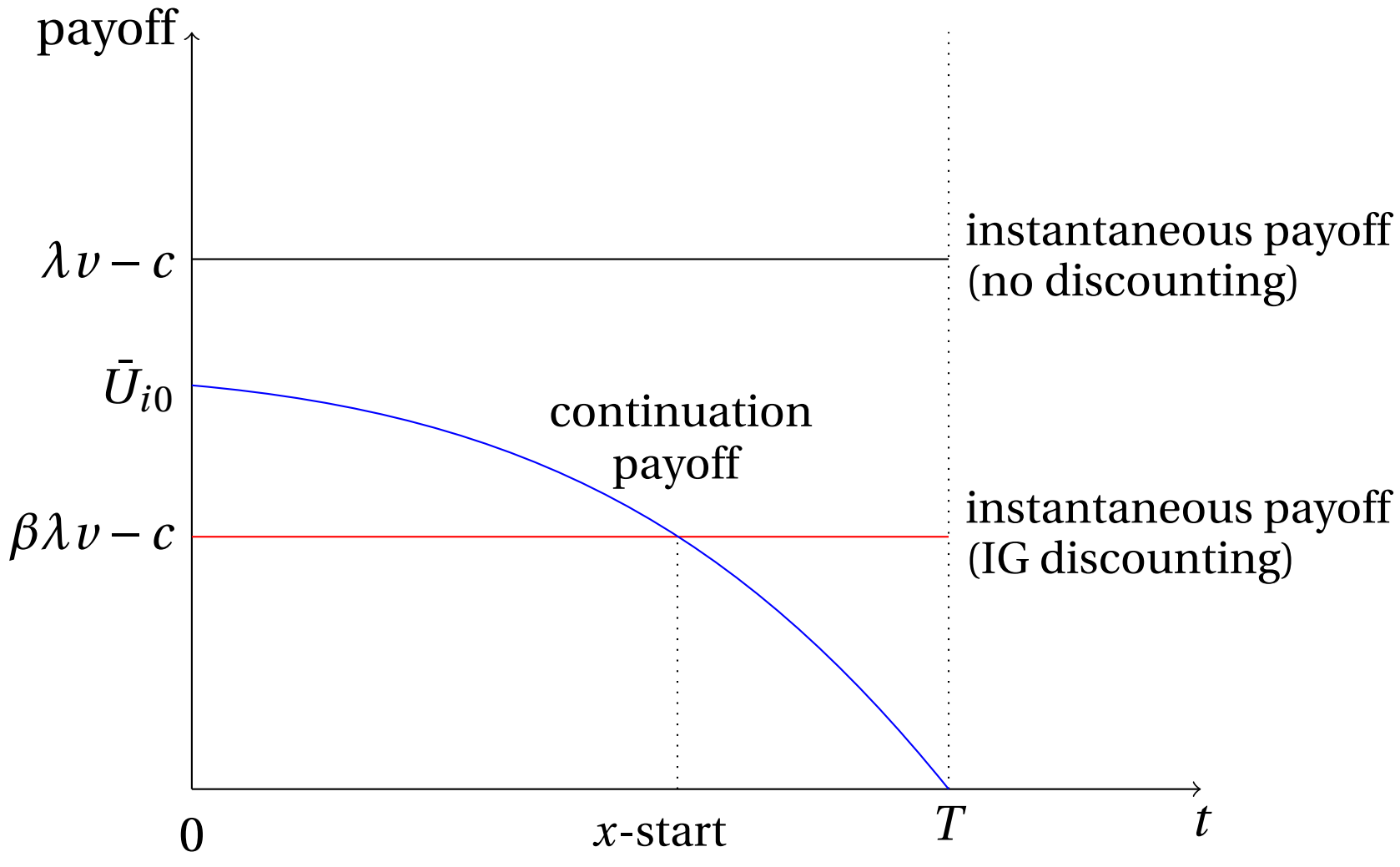

Figure 1: The appearance of $x$-start strategy with instantaneous-gratification discounting.

Although the presence of an optimal stopping strategy in dynamic contests with a predetermined deadline has been extensively studied in the literature since Taylor (1995), our characterization of the optimal starting strategy still makes a novel contribution to the contest literature.[7] We illustrate the intuition behind this optimal starting strategy here, and defer the formal analyses to the subsequent sections. As illustrated in Figure 1, since an agent's continuation payoff can also be treated as the opportunity cost at each time instant, the optimal starting strategy emerges at time $x \in (0, T)$, only when an agent's instantaneous payoff and her corresponding continuation payoff has a crossover before the fixed deadline $T$;

[7] To our knowledge, the phrase "optimal starting time" has also been employed in Seel (2018), albeit in a thoroughly distinct context. Briefly, while our analytic focus is the contestant (leader)'s strategic effort choices, Seel (2018) discusses the optimal starting time from the perspective of the contest designer (i.e., when the contest should start) under various reward schemes and success functions. Also, we sometimes use the two terms the *optimal starting strategy* and the *x-start strategy* interchangeably to represent the leader's equilibrium strategy in the chasing contest, if no confusion arises.

otherwise, the agent chooses to either start from the beginning if her instantaneous payoff surpasses the corresponding continuation payoff at time 0, or shirk throughout the contest (i.e., the continuation payoff that shirking at each instant of time exceeds the corresponding instantaneous payoff throughout the entire contest). This explains why the structure of an optimal starting strategy is absent in the existing literature on dynamic contests: unlike the leader with instantaneous-gratification preference in our chasing contest, a time-consistent agent will not disproportionately discount her instantaneous payoff (see Figure 1).

We then compare the effects of information provision between the public chasing and its hidden chasing counterpart. When the two contestants adopt their own $x$-start and $y$-stop equilibrium strategies, our Proposition 5 implies that the chaser will never stop earlier in the hidden chasing contest. The intuition behind this result is that once the chaser's first breakthrough is announced in the public chasing contest, an earlier starting time of the leader reduces the chaser's marginal reward for working for her second breakthrough; in contrast, without the immediate disclosure in the hidden chasing contest, the chaser has a larger marginal reward from exerting efforts for the second breakthrough when the leader is still shirking. However, Proposition 6 and 7 show that the two disclosure policies impose ambiguous incentives on the leader's starting time. We therefore conduct numerical analyses (see online Appendix B.1), and show that when the predetermined deadline $T$ is not too short, the leader will always start earlier in the public chasing contest.

Section 6 further explores the emergence of the $x$-start and $y$-stop equilibrium structure under some variations of the chasing contest. Our discussions indicate that although the presence of present bias is not indispensable, the assumption of a predetermined deadline is necessary for the appearance of the $x$-start and $y$-stop equilibrium structure in a continuous-time dynamic contest game. Intuitively, the emergence of the optimal $x$-start strategy necessitates a disproportionate discounting of an agent's instantaneous reward, which essentially pulls down her instantaneous payoff below the corresponding continuation payoff at time 0; such force, nevertheless, can only be fulfilled under certain conditions. Last, as discussed in online Appendix B.4 and B.5, neither introducing a present-biased chaser nor altering the level of sophistication will compromise the equilibrium structure of the chasing contest.

### 1.2 Related Literature

The theoretical exploration of intertemporal choice of a present-biased decision maker (agent) can be traced back to the seminal works of Strotz (1955) and Phelps and Pollak (1968), with mathematical refinements provided by some notable studies such as Laibson (1997) and O'Donoghue and Rabin (1999). The instantaneous-gratification time preference (Harris and Laibson, 2013) utilized in our chasing contest model generalizes the present bias with discrete quasi-hyperbolic discounting (Barro, 1999; Luttmer and Mariotti, 2003) in continuous time. This approach provides several desirable modeling attributes, including smooth policy functions, equilibrium uniqueness, as well as a single value function for welfare evaluation.

Carrillo and Mariotti (2000) and Bénabou and Tirole (2002) make two early contributions on the role of information on a present-biased agent, which study the optimal information acquisition choices of a time-inconsistent agent. More recent literature builds upon the Bayesian persuasion framework initiated by Rayo and Segal (2010) and Kamenica and Gentzkow (2011), exploring problems such as persuading a time-inconsistent agent to restrain from impulsive consumption behaviors (Lipnowski and Mathevet, 2018; Mariotti et al., 2022), or to exert costly effort (Habibi, 2020). Instead of focusing on disclosing information to a single present-biased agent who acts independently, our work addresses the externality of information disclosure in a contest game with behavioral elements.

While static (or reduced-form) contest games under different information structures and design policies have been widely studied in the literature (see Corchón and Serena (2018) and Fu and Wu (2019) for two recent and comprehensive surveys), our chasing contest model is more closely related to the literature on dynamic research contests. In a seminal early work, Taylor (1995) studies a fixed-prize dynamic research contest, in which agents conduct costly effort to obtain an innovation with a random arrival process. Benkert and Letina (2020) show that the first-best outcome is implementable in Taylor's framework with an interim transfers scheme and an endogenous termination date. On the role of information disclosure, Halac et al. (2017) consider a continuous-time dynamic research contest built in the experimentation framework of Keller et al. (2005), in which agents compete to obtain a single innovation with uncertain feasibility.[8] Chen et al. (2022) identify the optimal disclosure policies of agents' submissions in a research contest with either finite or infinite horizons. Ely et al. (2022) propose a cyclic feedback policy that informs agents about their rivals' successes at the end of each fixed-length cycle. Nevertheless, these studies have overlooked the potential inclination towards procrastination that may arise once an agent has secured certain advantages in research contests, especially when exerting efforts might also be psychologically burdensome.

Our paper also contributes to the literature on information disclosure in multistage contests. Lizzeri et al. (1999) and Yildirim (2005) are among the first to examine endogenous feedback in contests using a two-period, two-agent framework. Under a similar framework, Aoyagi (2010), Ederer (2010), and Goltsman and Mukherjee (2011) characterize the optimal interim public disclosure policies, while the possibilities of private disclosure and agents' voluntarily disclose are analyzed by Mihm and Schlapp (2019). Bimpikis et al. (2019) consider a continuous-time two-stage dynamic contest with learning, and find that the principal can benefit from disclosing information after an initial silent period. When there is no initial uncertainty on innovation feasibility, Khorasani et al. (2024) reveal that the optimal design features an initial period with no disclosure and a gradually increasing prize followed by a period of probabilistic disclosure policy. However, the two-dimensional heterogeneity—different time preferences and number of breakthroughs—between the two agents in our chasing contests are absent in these studies.

[8] We deliberately exclude the learning process in our chasing contest for two reasons: First, we can treat the realization of the leader's first breakthrough as the signal for feasibility, which has been adopted in the two-stage contest model of Bimpikis et al. (2019); it is thus unnecessary to further complicate our chasing contest model which has contained both heterogeneous contestants and a predetermined deadline. Second, as illustrated in the last paragraph of Section 6, although the leader will no longer follow an optimal starting strategy, allowing experimentation will not qualitatively overturn the main economic reasoning behind the two contestants' optimal effort choices in our chasing contest without learning.

The rest of the paper is structured as follows. Section 2 formalizes the chasing contest game. Section 3 and 4 characterize the equilibrium strategies of the two contestants under two distinct information policies, respectively. Section 5 further compares the incentive effects of the two information policies. Section 6 discusses the possible appearance of the optimal starting strategy under alternative setups, and Section 7 concludes. All proofs are in the online Appendix.

## 2 The Model

A principal hosts a research contest with a fixed deadline $T < \infty$ and a fixed prize $v > 0$. The contest requires completing two sequential breakthroughs to win the prize (e.g., acquiring an innovation). There are two risk-neutral contestants (he), $l$ (*leader*, *he*) and $c$ (*chaser*, *she*). At time $t = 0$, the leader is endowed with a breakthrough, while the chaser has none.[9] The contest is terminated immediately after either of the two contestants achieves two breakthroughs, or no contestant achieves two breakthroughs until time $T$. If there is a single winner, she receives $v$ and her rival gets 0; if both contestants achieve two breakthroughs at the same time, each of them is paid $v/2$; if no contestant wins at time $T$, both get 0.

If the contest has not been terminated at time $t \in (0, T)$, contestant $i \in \{l, c\}$ can choose $a_{it} \in \{0, 1\}$,[10] with instantaneous cost $ca_{it}dt$, where $c > 0$ is the (constant) cost for conducting research at each unit of time. The arrival of a breakthrough for a contestant with $a_{it} = 1$ follows a homogeneous Poisson process with the arrival rate $\lambda > 0$ at each instant time $t$; that is, conditional on no one has completed two breakthroughs by $t$, effort for an additional duration $dt$ produces a success with probability $\lambda dt$.[11]

[9]Alternatively, we can consider a two-stage contest (Stage $A$ and Stage $B$), in which the present-biased leader has already completed Stage $A$, while the time-consistent chaser needs to complete both stages to win the contest.

[10]There is no loss of generality in assuming the discrete action set $\{0, 1\}$, as the two contestants will still optimally choose the two boundary points of the interval $[0, 1]$ for shirking or exerting efforts given their linear payoff functions $u_{ct}$ and $u_{lt}$.

Assume that exit is *irreversible*. Each contestant $i$ first selects a starting time and then a stopping time for exerting effort; each contestant can perfectly observe her own efforts and immediately knows the arrival of her own success. The *terminal history* of contestant $i$ starting at time $t$ can thus be identified as $h_{it}^T = \left(\{a_{i\tau}\}_{\tau\in[t,T]}, \tilde{v}_i\right) \in H_{it}$ for both $i \in \{c, l\}$, where $\tilde{v}_i \in \{0, v/2, v\}$ is the realization of contestant $i$'s payoff at $T$, and $H_{it}$ is the collection of all contestant $i$'s possible terminal histories starting from time $t$.

We assume that there is no (exponential) discounting and risk aversion for simplicity.[12] However, although the chaser has no further discount for the future, the leader suffers from the present bias in term of Harris and Laibson (2013)'s *instantaneous-gratification* model (i.e., the hazard rate of transitions from the present to the future goes to infinity). Therefore, for any terminal history $h_{it}^T$, denote

$$\tilde{u}_t(h_{it}^T) = \begin{cases} -c & a_{it} = 1 \\ 0 & a_{it} = 0 \end{cases}, \quad \text{and} \quad \tilde{V}_t(h_{it}^T) = \tilde{v}_i - \int_t^T c a_{i\tau} d\tau,$$

the *present* and *future* payoffs of contestant $i$ at any $t \in [0, T]$, respectively. Thus, given terminal history $h_i^T$, contestant $i$'s payoff at date $t$ is

$$u_{it}(h_{it}^T) = \tilde{u}_t(h_{it}^T) + \beta_i \tilde{V}_t(h_{it}^T),$$

where $\beta_i$ represents agent $i$'s uniform discounting factor for future periods, with $\beta_c = 1$ for the time-consistent chaser, and $\beta_l = \beta < 1$ for the present-biased leader. As the preference of contestant $l$ changes over each instant of time, we say the payoff function $u_{lt}$ represents the preference of *self-$t$* of the leader. Also, both contestants are expected utility maximizers.

Once the chaser has made her first breakthrough, the chasing contest resembles a conventional continuous-time research contest, but with one-dimensional asymmetry coming from the leader's present-biased time preference. Denote $\tilde{t}_1 \in (0, T)$ the random time at which the chaser makes her first breakthrough, and $\tilde{t}_2 \in (0, T)$ the time instant that the leader realizes his second breakthrough. We term the time interval $[0, \tilde{t}_1)$ the *chasing stage* of the chasing contest, and $[\tilde{t}_1, T]$ as the *contest stage*. Notably, the contest may terminate without entering the contest stage, when either the leader's first breakthrough arrives before

[11] We assume the same arrival rate for the two breakthroughs to simplify the model, i.e., there is no knowledge accumulation over time.

[12] Green and Taylor (2016) argue that: *"... It is fairly straightforward to incorporate a common discount rate into the model. However, with discounting, closed-form solutions are no longer available in the multistage setting..."*.

$\tilde{t}_1$ or the chaser makes no breakthrough until $T$. Also, given the fixed deadline $T$, we introduce $\tilde{T}$ as the (potentially) random deadline at which the contest actually terminates, i.e., $\Pr(\tilde{T} \leq T) = 1$.

The principal can either disclose real-time information regarding the two contestants' progress (i.e., *full disclosure*, FD), or offer no information until the terminus (i.e., *no disclosure*, ND). Note that since the leader already has one breakthrough in hand, the full disclosure policy essentially resembles an immediate announcement of the chaser's first breakthrough. Also, the principal makes a credible commitment to the disclosure policy $\Phi \in \{FD, ND\}$. The primitives of the chasing contest are common knowledge. Its structure is depicted in Figure 2.

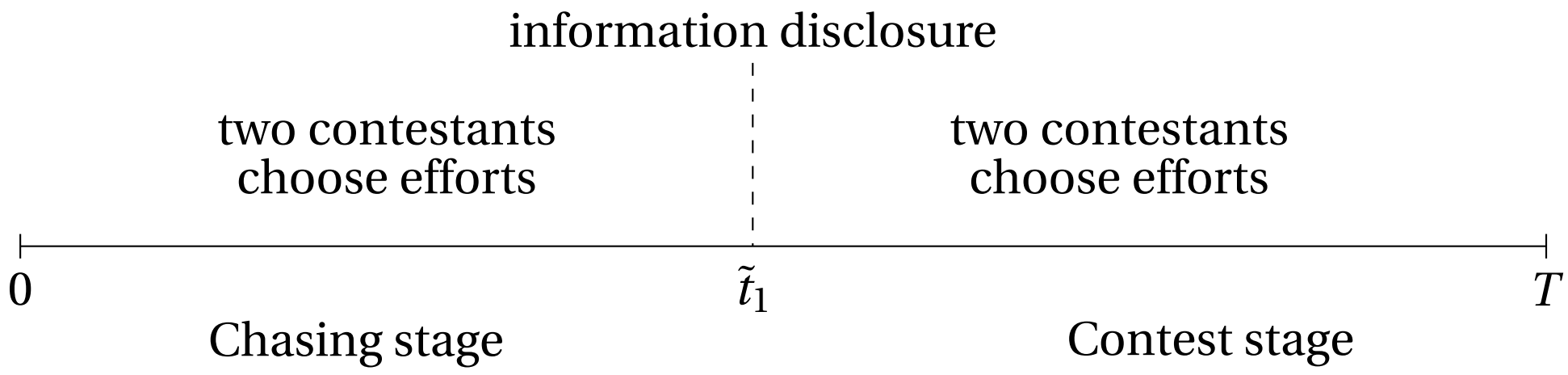

Figure 2: Structure of the chasing contest with termination time after $\tilde{t}_1$.

Let $s_t = (s_t^1, s_t^2) \in S$ be the current state of the contest, where $s_t^1 \in \{0,1\}$ indicates whether the chaser has made her first breakthrough ($s_t^1 = 1$) or not ($s_t^1 = 0$), and $s_t^2 \in \{0, l, c\}$ represents the corresponding state of the second breakthrough with $s_t^2 = 0$ as no one has made the second breakthrough and terminated the contest yet, while $s_t^2 = i$, $i \in \{l, c\}$, represents that contestant $i$ has completed the second breakthrough and terminated the contest. Thus, $S \equiv \{(1,0), (0,0), (1,c), (1,l), (0,l)\}$.

Denote $I_{it} \subseteq S$ the information that contestant $i \in \{l, c\}$ has at time $t$. As only the chaser has information about her own progress under the ND policy, we have $I_{ct} = \{s_t\}$ for the chaser. The information of the leader is assumed to be jointly determined by the disclosure policy $\Phi \in \{FD, ND\}$ and state $s_t = (s_t^1, s_t^2) \in S$. Under the full disclosure policy ($\Phi = FD$), since there is no asymmetric information between the two contestants, we have $I_{lt} = I_{ct} = \{s_t\}$. If $\Phi = ND$, the leader nevertheless cannot observe his rival's current progress unless the chaser wins the contest, that is, $I_{lt} \in \{\{(1,0),(0,0)\}, \{(1,c)\}, \{(1,l),(0,l)\}\}$.

We focus on the Markov perfect strategies on both contestants, $\sigma_{it}(\cdot)$: $2^S/\emptyset \to \{0,1\}$, which maps from the current information set $I_{it}$ into an action $a_{it}$ at each instant $t \in [0, T]$. Thus, for contestant $i$, the strategy profile $\sigma = \{(\sigma_{ct}, \sigma_{lt})\}_{t\in[0,T]}$ and the information $I_{it}$ is sufficient to determine a specific probability distribution of the terminal history at time $t$, which belongs to the collection of all the probability distributions of terminal histories. Let $E_t(\cdot | I_{it}, \sigma)$ be $i$'s expectation conditional on the strategy profile $\sigma$ and information $I_{it}$.

Our solution concept is the Markov perfect equilibrium (MPE) between the chaser and a continuum of "time-inconsistent selves" of the leader, in which the two contestants hold Markovian beliefs and optimally choose their own efforts according to $\sigma_{it}$ in each state $I_{it}$.[13] Formally, for any strategy profile $\sigma = \{(\sigma_{ct}, \sigma_{lt})\}_{t\in[0,T]}$, denote $\sigma'(\sigma, I_{it}, a)$ an alternative strategy profile which substitutes $\sigma_{it}(I_{it})$ by $a \in \{0,1\}$ at time $t \in [0,T]$ for contestant $i \in \{c, l\}$. The strategy profile $\sigma^* = \left\{(\sigma^*_{ct}, \sigma^*_{lt})\right\}_{t\in[0,T]}$ constitutes an MPE, if for any time $t$ and information $I_{it}$,

$$\sigma^*_{it}(I_{it}) \in \arg\max_{a\in\{0,1\}} E_t\left[u_{it}(\cdot)|\, I_{it}, \sigma'(\sigma^*, I_{it}, a)\right].$$

Throughout our analysis, we employ the following assumption to avoid trivialities.

**Assumption 1.** *$\delta \in (0,1)$ and $\varphi \in (0,1]$, where $\delta = \frac{\beta\lambda v - c}{\beta\lambda v - \beta c}$, and $\varphi = \frac{c}{\lambda v - c}$.*

We introduce $\delta$ as a measure of the severity of the leader's bias for the present given that $d\delta/d\beta > 0$, while $\varphi$ measures the instantaneous cost of the two contestants as $d\varphi/dc > 0$. When either of the two conditions is violated, the corresponding contestant will choose not to exert any effort throughout the chasing contest.[14] Therefore, the two conditions can be equivalently interpreted as the respective participation constraints with $\beta\lambda v \geq c$ for the leader, and $\lambda v \geq 2c$ for the chaser.

# 3 Public Chasing Contest

Before formally analyzing the equilibrium strategies of the two contestants in the public chasing contest (i.e., any progress of the two contestants is immediately and fully disclosed), we introduce the following two premised results that describe the leader's incentive compatibility condition and his starting time upon the chaser's absence.

**Lemma 1.** *Let $U^B_{lt}(\sigma)$ be the continuation payoff at time $t$ when only one breakthrough is left to the leader and the strategy profile $\sigma = \{(\sigma_{ct}, \sigma_{lt})\}_{t\in[0,T]}$ is played. The leader chooses $a_{lt} = 1$ at time $t$ if and only if*

$$\beta\lambda v \geq c + \beta\lambda U^B_{lt}(\sigma). \tag{1}$$

Lemma 1 delineates the leader's incentive compatibility condition for exerting effort, which extends the parallel condition in the single-agent setting of Green and Taylor (2016) to a dynamic contest environment with competition. It implies that given his present bias

[13] Noticeably, the solution concept implicitly assumes that the present-biased leader is *sophisticated* about her time inconsistency.

[14] When $\delta < 0$, the leader will not exert effort from the beginning with $\beta\lambda v < c$; on the other hand, if $\varphi > 1$, the chaser has no incentive to work in the chasing stage given that $\lambda v < 2c$.

$\beta$ and the arrival rate $\lambda$, the leader will start to work from the current instant $t$, if and only if his instantaneous reward from winning the contest (i.e., achieving his second breakthrough) exceeds the sum of his instantaneous cost $c$ and the corresponding continuation payoff (or equivalently, the opportunity cost of starting at time $t$) $\beta\lambda U^B_{l,t}(\sigma)$.

**Lemma 2.** *For all $t \in [0, \tilde{T}]$, without the chaser (i.e. $a_{c\tau} \equiv 0$ for all $\tau \in [t, \tilde{T}]$), the leader chooses $a_{lt} = 1$ if and only if*

$$t \geq \hat{x}_0 := \max\left\{0, T - \frac{1}{\lambda}\ln\frac{1}{1-\delta}\right\}. \tag{2}$$

That is, if the chaser stops working at time $\tau$, the leader essentially enters a *one-person contest* in which he decides his optimal starting time $\hat{x}_0$. Since we impose no restriction on the choice of $T$ and $\lambda$, it is possible to have $T - \frac{1}{\lambda}\ln\frac{1}{1-\delta} \leq 0$. We thus introduce the corner solution $\hat{x}_0 = 0$ in Lemma 2 (i.e., the leader starts exerting efforts from the beginning of the contest) when $T - \frac{1}{\lambda}\ln\frac{1}{1-\delta} \leq 0$. This could happen when we have a short deadline (i.e., a relatively small $T$) that incentives the leader to start earlier, or a sufficiently large $\delta$. In either case, the leader will choose to exert effort from the beginning of the one-person contest, and results in $\hat{x}_0 = 0$. Therefore, to exclude uninteresting cases (i.e., the leader exerts effort from time 0 even without the chaser), we impose the following assumption.

**Assumption 2.** *$\delta < 1 - e^{-\lambda T}$, or equivalently, $\hat{x}_0 > 0$.*

In words, the leader's present bias is not so light that he would put in effort right from the beginning even without the chaser's presence.

Our analysis begins from the *contest stage* of the public chasing contest, and moves backward to the *chasing stage*. In brief, we show that the public chasing contest constitutes a unique *$x$-start and $y$-stop* MPE, in which the present-biased leader will shirk until a time $x < T$, and will exert non-stopping effort until the contest terminates. Meanwhile, the time-consistent chaser will start working at the very beginning of the contest, and stop exerting effort at a time $y < T$ if she has made no breakthrough until $y$ (i.e., she expects a low chance to make two breakthroughs and wins the contest at time $y$).

## 3.1 Contest Stage

We first characterize the equilibrium strategies of the two contestants once the chaser has made her first breakthrough.

**Proposition 1.** *Suppose the leader did not exert effort before entering the contest stage. The public chasing contest admits the following Markov perfect equilibrium strategies in the contest stage (i.e. $t \geq \tilde{t}_1$):*

*(i) The leader chooses* $a_{lt} = 1$ *if and only if* $t \geq x_1^F$*, where*

$$x_1^F \equiv \max\left\{0, T - \frac{1}{2\lambda}\ln\frac{1}{1-2\delta}\right\}. \tag{3}$$

*(ii) The chaser always chooses* $a_{ct} = 1$ *for any* $t \in [\tilde{t}_1, \tilde{T}]$.

In words, Proposition 1 implies that the leader will shirk until $x_1^F$ and starts exerting effort from $x_1^F$ to the end of the contest if $\tilde{t}_1 < x_1^F$; otherwise, he will immediately start working from $\tilde{t}_1$ (i.e., no shirking in the contest stage) if $\tilde{t}_1 \geq x_1^F$ until the contest is terminated. That is, we can treat $x_1^F$ as the *trigger time* for the leader to exert effort in the contest stage. Meanwhile, the chaser will exert non-stopping efforts until the contest ends once she has made the first breakthrough. Also, note that this result is restricted to the case that the leader did not work before entering the contest stage. We will analyze their corresponding equilibrium strategies in the next subsection, when the leader would have started in the chasing stage.

In the proof of Proposition 1, we specify that $x_1^F > 0$ if and only if the inequality $\delta < \frac{1}{2}\left(1 - e^{-2\lambda T}\right)$ holds, and switches to the corner solution $x_1^F = 0$ when $\delta \geq \frac{1}{2}\left(1 - e^{-2\lambda T}\right)$. Compared with $\hat{x}_0$ in Lemma 2, Proposition 1 implies that the presence of the chaser, especially the arrival of the chaser's first breakthrough, clearly incentivizes the leader to start earlier, i.e., $x_1^F \leq \hat{x}_0$. Also, note that since we impose no restriction on the order of the arrival time of the chaser's first breakthrough $\tilde{t}_1$ and the leader's second breakthrough $\tilde{t}_2$, it is plausible that we have $\tilde{t}_2 < \tilde{t}_1$, and consequently the chasing contest will be terminated in the chasing stage before entering the contest stage.

## 3.2 Chasing Stage

We now move backward to the chasing stage in which the chaser has not attained her first breakthrough. Note that although the arrival of the first breakthrough properly incentivizes the chaser to work until the end of the contest as illustrated in Proposition 1, we cannot ensure that the chaser will exert non-stopping effort, if she has not achieved her first breakthrough by a time $t$ approaching the deadline $T$. In much of the existing literature on dynamic research contests, a contestant usually adopts a strategy to stop effort after time $t$, if she anticipates a low chance of winning after that. It is straightforward to conjecture an analogous stopping strategy of chaser in our chasing contest if she has not made her first breakthrough by time $t$. Conversely, the leader may further postpone his starting time should he foresee the earlier withdrawal of the chaser.

Therefore, we conjecture that the leader and the chaser will adopt an $x$*-start* and a $y$*-stop* strategy in the chasing stage, respectively. Specifically, the leader exerts non-stopping efforts to the end of the contest once he begins working from time $x$, whereas the chaser decides to cease working at time $y$ if she anticipates a lower chance of winning after $y$. The Markov perfect equilibrium (MPE) in the chasing stage is formally defined as follows.

**Definition 1.** *An equilibrium strategy profile* $\{(\sigma_{ct}, \sigma_{lt})\}_{t\in[0,T]}$ *constitutes an* $x$-start and $y$-stop *MPE in the chasing stage of the public chasing contest, if there exists an* $x_0^F \in [0,T]$ *and a* $y^F \in [0,T]$*, such that for any time* $t < \tilde{t}_1$ *(i.e., the chaser has not made the first breakthrough at time* $t$*):*

*(i) The leader chooses* $a_{lt} = 1$*, if and only if* $t \in [x_0^F, \tilde{T}]$*.*

*(ii) When the chaser made no breakthrough at time* $t$*, she chooses* $a_{ct} = 1$ *if and only if* $t \in [0, \min\{y^F, \tilde{T}\}]$*; when her first breakthrough has arrived before time* $t$*, she chooses* $a_{ct} = 1$ *if and only if* $t \in (0, \tilde{T}]$*.*

We are now ready to characterize the equilibrium strategies of the two contestants in the chasing stage.

**Proposition 2.** *The public chasing contest admits a unique* $x$*-start and* $y$*-stop MPE in the chasing stage (i.e.* $t < \tilde{t}_1$*):*

*(i) If* $\delta < \varphi$*, we have*

$$y^F = y_L^F = \max\left\{0, x_1^F - \frac{1}{\lambda}\ln\frac{1-\delta}{1-\varphi}\right\},$$

*and* $x_0^F = x_H^F \equiv \hat{x}_0$*.*

*(ii) If* $\varphi \le \delta < 1 - \sqrt{1-2\varphi}$*, we have*

$$y^F = y_H^F \equiv \left\{0, T - \frac{1}{2\lambda}\ln\frac{1}{1-2\varphi}\right\},$$

*and* $x_0^F = x_H^F \equiv \hat{x}_0$*.*

*(iii) If* $1 - \sqrt{1-2\varphi} \le \delta$*, we have* $y^F = y_H^F$ *and* $x_0^F = x_L^F$*, where* $x_L^F \in [x_1^F, x_H^F]$ *is the unique solution to the equation*

$$\frac{1}{4}\left[3 - 2\lambda(y_H^F - x_L^F)e^{-2\lambda(T-x_L^F)} + e^{-2\lambda(y_H^F - x_L^F)} - 4e^{-\lambda(T-x_L^F+y_H^F)}\right] = \delta,$$

*when* $3 - \left(4\sqrt{\frac{1}{1-2\varphi}} + 2\lambda T - \frac{1}{1-2\varphi} - \ln\frac{1}{1-2\varphi}\right)e^{-2\lambda T} > 4\delta$*; and otherwise,* $x_L^F = 0$*.*

We group the characterization results of Proposition 2 into three cases based on the compound effects of the severity of the leader's present bias $\delta$ and the instantaneous cost of effort $\varphi$ (see Figure 3). Note that when the chaser has made no breakthrough, the leader conducts the same starting strategy, $x_0^F = x_H^F$, in both Case (i) and (ii), which is identical to $\hat{x}_0$ of Lemma 2; that is, when the leader bears a relatively severe present bias (i.e., with $\delta < 1 - \sqrt{1-2\varphi}$), he always prefers to shirk until time $x_H^F \equiv \hat{x}_0$ in the chasing stage as in a one-person contest if the chaser makes no breakthrough and quits before $\hat{x}_0$ (i.e., $y_L^F < \hat{x}_0$ in Case (i) and $y_H^F < \hat{x}_0$ in Case (ii)). When the leader's present bias becomes relatively mild (i.e., with $1 - \sqrt{1-2\varphi} \le \delta$) in Case (iii), we can see that the presence of the chaser does incentivize the leader to exert more effort, which results in an earlier starting time $x_L^F < \hat{x}_0$ in the chasing stage. Note that, by Proposition 1, the leader has no further shirking in the contest stage if the arrival time of the chaser's first breakthrough is later than $x_1^F$ (i.e., $a_{lt} = 1$, $\forall t \in [\tilde{t}_1, \tilde{T}]$, when $x_1^F \le \tilde{t}_1$). Since the chaser stops before $x_1^F$ in Case (i), the leader will always procrastinate until $x_1^F$ in the contest stage. By contrast, when the leader is less biased for the present in Case (ii) and (iii), the immediate notification of the chaser's progress becomes more effective in the sense that the leader starts immediately when $x_1^F < \tilde{t}_1 < \min\{y_H^F, \hat{x}_0\}$.[15] Also, $x_0^F$ is strictly larger than $x_1^F$ for all the possible realized $\delta$ and $\varphi$, i.e., the leader always starts working earlier once the chaser has made her first breakthrough.

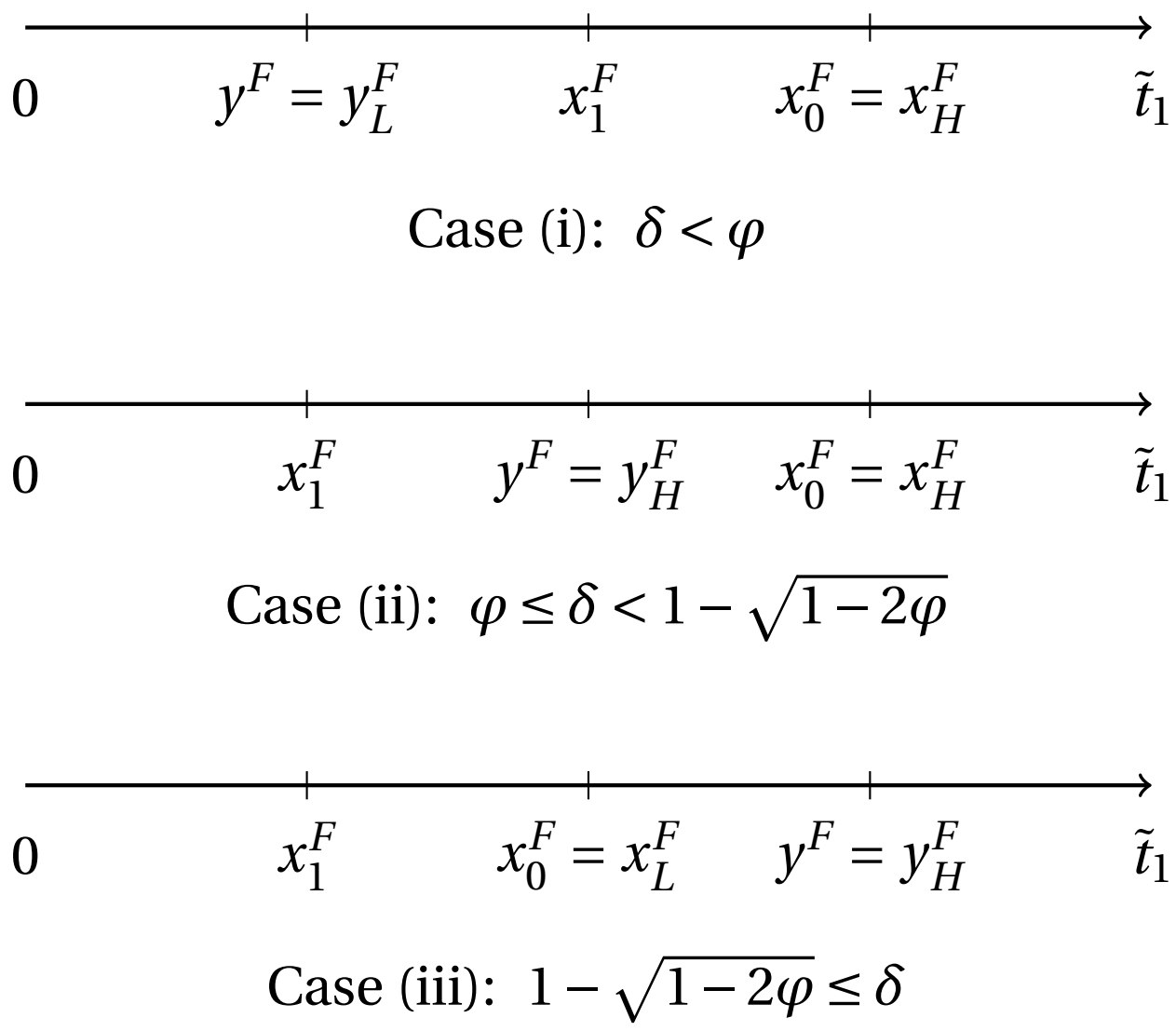

Figure 3: *x-start and y-stop* equilibrium structure in the chasing stage

Since the chaser has no present bias, we can alternatively interpret the changes of her equilibrium strategies in the chasing stage by our measure of the effort cost $\varphi$. Therefore, when the chaser bears a relatively higher effort cost in Case (i) (i.e., $\varphi > \delta$), she stops even before the leader's anticipated starting time in the contest stage with $y_L^F < x_1^F$. When making effort becomes less costly for the chaser in Case (ii) and Case (iii), she will alter her stopping strategy to $y_H^F$ and stop exerting efforts after $x_1^F$ in Case (ii) and the leader's anticipated

[15]We introduce $\min\{y_H^F, \hat{x}_0\}$ because $y_F^H < x_F^H \equiv \hat{x}_0$ in Case (ii), while $x_F^L < \hat{x}_0 < y_F^H$ in Case (iii).

starting time in the chasing stage $x_0^F$ in Case (iii), respectively.[16]

Also, note that the corner solution $y_L^F = 0$ in Case (i) occurs when either $x_1^F = 0$, or $T - \frac{1}{2\lambda}\ln\frac{1}{1-2\delta} - \frac{1}{\lambda}\ln\frac{1-\delta}{1-\varphi} < 0$. The latter situation could be induced by a sufficiently higher $\varphi$ (i.e., very close to 1) that makes exerting efforts too costly for the chaser, or a relatively short deadline $T$ which makes it almost impossible for the chaser to complete two breakthroughs before the deadline. In either case, the chaser will choose not to exert effort from the beginning of the chasing stage and consequently, make no effort for the whole chasing contest. We can interpret the corner solution of $y_H^F = 0$ emerged in Case (ii) and (iii) in a similar fashion, when the chaser embodies a relatively higher cost of effort with $\varphi \geq \frac{1}{2}\left(1 - e^{-2\lambda T}\right)$.

Next, we proceed to illustrate the two contestants' best responses to uncover the two contestants' strategic incentives behind the equilibrium structure of Proposition 2. Denote $BR_l^F(y)$ the leader's best response (i.e., his starting time) to the chaser's stopping time $y$ in the chasing stage, and $BR_c^F(x_0, x_1)$ the chaser's best response (i.e., her stopping time) to the leader's starting time $x_0$ (when the chaser makes no breakthrough) and $x_1$ (when the chaser holds one breakthrough). The following corollary further exhibits how the two contestants respond to the changes in their rivals' effort choices.

**Corollary 1.** *In the chasing stage of the public chasing contest:*

*(i)* $BR_l^F(y)$ *is constant with respect to* $y$ *when* $y < x_H^F$*, and strictly decreases with* $y$ *when* $y > x_H^F$.

*(ii)* $BR_c^F(x_0, x_1)$ *is constant with respect to* $x_0$*; it is constant with respect to* $x_1$ *when* $x_1 \leq y_H^F$*, and strictly increases with* $x_1$ *otherwise.*

Part (i) of Corollary 1 shows that the leader will not adjust his starting time $BR_l^F(y)$ in the chasing stage when the chaser stops before $x_H^F \equiv \hat{x}_0$. As is illustrated in Lemma 2, the leader essentially enters a one-person contest (without the chaser) once his rival stops at $y$, he will not revise his optimal starting time as long as the chaser stops before $\hat{x}_0$. However, when the chaser stops after $\hat{x}_0$, the leader will start earlier in response to the chaser's later stopping time. This result helps to explain the structure of the $x$-start strategy as illustrated in Figure 1: the leader will respond with a constant starting strategy $x_0^F = x_H^F \equiv \hat{x}_0$ when the chaser stops before his equilibrium starting time $x_0^F$ in Case (i) and (ii) of Proposition 2, and adjusts his equilibrium starting time to $x_0^F = x_L^F$ with the chaser's equilibrium stopping time $y_H^F$ in Case (iii).

[16] It is worth mentioning that we cannot claim that the chaser prolongs her working duration in Case (ii) compared to in Case (i), as there is no monotonic relation between $y_H^F$ and $y_L^F$ under different realizations of $\varphi$ and $\delta$; i.e., we have $y_L^F > y_H^F$ when $\delta < \varphi$ in Case (i), and $y_L^F < y_H^F$ when $\delta > \varphi$ and $\delta > 1 - \sqrt{1-2\varphi}$ in Case (ii) and (iii), respectively. Also, the inequality $\varphi \leq \delta < 1 - \sqrt{1-2\varphi}$ can be equivalently expressed in term of the chaser's effort cost $\varphi$ as $0.5(1-(1-\delta)^2) < \varphi \leq \delta$ for Case (ii), and the parallel expression $\varphi \leq 0.5(1-(1-\delta)^2)$ for the inequality $1 - \sqrt{1-2\varphi} \leq \delta$ in Case (iii).

Part (ii) of Corollary 1 first implies that the choices of the chaser's stopping time is irrelevant with $x_0$. Intuitively, the chaser is inclined to adjust her stopping time from $y$ to $y+dt$, only if she could receive a positive instantaneous payoff from doing so. However, the revision of the leader's starting time from $x_0$ to $x_0+dt$ (instead of $x_1$) has evidenced that the chaser's first breakthrough will not arrive at $y+dt$, which generates no incentive for the chaser to adjust her stopping time accordingly.

On the other hand, the chaser's stopping time will respond to the leader's starting time in the contest stage $x_1$, when $x_1 \geq y_H^F$. This is because the chaser knows that she will make non-stopping effort until the terminus if her first breakthrough arrives before $y_H^F$, and stops at $y_H^F$ otherwise. She clearly has no incentive to adjust her stopping time to any changes of $x_1$ when $x_1 < y_H^F$. However, when $x_1 > y_H^F$, since the leader will start after $y_H^F$, exerting effort to the next instant after $y_H^F$ will generate a positive instantaneous reward to the chaser. This evidently motivates the chaser to prolong her working duration in response to the changes of the leader's starting time from $x_1$ to $x_1+dt$. This result also helps to understand the structure of the $y$-stop strategy as illustrated in Figure 1: the chaser will respond with a constant starting strategy $y^F = y_H^F$ when the leader starts before her equilibrium stopping time $y_H^F$ in Case (ii) and (iii) of Proposition 2, and adopts the equilibrium stopping time $y_L^F$ which is positively correlated with the leader's equilibrium starting time $x_1^F$ in Case (i).

## 4 Hidden Chasing Contest

This section analyzes the equilibrium effort choices of the two contestants in the hidden chasing scenario, i.e., the principal follows the non-disclosure policy which only announces the final winner who has completed two breakthroughs before $T$. Note that as we assume that both contestants can perfectly observe their own progress while the leader already has one breakthrough in hand, the chaser is still fully aware of the progress of the hidden chasing contest. However, since there is no immediate disclosure of the chaser's progress in the hidden chasing contest, the leader now needs to consistently update his belief, given that the contest has not yet terminated at the current instant.

Since there is no public announcement of the chaser's first breakthrough to divide the hidden chasing contest into the chasing and the contest stage. We first alter our *$x$-start and $y$-stop* MPE structure to accommodate such change in the hidden chasing contest as follows.

**Definition 2.** *An equilibrium strategy profile $\{(\sigma_{ct}, \sigma_{lt})\}_{t\in[0,T]}$ constitutes an $x$-*start and *$y$-*stop *MPE in the hidden chasing contest, if there exists an $x^N \in [0,T]$ and a $y^N \in [0,T]$, such that for any time $t \in [0,T]$:*

*(i) The leader chooses $a_{lt} = 1$, if and only if $t \in [x^N, \tilde{T}]$.*

(ii) *When the chaser made no breakthrough before $t$, she chooses $a_{ct} = 1$ if and only if $t \in [0, \min\{y^N, \tilde{T}\}]$. When her first breakthrough has arrived before time $t$, she chooses $a_{ct} = 1$ if and only if $t \in (0, \tilde{T}]$.*

Accordingly, the belief that the chaser has made her first breakthrough at time $t$ is

$$p_{lt} = \begin{cases} 1 - e^{-\lambda t} & t < y^N, \\ 1 - e^{-\lambda y^N} & t \geq y^N, \end{cases} \tag{4}$$

which decreases with $t$ up to $y^N$ and remains constant afterward. Also, define

$$\Gamma(\varphi) \equiv \varphi + \left( \frac{1-\varphi}{\sqrt{1-2\varphi}} - 1 \right) e^{-\lambda T}. \tag{5}$$

The following two propositions present our equilibrium characterizations for the hidden chasing contest.

**Proposition 3.** *Suppose $\delta < \Gamma(\varphi)$. The hidden chasing contest admits a unique $x$-start and $y$-stop MPE characterized by $(x^N, y^N)$ with $y^N < x^N$, such that:*

(i) *If $\varphi > 1 - \frac{1-\delta+0.5\delta^2}{1-\delta} e^{-\lambda T}$, we have $y^N = 0$ and $x^N = x_H^N \equiv x_H^F$.*

(ii) *If $\varphi \leq 1 - \frac{1-\delta+0.5\delta^2}{1-\delta} e^{-\lambda T}$, we have $y^N = y_L^N \equiv \frac{1}{\lambda} \ln \left[ \frac{(1-e^{-2\lambda(T-x_H^N)})^2}{2\delta - 1 + e^{-2\lambda(T-x_H^N)}} \right]$, where $x_H^N$ is the unique real solution solved by*

$$e^{\lambda x_H^N} = \frac{(1-e^{-\lambda(T-x_H^N)})^2}{2(1-\varphi)} \cdot \frac{1-\frac{1}{2}\left(1-e^{-2\lambda(T-x_H^N)}\right)}{\delta - \frac{1}{2}\left(1-e^{-2\lambda(T-x_H^N)}\right)}. \tag{6}$$

Part (i) of Proposition 3 shows that when the leader has a relatively severe present bias (i.e., $\delta \leq \Gamma(\varphi)$), the chaser will not participate in the contest from the beginning when exerting effort is relatively costly (i.e., $y^N = 0$ if $\varphi > 1 - \frac{1-\delta+\delta^2}{1-\delta} e^{-\lambda T}$); meanwhile, the leader will procrastinate until the latest possible starting time $x_H^F$ which has been characterized in Proposition 2 of the public chasing contest. Part (ii) indicates that the chaser chooses to work for her first breakthrough until time $y^N = y_L^N$ when $\varphi \leq 1 - \frac{1-\delta+\delta^2}{1-\delta} e^{-\lambda T}$, while the leader will exert non-stopping efforts at any instant of time $t \in [x_H^N, \tilde{T}]$ given that $\delta < \Gamma(\varphi)$.

**Proposition 4.** *Suppose $\delta \geq \Gamma(\varphi)$. The hidden chasing contest admits a unique $x$-start and $y$-stop MPE characterized by $(x^N, y^N)$, with $x^N \leq y^N$, where $y^N = y_H^N \equiv y_H^F$, and $x^N = x_L^N \in$*

$(x_1^F, y_H^N)$ *is implicitly determined by*

$$\delta = (1 - e^{-\lambda x_L^N}) \cdot \frac{1}{2}\left(1 - e^{-2\lambda(T - x_L^N)}\right) + e^{-\lambda x_L^N}\left[\frac{3}{4}\left(1 - e^{-2\lambda(y_H^N - x_L^N)}\right) - \frac{1}{2}\lambda(y_H^N - x_L^N)e^{-2\lambda(T - x_L^N)} + e^{-2\lambda(y_H^N - x_L^N)}\left(1 - e^{-\lambda(T - y_H^N)}\right)\right]. \quad (7)$$

Proposition 4 further characterizes the equilibrium strategies of the two contestants when the chaser competes with a less present-biased leader (i.e., $\delta \geq \Gamma(\varphi)$). It reveals that the chaser will stop working for her first breakthrough at $y_H^N$ which is identical to her optimal stopping $y_H^F$ when faced with a mildly present-biased leader (with $\delta \geq \varphi$) in the public chasing contest; meanwhile, the leader will exert non-stopping efforts from the instant of time $x_L^N \in (x_1^F, y_H^N)$ if the contest is not terminated at $t$.

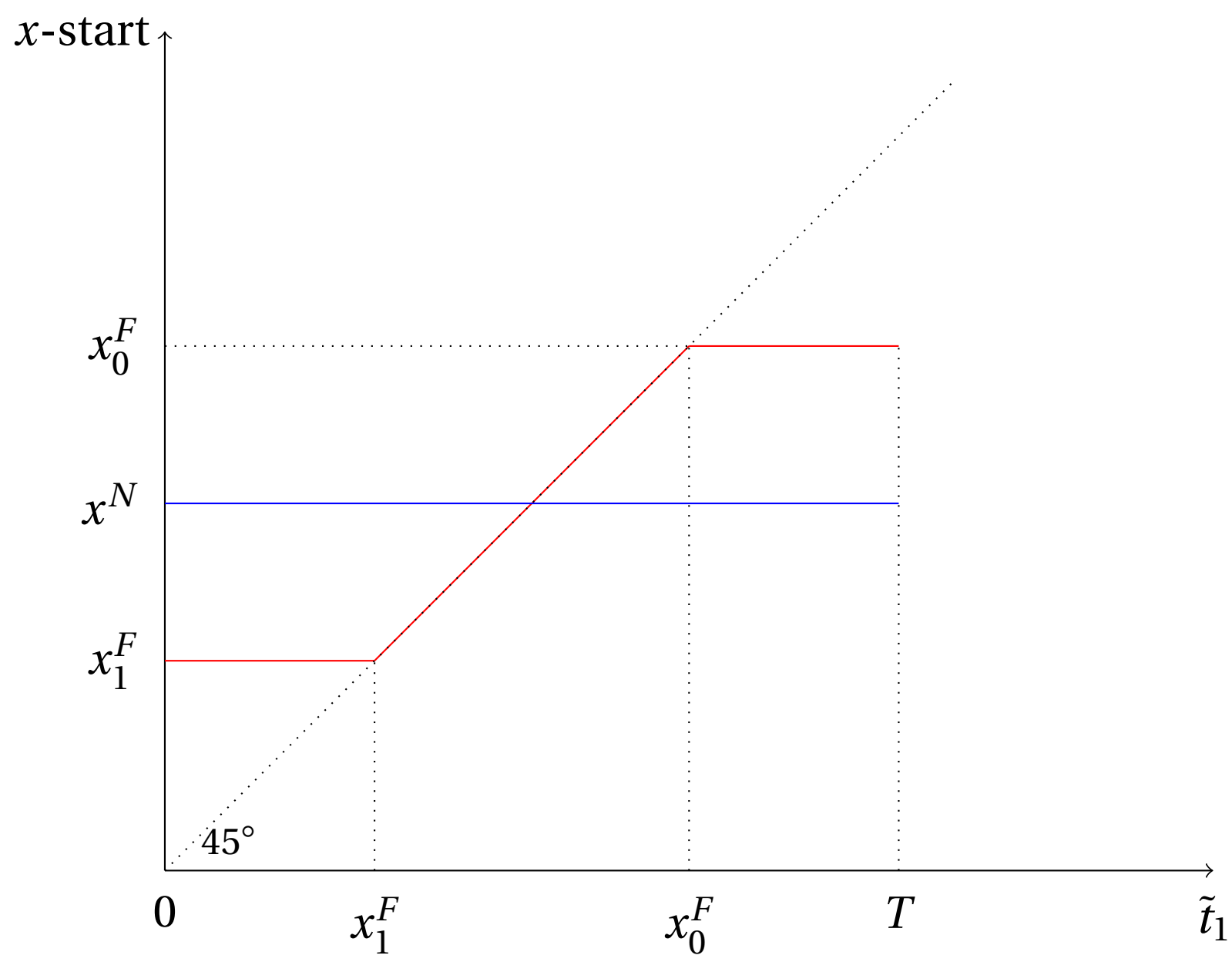

Figure 4: Comparing the leader's starting time in the public and the hidden chasing contests.

**Corollary 2.** $x^N \in (x_1^F, x_0^F]$.

Corollary 2 indicates that $x^N$ always lies between $x_1^F$ in the contest stage and $x_0^F$ in the chasing stage of the public chasing contest (see Figure 4); that is, the effects of the information provision in the public chasing contest depend on the realization of the chaser's first breakthrough. Intuitively, since *no news is good news* for the leader in the public chasing contest, he can safely wait until $x_0^F$ if the chaser has not yet achieved her first breakthrough. However, if there is no immediate disclosure of his rival's progress, he becomes more skeptical that the chaser could have made a breakthrough as the time approaches $x^N$. Therefore, the leader is inclined to exert efforts at $x^N$ instead of procrastinating until $x_0^F$ in the hidden chasing contest; meanwhile, the present-biased leader becomes more comfortable to shirk later than $x_1^F$, if there is no real-time alarm about the arrival of the chaser's first breakthrough to provide incentives.

Denote $BR_c^N(x^N)$ (resp. $BR_l^N(y^N)$) the chaser (resp. leader)'s best response to her (resp. his) rival's starting (resp. stopping) strategy in the hidden chasing scenario. The following corollary illustrates the changes in the best responses of the two contestants toward their rivals' effort choices.

**Corollary 3.** *In the hidden chasing contest:*

*(i)* $BR_l^N(y)$ *is decreasing with* $y$*.*

*(ii)* $BR_c^N(x)$ *is constant with respect to* $x$ *when* $x \leq y_H^N$*, and is strictly increasing with* $x$ *otherwise.*

Corollary 3 implies that the leader will always move forward his starting time in response to a later stopping time of the chaser; meanwhile, the chaser is willing to extend her optimal stopping time to the next instant after $y$, only when the anticipated starting time of the leader in the hidden chasing contest $x^N$ is later than $y$. It follows an analogous interpretation of the two contestants' best responses as in Corollary 1 for the public chasing contest. The only difference comes from the monotonic decrease of the leader's starting time in response to a later stopping time of his rival in the hidden chasing contest. This is because when there is no real-time feedback about the chaser's progress, the leader's belief that the chaser could possibly have made one breakthrough is monotonically increasing with her stopping time. A higher $y$, therefore, clearly induces a stronger incentive for the leader to bring forward his starting time.

# 5 Comparison

This section compares the effects of information provision between the public chasing contest and its hidden chasing counterpart. Note that when the two contestants adopt the *$x$-start and $y$-stop* equilibrium strategy illustrated in the preceding two sections, the principal essentially prefers a contest mechanism with an earlier starting time for the leader as well as a later stopping time for the chaser. Therefore, the comparisons are conducted between the two contestants' equilibrium starting and stopping times in the public chasing contest and its hidden chasing counterpart, respectively.

## 5.1 The Chaser

We begin our analysis by comparing the chaser's unique stopping time $y^F$ and $y^N$ under the two contest mechanisms.

**Proposition 5.**

*(i) When* $\delta \geq \Gamma(\varphi)$, $y^F = y^N$.

*(ii) When* $\delta < \Gamma(\varphi)$, $y^F < y^N$.

In words, Proposition 5 implies that the chaser will never stop earlier in the hidden chasing contest compared to its public chasing counterpart. To see this , note that a larger marginal reward can thus support a longer chasing time. Therefore, once the chaser's first breakthrough is announced in the public chasing contest, an earlier starting time of the leader ($x_1^F < x^N$ of Corollary 2) reduces the chaser's marginal reward for her second breakthrough. In contrast, without the immediate disclosure in the hidden chasing contest, the chaser has a larger marginal reward from exerting efforts for the second breakthrough when the leader is still shirking.[17]

## 5.2 The Leader

However, given the leader's contingent starting time $x_1^F$ or $x_0^F$ conditional on the realization of the chaser's first breakthrough in the public chasing contest, we cannot directly compare the leader's starting time $x^N$ in the hidden chasing contest with either $x_1^F$ or $x_0^F$ as is the case for the chaser (see Proposition 5). In this regard, we introduce the leader's *expected* starting time $\bar{x}^F$, and compare $\bar{x}^F$ with $x^N$ for the leader. Given the distinct combined realizations of $\delta$ and $\varphi$ of Proposition 2, we formally define $\bar{x}^F$ as follows.

*Case (i):* when $\delta < \varphi$, the leader starts at either $x_1^F$ (if the chaser makes the first breakthrough before $y_L^F$) or $x_H^F$ (if the chaser has not made the first breakthrough before $y_L^F$). Thus,

$$\bar{x}^F = \int_0^{y_L^F} \lambda e^{-\lambda t} x_1^F dt + e^{-\lambda y_L^F} x_H^F = (1 - e^{-y_L^F}) x_1^F + e^{-y_L^F} x_H^F. \tag{8}$$

*Case (ii):* when $\varphi \leq \delta < 1 - \sqrt{1 - 2\varphi}$, the leader starts at either any $t \in [x_1^F, y_H^F]$ (if the chaser makes the first breakthrough before $y_H^F$) or $x_H^F$ (if the chaser has not made the first

[17]This observation, neverthelss, is not limited in our stylized chasing contest with a present-biased leader. In many real-world business cases, a newcomer often maintains a low profile to prevent hindrances from incumbents in the early stage of competition. For instance, many tech startups, especially those in the software and app development industry, tend to operate stealthily during their early stages. This approach allows them to evaluate their products within a selected group of users (or within a specific geographic area), enhance their products according to feedback received, and refrain from extensive marketing or publicity promotion until they are ready for the full-scale launch. A recent example is the groundbreaking ChatGPT developed by OpenAI. The company, founded in 2015, decided to publicize its product only after attaining considerable competition advantage over powerful incumbents such as Google or Facebook.

breakthrough before $y_H^F$).

$$\bar{x}^F = \int_0^{x_1^F} \lambda e^{-\lambda t} x_1^F dt + \int_{x_1^F}^{y_H^F} \lambda e^{-\lambda t} t dt + e^{-\lambda y_H^F} x_H^F = x_1^F + \frac{1}{\lambda}\left(e^{-\lambda x_1^F} - e^{-\lambda y_H^F}\right) + e^{-\lambda y^F}(x_H^F - y_H^F). \tag{9}$$

*Case (iii):* when $1-\sqrt{1-2\varphi} \leq \delta$, the starting time can be any $t \in [x_1^F, x_L^F]$. Thus,

$$\bar{x}^F = \int_0^{x_1^F} \lambda e^{-\lambda t} x_1^F dt + \int_{x_1^F}^{x_L^F} \lambda e^{-\lambda t} t dt + e^{-\lambda x_L^F} = x_1^F + \frac{1}{\lambda}\left(e^{-\lambda x_1^F} - e^{-\lambda x_L^F}\right). \tag{10}$$

Obviously, when the lengths of the chaser's working time are different under the two contest mechanisms (Part (ii) of Proposition 5), the leader will alter his optimal starting time as a strategic response (see Corollary 1 and 3). However, even when the lengths of the chaser's effort are identical under the two contest mechanisms (Part (i) of Proposition 5), the leader's expected starting time in the public chasing contest $\bar{x}^F$ still could be different from his corresponding optimal starting time in the hidden chasing contest $x^N$. This is because if the chaser makes her first breakthrough early in the public chasing contest, the immediate disclosure will alert the leader to start on time; on the other hand, if the chaser is unable to achieve her first breakthrough early, the commitment of immediate disclosure will make the leader more comfortable to shirk since *no news is good news* for him. Such strategic trade-off between the two contingent starting times in the public chasing contest, nevertheless, does not occur in its hidden chasing counterpart with no information updating.

We introduce an intermediate variable $BR_l^N(y^F)$, which represents the leader's best response to $y^F$ (the chaser's optimal stopping time in the public chasing contest) in the hidden chasing contest, and decompose the difference between $x^N$ and $\bar{x}^F$ as the *competition effect* and the *beep effect*.[18] As a result, we have

$$x^N - \bar{x}^F = \underbrace{x^N - BR_l^N(y^F)}_{\text{competition effect}} + \underbrace{BR_l^N(y^F) - \bar{x}^F}_{\text{beep effect}}. \tag{11}$$

The competition effect measures how the leader reacts differently when he expects that the chaser stops at $y^F$ (with $BR_l^N(y^F)$) instead of $y^N$ (with $BR_l^N(y^N) \equiv x^N$) in the hidden chasing contest; it is positive if the leader starts working earlier when the chaser's expected stopping time is shifted to $y^F$ in the hidden chasing contest. Meanwhile, the beep effect represents the difference in the leader's best response to an identical stopping time $y^F$ under the two information policies, which resembles the effect of the *beep* signal of Ely (2017). It is positive

[18] Alternatively, we can interpret $BR_l^N(y^F)$ as the leader's starting time in the hidden chasing contest when there is no competition effect.

(resp. negative) if the leader starts earlier (resp. later) in the public chasing contest for a given $y^F$.

**Proposition 6.**

*(i) When $\delta \geq \Gamma(\varphi)$, there is no competition effect, i.e., $x^N = BR_l^N(y^F)$.*

*(ii) When $\delta < \Gamma(\varphi)$, the competition effect is negative, i.e., $x^N < BR_l^N(y^F)$.*

Recall Corollary 3, we know that if the chaser works longer in the hidden chasing contest, the leader will respond with an earlier starting time. Since the chaser does work longer in equilibrium in the hidden chasing contest when $\delta < \Gamma(\varphi)$ (Part (ii) of Proposition 5), the competition effect is negative as the leader starts earlier in the hidden chasing contest when the chaser has a later stopping time, i.e., $x^N \equiv BR_l^N(y^N) < BR_l^N(y^F)$ when $y^N > y^F$. Also, the chaser's stopping times will not differ ($y^F = y^N$) when $\delta \geq \Gamma(\varphi)$, i.e., there is no competition effect.

However, different from the non-positive competition effect for any possible realizations of $\delta$, $\varphi$ and $T$, the following proposition implies that the directions of the beep effect largely depend on the ranges of the parameters.

**Proposition 7.**

*(i) When $\delta < \varphi$, the beep effect is positive, i.e., $BR_l^N(y_L^F) > \bar{x}^F$.*

*(ii) When $\delta \geq \varphi$, the direction of the beep effect is ambiguous.*

Although the direction of the beep effect becomes unclear when $\delta \geq \varphi$, we conduct numerical analysis in online Appendix B.2.1 to identify its directional pattern. In short, our numerical results show that the beep effect remains negative with a relatively small $\lambda T$ (i.e., an urgent deadline) and a sufficiently large $\delta$.[19] Conversely, when $\lambda T$ is greater than 0.7, the beep effect becomes positive for all possible combinations of $\delta$ and $\varphi$. Table 1 summarizes the directions of the competition effect and the beep effect under different combinations of the parameters.

Intuitively, both a less pressing deadline and a more severe bias for the present will incentivize the leader to further postpone his starting time. Announcing the chaser's progress in the public chasing contest becomes more effective as it eliminates the leader's belief that the chaser has made no progress yet. It thus induces a relatively smaller expected starting time $\bar{x}^F$ compared to his uninformative starting time $BR_l^N(y)$ in the hidden chasing contest.

[19] Given that the Poisson rate $\lambda$ is fixed, we use $\lambda T$ to approximately represent the deadline of the contest; for instance, without loss of generality, we can normalize $\lambda$ to 1.

| | Competition effect | Beep effect |
|---|---|---|
| $\delta < \varphi$ | Negative | Positive |
| $\varphi \le \delta < \Gamma(\varphi)$ | Negative | Positive with a large $\lambda T$; negative with a small $\lambda T$ and a relatively large $\delta$ |
| $\Gamma(\varphi) \le \delta$ | None | Positive with a large $\lambda T$; negative with a small $\lambda T$ and a relatively large $\delta$ |

Table 1: The directions of motivation and beep effect

On the other hand, the negative beep effects come from the resultant force of a larger $\delta$ and a smaller $T$. This is because, first, both a lower bias for the present and an urgent deadline incentivize the leader to exert efforts earlier even without any immediate disclosure in the hidden chasing contest, which results in a smaller $BR_l^N(y)$. The chaser is thus less likely to make her first breakthrough before $BR_l^N(y)$ in the hidden chasing scenario. This neutralizes the effect of immediate disclosure in its public chasing counterpart, and induces a larger $\bar{x}^F$. In addition, when the deadline approaches, the leader also becomes more confident to further defer his starting time in the public chasing contest, if there is no alarm about the arrival of the chaser's first breakthrough, which generates a larger $\bar{x}^F$.

Consequently, the direction of the motivation and beep effects aligns only when: (i) $\varphi \le \delta < \Gamma(\varphi)$ and the contest environment embodies a short deadline and a less present-biased leader, in which the leader will start working earlier in hidden chasing scenario with $x^N < \bar{x}^F$; (ii) $\delta \ge \Gamma(\varphi)$, in which the direction of the total effect is identical to that of the beep effect. For all the rest realizations of $\varphi$, $\delta$ and $T$, the competition effect and the beep effect move in opposite directions. Our numerical exercises (in online Appendix B.2.2) suggest that when $\lambda T$ becomes sufficiently large, the positive beep effect surpasses the non-positive competition effect, which results in a positive total effect, i.e., an earlier starting time in the public chasing contest, for any feasible combination of $\delta$ and $\varphi$.

# 6 Discussions

This section discusses the possible appearance of the $x$-start strategy in some alternative settings different from our chasing contest. We summarize the main observations here, and leave the analytic details in online Appendix B.3. In short, our analysis implies that although the behavioral abnormality assumption (of present bias) is not indispensable, a predetermined deadline is necessary for the presence of the optimal $x$-start strategy structure.

First, note that a nonzero optimal starting time could also arise if the leader has no present bias but possesses a substantial *fixed cost* for initiating effort, which has been com-

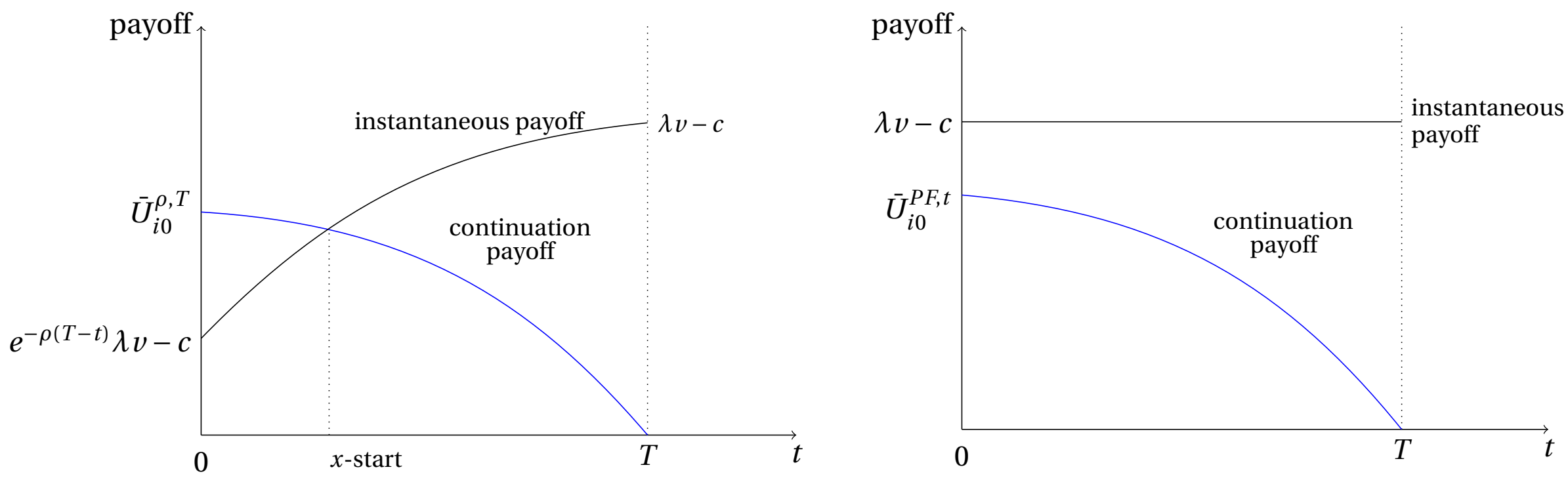

(a) exponential discounting (with postponed reward)

(b) present-future discounting

Figure 5: The occurrence of $x$-start strategies in alternative settings

monly interpreted as the *organizational inertia* phenomenon in the literature of organizational behavior (see, Hannan and Freeman (1984), Kelly and Amburgey (1991), and Gilbert (2005), among others). In this sense, the inertial leader's optimal starting strategy is behaviorally indistinguishable with the present-biased leader's $x$-start strategy in our chasing contest. To see this, consider a scenario where, following a period of intensive exertion of effort, a *time-consistent* agent decides to take a break. Therefore, by the reasoning of Lemma 1, at the moment he decides to resume exerting effort, his instantaneous payoff is given by $\lambda v - c - FC$, where $FC$ represents his fixed cost for initiating effort. As long as $\lambda v - c - FC$ is less than $\bar{U}_{i0}$, the agent will also opt for a delayed starting time until his instantaneous payoff surpasses the corresponding continuation payoff as is the IG discounting case in Figure 1.

We then reintroduce the exponential discounting back to a conventional dynamic research contest with a set of time-consistent contestants who are ex-ante homogeneous. The contest is terminated either when a publicly announced breakthrough occurs at time $t \in (0, T)$ or upon the predetermined deadline $T$ is met. However, rather than awarding the winning contestant immediately upon achieving the breakthrough, we require that the prize $v$ is deferred until the predetermined deadline.[20] Our analysis implies that once the payday is sufficiently distant from the present, an agent's discounted instantaneous payoff from exerting effort at the beginning of the contest cannot compensate her corresponding opportunity cost (of exerting effort) at present, the agent will defer her starting time for exerting effort until an instant $t$ at which her discounted instantaneous payoff surpasses the corresponding continuation payoff at $t$ (see Panel (a) of Figure 5). In other words, even without assuming behavioral abnormality, agents may still opt for an $x$-start strategy which has not been explored in other dynamic contest literature. However, the presence of the $x$-start strategy in this continuous-time symmetric contest comes from a gradually increasing instantaneous payoff function when $t$ approaches the deadline, i.e., a less discounted instantaneous payoff over time, which is different from the reasoning illustrated for the presence of the $x$-start strategy in our chasing contest (see Figure 1).

[20] Such postponed payment could happen, for instance, when a company awards the annual bonus to its top salesperson by the end of the year. Also, even if one candidate has secured a substantial lead in a political race, the final result will remain uncertain until every ballot has been tallied.

Next, we remove the predetermined deadline in the chasing contest; that is, instead of imposing a fixed deadline $T$, we assume that the present-biased leader competes with the time-consistent chaser over an infinite horizon. Our analysis indicates that instead of performing their respective *$x$-start and $y$-stop* equilibrium strategies, the two contestants utilize a stationary MPE as they face a stationary optimization problem at each instant of time until infinite.

We also discuss the scenario in which the present bias appears in the form of the *present-future* (PF) model of Harris and Laibson (2013). That is, although an agent's present self still discounts the values of her future selves by a constant discount factor $\beta < 1$; however, unlike the instantaneous-gratification (IG) model, the transition from the present to the future occurs with a finite constant hazard rate $\eta > 0$.[21] To avoid unnecessary technical complication that contributes limited economic insights, we consider a one-person contest in which a single agent with the PF preference only needs to complete one breakthrough. Our analysis implies that once the agent embodies a less drastic time discounting (i.e., there is no discontinuous change of discounting rate between present and next instant of time as in the IG model), the agent will no longer follows an $x$-start strategy since his continuation payoff has no intersect with her instantaneous payoff (see Panel (b) of Figure 5). In other words, the agent will either choose to work from the beginning if her instantaneous payoff is always higher than the corresponding continuation payoff, and shirks throughout the contest otherwise.

Last, we consider an infinite-horizon chasing contest with initial uncertainty on the feasibility of breakthroughs. Halac et al. (2017) and Bimpikis et al. (2019) have considered such continuous-time dynamic contest with experimentation, in which each agent exerts costly effort until her posterior belief about the feasibility of success can no longer elicit her effort. Clearly, given their diminishing confidence in the feasibility of success, both the leader and the chaser will opt for an optimal stopping time in this infinite-horizon chasing contest with experimentation; however, the leader will not utilize an optimal $x$-start strategy anymore. To see this, we employ a simplified one-person contest in which a present-biased (in term of instantaneous-gratification) agent needs to complete one (possibly infeasible) breakthrough with no predetermined deadline. Intuitively, the agent needs to continuously work over a period of time until her posterior belief about success drops to the threshold for exerting effort. In other words, the agent faces with the same optimization problem at time 0 and any delayed starting time $x > 0$ in this infinite-horizon one-person contest. If her incentive compatibility condition fails to hold at the beginning of the contest, she will also lack motivation to exert effort at any later time $x$.[22] Consequently, a present-biased agent will not adopt an optimal $x$-start strategy in an infinite-horizon chasing contest involving experimentation.[23]

[21] In other words, the IG discounting can be considered as the extreme case of the PF discounting in which $\eta \to \infty$. Harris and Laibson (2013) have shown that the IG model can be treated as a good approximation to measure a present-biased agent whose present collapses into the future very quickly.

# 7 Conclusion

This paper proposes a dynamic research contest, the *chasing contest*, which requires achieving two breakthroughs of the project to win a fixed prize before a predetermined deadline. There are two asymmetric contestants: a present-biased leader who has already held a breakthrough in hand, and a time-consistent chaser who needs to achieve two breakthroughs to win the contest. The principal can either immediately disclose the arrival of the chaser's first breakthrough (i.e., public chasing contest), or only announce the contest result until the terminus (i.e., hidden chasing contest). We assume that the two contestants' beliefs are Markovian, and focus on the Markov perfect strategies which involve both the *interpersonal* strategies (of the two contestants) as in a standard contest game and the *intrapersonal* strategies (for the time-inconsistent leader) as in a decision game.

We fully characterize the unique *$x$-start and $y$-stop* Markov perfect equilibrium of the public chasing contest and its hidden counterpart, and compare the respective starting time and stopping time of the two contestants under the two disclosure policies. Our results imply that the chaser will never stop earlier in the hidden chasing contest compared to its public chasing counterpart, whereas the leader starts earlier in the public chasing contest if the predetermined deadline is not too urgent.

The current paper is, to our knowledge, the first to analyze strategic interactions among agents in a contest environment with both different time preferences and distinct information structures. One novel insight of our findings is the characterization of the leader's optimal starting time, which is absent in the literature on dynamic contests. We demonstrate that the emergence of the optimal $x$-start strategy essentially demands a significant discounting of an agent's instantaneous payoff below his corresponding continuation payoff at time 0, which clearly is not limited to our stylized chasing contest with a present-biased leader. Overall, our analysis confirms that instead of imposing discrimination instruments to level the playing field, the designer can incorporate proper information disclosure policies to incentivize participants with behavioral issues in a dynamic research contest.

[22] For instance, if the agent decides to shirk from the beginning of the contest up to time $x$ (with $a_t = 0$ for $t \in [0, x)$), her posterior belief at $x$ will remain unchanged from her prior belief at time 0 (i.e., $p_x = p_0$). As a result, she will parallelly postpone her optimal stopping time from $y$ to $y + x$, where $y$ is her optimal stopping time if she starts working from the beginning of the one-person contest.

[23] We also briefly discuss the case of experimentation with a predetermined deadline in online Appendix B.3.4. Even though the agent behaves differently as she might have choose an optimal starting time instead, she essentially follows the same strategic reasoning discussed in the preceding infinite-horizon case.

# Appendix to “Chasing Contests” (Not for Publication)

Zhuo Chen[*] and Yun Liu[†]

January 6, 2025

## Contents

[*]Center for Economic Research, Shandong University, Jinan 250100, China. Email: zhuochen@sdu.edu.cn.

[†]Corresponding author. Center for Economic Research, Shandong University, Jinan 250100, China. Email: yunliucer@sdu.edu.cn.

# Appendix A Proofs

We first introduce some notations. The second-order infinitesimal term $(dt)^2$ is omitted for simplicity in the appendices (except for its first appearance in the Proof of Lemma 1).

- $\delta$: A parameter equals $\frac{\beta\lambda v-c}{\beta\lambda v-\beta c}$, which measures the level of the leader's present bias.
- $\varphi$: A parameter equals $\frac{c}{\lambda v-c}$, which measures the two contestants' effort costs per unit of time.
- $\tilde{t}_1$: The random time at which the chaser makes her first breakthrough.
- $\tilde{T}$: The random time at which the chasing contest is terminated when either one contestant wins the contest or the deadline $T$ is reached.
- $U_{it}^A$: The continuation payoff of contestant $i \in \{l, c\}$ at time $t$ when there are two breakthroughs left to be completed. Since it, as well as $U_{it}^B$ below, will depend on the particular strategies employed after time $t$ ,we will specify on a case-by-case basis when $U_{it}^A$ and $U_{it}^B$ appear in the proofs.
- $U_{it}^B$: The continuation payoff of contestant $i \in \{l, c\}$ at time $t$ when there is only one breakthrough left to be completed.
- $\bar{U}_{it}$: The continuation payoff of contestant $i \in \{l, c\}$ at time $t$ when there is only one breakthrough left to be completed and the contestant continuously works from $t$ to the end of the game.
- $x_1^F$: The leader's optimal starting time in the public chasing contest if the chaser has made her first breakthrough.
- $x_0^F$: The leader's optimal starting time in the public chasing contest if the chaser has not made her first breakthrough.
- $y^F$: The chaser's optimal stopping time in the public chasing contest when she has not made the first breakthrough.
- $x^N$: The leader's optimal starting time in the hidden chasing contest.
- $y^N$: The chaser's optimal stopping time in the hidden chasing contest if the chaser has not made her first breakthrough.
- $\bar{x}^F$: The leader's expected starting time in the public chasing contests.

- $$\Gamma(\varphi) \equiv \varphi + \left(\frac{1-\varphi}{\sqrt{1-2\varphi}} - 1\right) e^{-\lambda T},$$

  the upper bound of $\delta$ such that $x^N \geq y^N$ in the equilibrium of the hidden chasing contest.

Also, we introduce the following three characterization results, which are applicable in both the public and the hidden chasing contest.

**Lemma A.1.** *Suppose the leader started before time $t$ and has not stopped yet. Then the leader will not stop until the contest terminates if $\bar{U}_{l\tau} \geq 0$ for all $\tau \geq t$.*

*Proof.* Suppose the leader is working at time $\tau \geq t$ and anticipates that his future selves will not stop until the contest terminates, and then his payoff from stopping at time $\tau$ is 0. Also, the payoff of continuing to work is

$$U^B_{l\tau} = \beta\lambda dt \cdot v - c dt + \beta(1 - \lambda dt - q_t dt)\bar{U}_{i,\tau+dt},$$

where $q_t dt$ is the probability that the chaser wins the contest in the interval $[\tau, \tau + dt)$. If $\bar{U}_{i,\tau+dt} \geq 0$, and since $\beta\lambda v - c \geq 0$ by Assumption 1, then $U^B_{l\tau}$ is greater than 0, which implies that the leader will not stop until $\tilde{T}$. □

Lemma A.1 provides a sufficient working condition for the leader, who will not stop as long as his continuation payoff $\bar{U}_{lt}$ for exerting non-stopping effort until the terminus is non-negative. Lemma A.2 supposes her opponent utilizes an $x$-start strategy, and further presents the contestant's continuation payoff under a particular strategy.

**Lemma A.2.** *Suppose the chaser's first breakthrough arrives at $t_1$. For any time $t$ and $t'$ with $t_1 < t < t'$, suppose $a_{j\tau} = 0$, $\forall \tau \in [t, t')$, and $a_{j\tau} = 1$, $\forall \tau \in [t', \tilde{T}]$, $j \in \{l, c\}$. Then for her opponent $i$ ($i \in \{l, c\}$ and $i \neq j$), the payoff that she works from time $t$ until the contest terminates is given by*

$$\bar{U}_{it} = \left[1 - e^{-\lambda(t'-t)} + e^{-\lambda(t'-t)} \cdot \frac{1}{2}\left(1 - e^{-2\lambda(T-t')}\right)\right] \cdot \left(v - \frac{c}{\lambda}\right). \tag{1}$$

*Proof.* We show the lemma in two steps: (i) $t = t'$ and (ii) $t < t'$.
*Step (i):* If $t = t'$, $i$'s opponent $j$ works from $t$ until the end, and the expected reward of $i$ when she also works from $t$ until the end of the contest is

$$\int_t^T \lambda e^{-2\lambda(\tau-t)} v \, d\tau = \frac{1}{2}\left(1 - e^{-2\lambda(T-t)}\right) v,$$

where $\lambda e^{-2\lambda(\tau-t)}dt$ is the probability that $i$ wins at time $\tau$. Accordingly, $i$'s expected cost from $t$ to $T$ is

$$\int_t^T 2\lambda e^{-2\lambda(\tau-t)}(\tau-t)\cdot c d\tau + e^{-2\lambda(T-t)}(T-t)\cdot c = \frac{1}{2}\left(1-e^{-2\lambda(T-t)}\right)\cdot\frac{c}{\lambda},$$

where $2\lambda e^{-2\lambda(\tau-t)}dt$ is the probability that the chasing contest terminates at time $\tau$, and $e^{-2\lambda(T-t)}$ is the probability that no one wins until the contest terminates at time $T$. And $\bar{U}_{It}$ is the difference between the two preceding expressions.

*Step (ii):* If $t < t'$, contestant $i$ works alone in the interval $(t, t')$. By step (i), we know that $i$'s continuation payoff at time $t'$ by working from $t$ until the end of the contest is

$$U^B_{it'} = \bar{U}_{it'} = \frac{1}{2}\left(1-e^{-2\lambda(T-t')}\right)\left(v-\frac{c}{\lambda}\right).$$

Contestant $i$'s continuation payoff of the strategy above at $t$ is given by the summation of $U^B_{it'}$ (weighted by the probability that the game has not been terminated until $t'$) and the integral of the instantaneous payoff flow from $t$ to $t'$. The expected reward if the game is terminated before $t'$ is given by

$$\int_t^{t'} \lambda e^{-\lambda(\tau-t)}\cdot v d\tau = \left(1-e^{-\lambda(t'-t)}\right)\cdot v,$$

with the corresponding expected cost as

$$\int_t^{t'} \lambda e^{-\lambda(\tau-t)}(\tau-t)\cdot c d\tau + e^{-\lambda(t'-t)}(t'-t)\cdot c = \left(1-e^{-\lambda(t'-t)}\right)\cdot\frac{c}{\lambda}.$$

Accordingly, we have

$$\begin{aligned} U^B_{it} &= \left(1-e^{-\lambda(t'-t)}\right)\cdot(v-\frac{c}{\lambda}) + e^{-\lambda(t'-t)}\frac{1}{2}\left(1-e^{-2\lambda(T-t')}\right)\left(v-\frac{c}{\lambda}\right) \\ &= \left[1-e^{-\lambda(t'-t)} + e^{-\lambda(t'-t)}\cdot\frac{1}{2}\left(1-e^{-2\lambda(T-t')}\right)\right]\cdot\left(v-\frac{c}{\lambda}\right). \end{aligned}$$

□

**Remark 1.** *The preceding two lemmata suffice the emergence of the leader's $x$-start strategy in equilibrium, given that the chaser utilizes her $y$-stop strategy in the chasing contest. To see this, first note that the probability that the chaser wins the contest before achieving her first breakthrough is zero, the leader can thus ignore the chaser's effort (i.e., with a shirking chaser) before entering the contest stage. Denote $F_t$ the distribution for the random time $\tilde{t}_1$ that the chaser's first breakthrough arrives, and $t_1$ the realization of $\tilde{t}_1$. At a time $t < t_1$, the payoff of* *the leader from working until the terminus can be expressed as*

$$\bar{U}_{lt} = \int_t^T \left[1 - e^{-\lambda(z-t)} + e^{-\lambda(z-t)} \cdot \frac{1}{2}\left(1 - e^{-2\lambda(T-z)}\right)\right] \cdot \left(v - \frac{c}{\lambda}\right) \cdot dF_t(z).$$

*Given that (1) is non-negative for any $t_1 < T$, it follows that $\bar{U}_{lt}$ is also non-negative.*[1] *In addition, for the chaser, it is clear that (1) can also represent the chaser's continuation payoff at time $t$ for exerting non-stopping effort in the contest stage, if the leader adopts his $x$-start strategy in the chasing contest.*

The following lemma establishes the structure of the chaser's $y$-stop strategy; that is, the chaser starts at time 0 and exerts effort until an instant $y < \tilde{T}$, if her first breakthrough does not come.

**Lemma A.3.** *Suppose the leader use the following strategy:*

- *if the chaser has made one breakthrough, the leader works at time $t$ if and only if $t \in [x_1, \tilde{T}]$;*
- *if the chaser has made no breakthrough, the leader works at time $t$ if and only if $t \in [\hat{x}_0, \tilde{T}]$, where $x_1 \le \hat{x}_0 < T$.*

*Also, suppose that the chaser will always work if she has made one breakthrough, and there exists $y \in [0, T]$ such that the chaser stops working for the first breakthrough when $t \ge y$. Then the chaser is willing to work if and only if $t \le \min\{\tilde{T}, y\}$.*

*Proof.* Given the structure of the $x$-start and $y$-stop strategy profile, the chaser is willing to work at time $t$ before the arrival of her first breakthrough, if and only if

$$\lambda dt \cdot \bar{U}_{c,t+dt} - cdt + (1 - \lambda dt - \lambda a_{lt} dt) U^A_{c,t+dt} \ge (1 - \lambda a_{lt} dt) U^A_{c,t+dt},$$

which gives us $\lambda(\bar{U}_{ct} - U^A_{ct}) \ge c$. Next, we show that $\lambda(\bar{U}_{ct} - U^A_{ct}) \ge c$ for any $t \in [0, y]$ in the following three cases.

*Case 1*: $y < x_1 \le \hat{x}_0$. In this case, if the contest enters the contest stage at $t$, by Lemma A.2 we have,

$$\bar{U}_{ct} = \left[1 - e^{-\lambda(x_1 - t)}\left(1 - \frac{1}{2}\left(1 - e^{-2\lambda(T - x_1)}\right)\right)\right] \cdot \left(v - \frac{c}{\lambda}\right).$$

This function decreases with $t$. Thus, in the chasing stage, since $y \le \hat{x}_0$, the continuation payoff that the chaser works for the first breakthrough from $t$ to $y$ is given by

$$U^A_{ct} = \int_t^y e^{-\lambda(\tau - t)} \left(\lambda \bar{U}_{c\tau} - c\right) d\tau. \tag{2}$$

[1] Note that if the chaser's first breakthrough arrives at $t'$, i.e., $t_1 = t'$, (1) essentially represents the leader's continuation payoff at time $t$.

Thus,

$$\begin{aligned}\lambda\left(\bar{U}_{ct}-U^A_{ct}\right)-c &= \lambda\bar{U}_{ct}-c-\int_t^y \lambda e^{-\lambda(\tau-t)}\left(\lambda\bar{U}_{c\tau}-c\right)d\tau \\ &\geq \lambda\bar{U}_{ct}-c-\int_t^y \lambda e^{-\lambda(\tau-t)}\left(\lambda\bar{U}_{ct}-c\right)d\tau \\ &= \left(\lambda\bar{U}_{ct}-c\right)e^{-\lambda(y-t)}\geq 0.\end{aligned} \tag{3}$$

*Case 2*: $x_1 \leq y \leq \hat{x}_0$. In this case, if the contest enters the contest stage at $t$, by Lemma A.2,

$$\bar{U}_{ct}=\begin{cases}\left[1-e^{-\lambda(x_1-t)}\left(1-\frac{1}{2}\left(1-e^{-2\lambda(T-x_1)}\right)\right)\right]\cdot\left(v-\frac{c}{\lambda}\right) & t\leq x_1,\\ \frac{1}{2}\left(1-e^{-2\lambda(T-t)}\right)\cdot\left(v-\frac{c}{\lambda}\right) & t>x_1.\end{cases}$$

This function is still decreasing with $t$. Thus, in the chasing stage, since $y\leq\hat{x}_0$, the continuation payoff that the chaser works for the first breakthrough from $t$ to $y$ is still given by (2). Since $\tilde{t}_1$ is still smaller than $\hat{x}_0$ as in Case 1, by (3), $\lambda(\bar{U}_{ct}-U^A_{ct})-c\geq 0$ is guaranteed.

*Case 3:* $x_1 \leq \hat{x}_0 < y$. In this case, if the contest enters the contest stage at $t$, by Lemma A.2, we still have

$$\bar{U}_{ct}=\begin{cases}\left[1-e^{-\lambda(x_1-t)}\left(1-\frac{1}{2}\left(1-e^{-2\lambda(T-x_1)}\right)\right)\right]\cdot\left(v-\frac{c}{\lambda}\right) & t\leq x_1,\\ \frac{1}{2}\left(1-e^{-2\lambda(T-t)}\right)\cdot\left(v-\frac{c}{\lambda}\right) & t>x_1.\end{cases}$$

This function is still decreasing with $t$. Thus, in the chasing stage, if $t>\hat{x}_0$, the continuation payoff that the chaser works for the first breakthrough from $t$ to $y$ is still given by (2), where $\bar{U}_{ct}=\frac{1}{2}\left(1-e^{-2\lambda(T-t)}\right)\cdot\left(v-\frac{c}{\lambda}\right)$. Thus,

$$\begin{aligned}\lambda\left(\bar{U}_{ct}-U^A_{ct}\right)-c &= \lambda\bar{U}_{ct}-c-\int_t^y \lambda e^{-2\lambda(\tau-t)}\left(\lambda\bar{U}_{c\tau}-c\right)d\tau \\ &\geq \lambda\bar{U}_{ct}-c-\int_t^y \lambda e^{-2\lambda(\tau-t)}\left(\lambda\bar{U}_{ct}-c\right)d\tau=\left(\lambda\bar{U}_{ct}-c\right)\left[1-\frac{1}{2}\left(1-e^{-2\lambda(y-t)}\right)\right],\end{aligned}$$

which is non-negative. That is, $\lambda(\bar{U}_{ct}-U^A_{ct})\geq c$ for all $t\geq\hat{x}_0$, and therefore we have $\lambda(\bar{U}_{c\hat{x}_0}-U^A_{c\hat{x}_0})\geq c$.

For $t\leq\hat{x}_0$, the continuation payoff that the chaser works for the first breakthrough from $t$ to $y$ is given by

$$U^A_{ct}=\int_t^{\hat{x}_0} e^{-\lambda(\tau-t)}\left(\lambda\bar{U}_{c\tau}-c\right)d\tau+e^{-\lambda(\hat{x}_0-t)}\int_{\hat{x}_0}^{y} e^{-2\lambda(\tau-\hat{x}_0)}\left(\lambda\bar{U}_{c\tau}-c\right)d\tau.$$

Then

$$\begin{aligned}
&\lambda\left(\bar{U}_{ct}-U_{ct}^{A}\right)-c\\
&=\lambda\bar{U}_{ct}-c-\int_{t}^{\hat{x}_0}e^{-\lambda(\tau-t)}\left(\lambda\bar{U}_{c\tau}-c\right)d\tau-e^{-\lambda(\hat{x}_0-t)}\int_{\hat{x}_0}^{y}e^{-2\lambda(\tau-\hat{x}_0)}\left(\lambda\bar{U}_{c\tau}-c\right)d\tau\\
&\geq\lambda\bar{U}_{ct}-c-\int_{t}^{\hat{x}_0}e^{-\lambda(\tau-t)}\left(\lambda\bar{U}_{ct}-c\right)d\tau-e^{-\lambda(\hat{x}_0-t)}\int_{\hat{x}_0}^{y}e^{-2\lambda(\tau-\hat{x}_0)}\left(\lambda\bar{U}_{ct}-c\right)d\tau\\
&=\left(\lambda\bar{U}_{ct}-c\right)\cdot e^{-\lambda(\hat{x}_0-t)}\left(1-e^{-2\lambda(y-\hat{x}_0)}\right)\geq 0,
\end{aligned}$$

which completes the proof. □

**Remark 2.** *Note that $\hat{x}_0$, $x_1$, and $y$ are treated as exogenous variables given in the derivations of Lemma A.3. We can utilize the corresponding expressions for the public chasing contest, by setting $\hat{x}_0 = x_0^F$, $x_1 = x_1^F$, $y = y^F$. In addition, although there is no real-time disclosure of the chaser's progress in the hidden chasing contest, expressions in the proof of Lemma A.3 are still applicable in our hidden chasing scenario, if we let $\hat{x}_0 = x_1 = x^N$ and $y = y^N$ accordingly.*

## A.1 Proof of Lemma 1

For the leader, suppose he did not start before date $t$. His payoff of the strategy profile $\sigma = \{(\sigma_{ct}, \sigma_{lt})\}_{t\in[0,T]}$ is

$$\beta\lambda a_{lt}dt\cdot(1-\lambda dt)v+\beta\lambda a_{lt}dt\cdot\lambda dt\cdot\frac{v}{2}-ca_{lt}dt+\beta(1-\lambda dt)(1-\lambda a_{lt}dt)U_{l,t+dt}^{B},$$

where $U_{l,t+dt}^{B}$ represents the leader's continuation payoff at $t+dt$ given strategy $\sigma$. Thus, the leader chooses to work instead of shirking at time $t$ if and only if

$$\begin{aligned}
\beta\lambda dt\cdot(1-\lambda dt)\cdot v+\beta\lambda^2(dt)^2\cdot\frac{v}{2}-cdt+\beta(1-\lambda dt)^2\cdot U_{l,t+dt}^{B}&\\
\geq\beta(1-\lambda dt)\cdot U_{l,t+dt}^{B}&.
\end{aligned}$$

Rearrange,

$$\beta\lambda\cdot(1-\lambda dt)\cdot v+\beta\lambda^2 dt\cdot\frac{v}{2}-c\geq\beta\lambda(1-\lambda dt)\cdot U_{l,t+dt}^{B}.$$

Let $dt\to 0$, we have $\beta\lambda v\geq c+\beta\lambda U_{lt}^{B}$.

## A.2 Proof of Lemma 2

If $a_{c\tau}=0$, $\forall\tau\in[t,T]$, by Lemma A.2, the continuation payoff of exerting continuous effort from $t$ to $\tilde{T}$ is given by

$$U_{lt}^{B}=\bar{U}_{lt}=\left(1-e^{-\lambda(T-t)}\right)\left(v-\frac{c}{\lambda}\right),$$

by letting $t' = T$. Substitute into

$$\beta\lambda v \geq c + \beta\lambda U^B_{lt}(\sigma) \tag{4}$$

of Lemma 1 and rearrange, the leader exerts effort at time $t$ if and only if $\left(1 - e^{-\lambda(T-t)}\right) \leq \delta$, where $\delta = \frac{\beta\lambda v - c}{\beta\lambda v - \beta c}$ is denoted in Assumption 1. Thus, we have $t \geq T - \frac{1}{\lambda}\ln\frac{1}{1-\delta}$.

Recall that $\hat{x}_0$ is denoted as the leader's optimal starting time in this one-person contest, which is essentially the earliest instant of time that can make the leader earns a positive payoff from making effort. Thus, let the above inequality hold as equality, we have

$$\hat{x}_0 \equiv \max\left\{0, T - \frac{1}{\lambda}\ln\frac{1}{1-\delta}\right\},$$

where $\hat{x}_0 = 0$ occurs when $T - \frac{1}{\lambda}\ln\frac{1}{1-\delta} < 0$, as we impose no restriction on the choice of $T$ and $\lambda$.

## A.3 Proof of Proposition 1

**Part (i):** Suppose the chaser will always work in the contest stage. Combining the result of Lemma 1 and A.2, if the leader has not worked before $t$, he starts at time $t$ if and only if

$$\beta\lambda v - c \geq \beta\lambda \cdot \frac{1}{2}\left(1 - e^{-2\lambda(T-t)}\right)(v - \frac{c}{\lambda}).$$

Recollect terms, we have $\frac{1}{2}\left(1 - e^{-2\lambda(T-t)}\right) \leq \delta$, where $\delta = \frac{\beta\lambda v - c}{\beta\lambda v - \beta c}$ is denoted in Assumption 1. Since $t \geq 0$, the left-hand side must be smaller than $\frac{1}{2}\left(1 - e^{-2\lambda T}\right)$; therefore, the inequality holds if $\delta \geq \frac{1}{2}\left(1 - e^{-2\lambda T}\right)$. We set $x_1^F = 0$ as the corner solution for $\delta \geq \frac{1}{2}\left(1 - e^{-2\lambda T}\right)$.

When $\delta < \frac{1}{2}\left(1 - e^{-2\lambda T}\right)$, since $\frac{1}{2}\left(1 - e^{-2\lambda(T-t)}\right)$ is monotonically decreasing in $t$ and $t \geq 0$, the leader works at time $t$ if and only if

$$t \geq T - \frac{1}{2\lambda}\ln\frac{1}{1-2\delta}. \tag{5}$$

**Part (ii):** For the chaser, if $t > x_1^F$, as the leader always chooses to work, the chaser faces an identical decision problem of the leader when $\beta = 1$. Thus, the chaser will work (i.e., $a_{ct} = 1$) when $t \geq x_1^F$. When $t < x_1^F$, by Lemma A.2, given that all future selves will work, the chaser is willing to work in $[t, T]$ if and only if

$$\lambda dt \cdot (1 - \lambda dt) \cdot v + \lambda^2 (dt)^2 \cdot v - c dt + (1 - \lambda dt)^2 \bar{U}_{c,t+dt} \geq (1 - \lambda dt) \cdot \bar{U}_{c,t+dt}.$$

Reorganize, we have

$$v - \frac{c}{\lambda} \geq \left[1 - e^{-\lambda(x_1^F - t)} + e^{-\lambda(x_1^F - t)} \cdot \frac{1}{2}\left(1 - e^{-2\lambda(T - x_1^F)}\right)\right] \cdot \left(v - \frac{c}{\lambda}\right),$$

which holds for all $t \in [0, x_1^F)$. Thus, the chaser keeps working from the beginning to the end.

## A.4 Proof of Proposition 2

**Part (i): The chaser.**

The following two lemmata delineate the chaser's best response in the chasing stage given the leader's optimal starting time.

**Lemma 1.** *Suppose $x_1^F > 0$. At time $t < x_1^F$ if the chaser anticipates her future selves will not work in the chasing stage, then the chaser is willing to select $a_{ct} = 1$ if and only if $t < y_L^F$.*

*Proof.* For $t < x_1^F$, by Proposition 1, if the chaser makes her first breakthrough (i.e., enters the contest stage) at the moment, the leader will wait until $x_1^F$ to start working. By Lemma A.2, when the chaser keeps working until the contest terminates, her continuation payoff at time $t$ is given by

$$U_{ct}^B = \bar{U}_{ct} = \left[1 - e^{-\lambda(x_1^F - t)} + e^{-\lambda(x_1^F - t)} \cdot \frac{1}{2}\left(1 - e^{-2\lambda(T - x_1^F)}\right)\right] \cdot \left(v - \frac{c}{\lambda}\right) \\ \left[1 - (1-\delta) e^{-\lambda(x_1^F - t)}\right] \cdot \left(v - \frac{c}{\lambda}\right). \tag{6}$$

By backward induction, in the chasing stage, suppose at time $t$ the chaser's future selves will not work if her current self cannot make the first breakthrough, and then $U_{c,t+dt}^A = 0$. Thus, $a_{ct} = 1$ if and only if

$$\lambda dt \bar{U}_{ct} + (1 - \lambda dt) U_{c,t+dt}^A - c dt \geq U_{c,t+dt}^A.$$

Substituting $\bar{U}_{ct}$ by (6) and reorganizing,

$$1 - e^{-\lambda(x_1^F - t)} + e^{-\lambda(x_1^F - t)} \cdot \frac{1}{2}\left(1 - e^{-2\lambda(T - x_1^F)}\right) \geq \frac{c}{\lambda v - c} = \varphi, \tag{7}$$

which holds if and only if $t = x_1^F - \frac{1}{\lambda} \ln \frac{1-\delta}{1-\varphi} = y_L^F$.

Indeed, if there exists an instant $y^F > 0$ such that the above condition holds if and only if $t \leq y^F$, then $y^F$ can be treated as the exact stopping time in the chasing stage. Otherwise, note that (7) is decreasing with $t$, the inequality cannot hold for all $t > 0$ and therefore the chaser will not work at all. Therefore, when $x_1^F - \frac{1}{\lambda} \ln \frac{1-\delta}{1-\varphi} < 0$, (7) cannot hold for any $t > 0$ by the monotonicity of the left-hand side of (7). Thus, we set $y^F = y_L^F = 0$. Otherwise, we have $y_L^F = x_1^F - \frac{1}{\lambda} \ln \frac{1-\delta}{1-\varphi}$. In sum,

$$y^F = y_L^F = \max\left\{0, x_1^F - \frac{1}{\lambda} \ln \frac{1-\delta}{1-\varphi}\right\}, \tag{8}$$

which completes the proof. $\square$

Note that at $t = x_1^F$,

$$1 - e^{-\lambda(x_1^F - x_1^F)} + e^{-\lambda(x_1^F - x_1^F)} \cdot \frac{1}{2}\left(1 - e^{-2\lambda(T - x_1^F)}\right) = \frac{1}{2}\left(1 - e^{-2\lambda(T - x_1^F)}\right) = \delta,$$

where $\delta = \frac{\beta\lambda v - c}{\beta\lambda v - \beta c}$ is denoted in Assumption 1. Since (6) is decreasing with $t$, we know that $y_L^F < x_1^F$ if and only $\delta < \varphi$. That is, if $\delta < \varphi$, the chaser stops at $y_L^F$ before $x_1^F$, and otherwise the chaser's stopping time must be later than $x_1^F$.

**Lemma 2.** *For any $t > x_1^F$, if the chaser anticipates her future selves will not work in the chasing stage, then she is willing to exert efforts at time $t$, $a_{ct} = 1$, if and only if $t < y_H^F$.*

*Proof.* For $t \geq x_1^F$, the leader will start working immediately once the chaser enters the contest stage. Thus, by Lemma A.2, since both contestants work until the end of the contest, the chaser's continuation payoff at time $t$ is given by

$$U_{ct}^B = \bar{U}_{ct} = \frac{1}{2}\left(1 - e^{-2\lambda(T-t)}\right) \cdot \left(v - \frac{c}{\lambda}\right).$$

Suppose at time $t$ the chaser anticipates that her future selves will stop working if her current self cannot make the first breakthrough, i.e., $U_{c,t+dt}^A = 0$. The chaser chooses $a_{ct} = 1$ if and only if

$$\begin{aligned}\lambda dt U_{ct}^B - c dt + (1 - \lambda dt) U_{c,t+dt}^A &= \lambda dt \bar{U}_{ct} - c dt + (1 - \lambda dt) U_{c,t+dt}^A \\ &= \left[\frac{1}{2}\left(1 - e^{-2\lambda(T-t)}\right) \cdot (\lambda v - c) - c\right] \cdot dt \geq U_{c,t+dt}^A = 0.\end{aligned}$$

Reorganize, we have $\frac{1}{2}\left(1 - e^{-2\lambda(T-t)}\right) \geq \varphi$, where $\varphi = \frac{c}{\lambda v - c}$ is denoted in Assumption 1. Note that for any $t \geq 0$, the left-hand side is smaller than $\frac{1}{2}\left(1 - e^{-2\lambda T}\right)$. Thus, when $\varphi \geq \frac{1}{2}\left(1 - e^{-2\lambda T}\right)$, the preceding inequality cannot never hold, and we set $y_H^F = 0$ as the corner solution when $\varphi \geq \frac{1}{2}\left(1 - e^{-2\lambda T}\right)$. When $\varphi < \frac{1}{2}\left(1 - e^{-2\lambda T}\right)$, since $\frac{1}{2}\left(1 - e^{-2\lambda(T-t)}\right)$ is monotonically decreasing in $t$ and $t \geq 0$, the leader works at time $t$ if and only if

$$t \geq T - \frac{1}{2\lambda} \ln \frac{1}{1 - 2\varphi}. \tag{9}$$

This completes the proof. $\square$

Note that $\frac{1}{2}\left(1 - e^{-2\lambda(T - x_1^F)}\right) = \delta$, where $\delta = \frac{\beta\lambda v - c}{\beta\lambda v - \beta c}$ is denoted in Assumption 1, and $x_1^F = T - \frac{1}{2\lambda} \ln \frac{1}{1-2\delta}$ is given by Proposition 1. Since $\frac{1}{2}\left(1 - e^{-2\lambda(T-t)}\right)$ is decreasing with $t$, $y_H^F \geq x_1^F$ if and only if $\delta \geq \varphi$. Summarizing the results of Lemma 1 and 2, the chaser stops at $y_L^F$ when $\delta \leq \varphi$ and stops at $y_H^F$ otherwise.

**Part (ii): The leader**

We start with the following lemma which characterizes the leader's starting time when the chaser stops before $\hat{x}_0$.

**Lemma 3.** *Suppose the chaser stops working for the first breakthrough at $y < \hat{x}_0$, the leader's starting time in the chasing stage is $\hat{x}_0$.*

*Proof.* By Lemma A.2, the continuation payoff at time $t$ if the leader starts at $t$ until the contest terminates is

$$\bar{U}_{it} = \left[1 - e^{-\lambda(y-t)} + e^{-\lambda(y-t)} \cdot \frac{1}{2}\left(1 - e^{-2\lambda(T-y)}\right)\right] \cdot \left(v - \frac{c}{\lambda}\right),$$

if $t < y$, and is

$$\bar{U}_{it} = \left(1 - e^{-\lambda(T-t)}\right) \cdot \left(v - \frac{c}{\lambda}\right)$$

if $t \geq y$. Both expressions are always non-negative by Assumption 1. Thus, by Lemma A.1, the leader will keep working once he starts. By Lemma 1, his starting time is given by

$$\inf\left\{t \geq 0 \middle| \beta\lambda v - c \geq \beta\bar{U}_{lt}\right\}.$$

Since $\bar{U}_{it}$ is decreasing with $t$ and $\beta\lambda v - c = \beta\bar{U}_{l,\hat{x}_0}$, the starting time is exactly $\hat{x}_0$. □

Next, we consider the case when the chaser stops after $\hat{x}_0$, where $\delta \geq 1 - \sqrt{1-2\varphi}$. We first introduce Lemma 4 which specifies the leader's continuation payoff $U^B_{lt}$, and verifies its monotonicity by Lemma 5. Recall that our characterization of $y^F_H$ from Lemma 2 is valid when $\delta \geq \varphi$, we then derive the leader's unique optimal starting time $x^F_0 = x^F_L$ as his best response to $y^F_H$.

**Lemma 4.** *When $y^F \geq \hat{x}_0$, the continuation payoff of the leader if he works from the $t$ ($t \leq y^F$) to the end of the contest is*

$$\bar{U}_{lt} = \frac{1}{4}\left[3 - 2\lambda(y^F - t)e^{-2\lambda(T-t)} + e^{-2\lambda(y^F-t)} - 4e^{-\lambda(T-2t+y^F)}\right] \cdot \left(v - \frac{c}{\lambda}\right). \tag{10}$$

*Proof.* If the leader starts working at a time before $y^F$, then at time $y^F$ there are three possible outcomes:

- *Case 1:* The leader makes his first breakthrough before the chaser stops, and he wins the game.

- *Case 2:* The chaser makes her first breakthrough before the leader stops, and the game enters the contest stage.

- *Case 3:* No one makes any breakthrough, and the chaser quits at $y^F$.

We begin with the case that the chaser has not made the first breakthrough at time $t > x_1^F$, where $x_1^F$ is the leader's earliest starting time once the chaser has made her first breakthrough. We show that there exists $x_L^F$ at which the leader starts in the chasing stage. Then we verify that $x_L^F \geq x_1^F$. In order to derive $x_L^F$, we first calculate the probability that the leader wins the contest if he works continuously from $t$ to the end of the contest conditional on the contest not been terminated at time $t$ in the above three cases, which are denoted by $\pi_1(t)$, $\pi_2(t)$ and $\pi_3(t)$, respectively.

*Case 1*: We have

$$\pi_1(t) = \int_t^{y^F} \lambda e^{-2\lambda(\tau - t)} d\tau = \frac{1}{2}\left(1 - e^{-2\lambda(y^F - t)}\right).$$

The corresponding expected cost in Case 1 is

$$C_1(t) = \int_t^{y^F} \lambda e^{-2\lambda(\tau - t)} \cdot (\tau - t) \cdot c d\tau = \left(1 - e^{-2\lambda(y^F - t)} - 2\lambda(y^F - t)e^{-2\lambda(y^F - t)}\right) \cdot \frac{c}{4\lambda}.$$

*Case 2*: Note that the probability that the chaser makes her first breakthrough in the interval $[t, y^F)$ at time $t$ is also $\lambda e^{-2\lambda(\tau - t)} dt$. Then, if $\tau \geq x_1^F$, both contestants work from $\tau$ to $\tilde{T}$, and the leader wins with probability $\frac{1}{2} e^{-2\lambda(T - \tau)}$. Thus,

$$\pi_2(t) = \int_t^{y^F} \lambda e^{-2\lambda(\tau - t)} \cdot \frac{1}{2}\left(1 - e^{-2\lambda(T - \tau)}\right) d\tau = \frac{1}{4}\left(1 - e^{-2\lambda(y^F - t)} - 2\lambda(y^F - t)e^{-2\lambda(T - t)}\right).$$

Also, when the game enters the contest stage at time $\tau > t$, the expected cost in the future is $\frac{1}{2\lambda}(1 - e^{-2\lambda(T - \tau)}) \cdot c$, while the cost before entering the chasing stage is $(\tau - t) \cdot c$. Thus, the expected cost in Case 2 is

$$\begin{aligned} C_2(t) &= \int_t^{y^F} \lambda e^{-2\lambda(\tau - t)} \left[\frac{1}{2\lambda}\left(1 - e^{-2\lambda}(T - \tau)\right) + \tau - t\right] d\tau \cdot c \\ &= \frac{1}{2\lambda}\left[1 - \lambda(y^F - t)e^{-2\lambda(T - t)} - e^{-2\lambda(y^F - t)}\left(1 + \lambda(y^F - t)\right)\right] \cdot c. \end{aligned}$$

*Case 3*: The probability that the leader wins conditional on the chaser quits at time $y^F$ is

$$\pi_3(t) = e^{-2\lambda(y^F - t)} \cdot \left(1 - e^{-\lambda(T - y^F)}\right).$$

The costs before and after $y^F$ are $(y^F - t) \cdot c$ and $\left(1 - e^{-\lambda(T - y^F)}\right) \cdot \frac{c}{\lambda}$, respectively. The ex-

pected cost in Case 3 is

$$C_3(t) = e^{-2\lambda(y^F - t)} \cdot \left[ \frac{1}{\lambda}\left(1 - e^{-\lambda(T-y^F)}\right) + y^F - t \right] \cdot c.$$

Combing the above three cases together, the total probability at time $t$ that the leader wins is

$$\pi(t) = \pi_1(t) + \pi_2(t) + \pi_3(t) = \frac{1}{4}\left[3 - 2\lambda(y^F - t)e^{-2\lambda(T-t)} + e^{-2\lambda(y^F-t)} - 4e^{-\lambda(T-2t+y^F)}\right]. \quad (11)$$

Accordingly, the total expected cost of working from $t$ to the end of the chasing stage is

$$C(t) = C_1(t) + C_2(t) + C_3(t) = \frac{c}{4\lambda} \cdot \left[3 - 2\lambda(y^F - t)e^{-2\lambda(T-t)} + e^{-2\lambda(y^F-t)} - 4e^{-\lambda(T-2t+y^F)}\right].$$

Thus, the payoff of working from $t$ to the end of the chasing stage is the difference of $v\pi(t)$ and $C(t)$, which is given by (10). □

Write $\pi(t)$ as a function of $x$, the leader's starting time, and $y$, the chaser's stopping time; that is,

$$\Pi(x, y) = \frac{1}{4}\left[3 - 2\lambda(y - x)e^{-2\lambda(T-x)} + e^{-2\lambda(y-x)} - 4e^{-\lambda(T-2x+y)}\right]. \quad (12)$$

**Lemma 5.** *For any $\delta \in (0,1)$, suppose $y \geq x_H^F$, then equation $\Pi(x, y) = \delta$ has at most one solution $\tilde{x}$ in the interval $[0, y]$. Also, if such a solution exists, then*

$$\left.\frac{\partial \Pi(x, y)}{\partial x}\right|_{x=\tilde{x}} < 0.$$

*Proof.* The first partial derivative of $\Pi(x, y)$ with respect to $x$ is given by

$$\frac{\partial \Pi(x, y)}{\partial x} = \frac{\lambda e^{-2\lambda(T-x+y)}}{2}\left(-4e^{\lambda(T+y)} + e^{2\lambda T} + e^{2\lambda y}(2\lambda(x - y) + 1)\right),$$

which is positive if and only if

$$x \geq y + \frac{4e^{\lambda(T-y)} - e^{2\lambda(T-y)} - 1}{2\lambda} \equiv x^*. \quad (13)$$

Thus, in the interval $[0, y]$, function $\Pi$ can be decreasing, increasing or U-shaped with respect to $x$, with no other possibilities. Given that $y \geq x_H^F$ by assumption, we have

$$1 - e^{-\lambda(T-y)} \leq 1 - e^{-\lambda(T-x_H^F)} = \delta.$$

Since in the interval $[x^*, y)$, $\partial\Pi(x,y)/\partial x > 0$, we have

$$\Pi(x,y) \le \Pi(y,y) = 1 - e^{-\lambda(T-y)} \le \delta,$$

which indicates that there is no solution to equation $\Pi(x,y) = \delta$ in $[x^*, y)$. □

Define

$$\begin{cases} \pi_0'(t) = 1 - e^{-\lambda(T-t)} \\ \pi_0(t) = \frac{1}{4}\left[3 - 2\lambda(y^F - t)e^{-2\lambda(T-t)} + e^{-2\lambda(y^F-t)} - 4e^{-\lambda(T-2t+y^F)}\right] \\ \pi_0''(t) = \frac{1}{2}\left(1 - e^{-2\lambda(T-t)}\right). \end{cases}$$

Here, $\pi_0'(\cdot)$ and $\pi_0''(\cdot)$ are both decreasing functions with $\pi_0'(t) \ge \pi_0''(t)$ for all $t$. Note that

$$\pi_0'(t) = \int_t^{y^F} \lambda e^{-2\lambda(\tau-t)} d\tau + \int_t^{y^F} \lambda e^{-2\lambda(\tau-t)} \cdot \left(1 - e^{-\lambda(T-\tau)}\right) d\tau + e^{-2\lambda(y^F-t)} \cdot \left(1 - e^{-\lambda(T-y^F)}\right),$$

and,

$$\pi_0(t) = \int_t^{y^F} \lambda e^{-2\lambda(\tau-t)} d\tau + \int_t^{y^F} \lambda e^{-2\lambda(\tau-t)} \cdot \frac{1}{2}\left(1 - e^{-2\lambda(T-\tau)}\right) d\tau + e^{-2\lambda(y^F-t)} \cdot \left(1 - e^{-\lambda(T-y^F)}\right).$$

Also,

$$\pi_0''(t) = \int_t^{y^F} \lambda e^{-2\lambda(\tau-t)} d\tau + \int_t^{y^F} \lambda e^{-2\lambda(\tau-t)} \cdot 0 d\tau + e^{-2\lambda(y^F-t)} \cdot \frac{1}{2}\left(1 - e^{-2\lambda(T-y^F)}\right).$$

Thus, for any $t$ we have $\pi_0'(t) \ge \pi_0(t) \ge \pi_0''(t)$. Since $\pi_0''(t) \ge 0$, (10), which is equal to $\pi_0(t)(v - c/\lambda)$, is positive for all $t$. By Lemma A.1, the leader will not stop once he starts.

It remains to specify the optimal starting time. By Lemma 1, the leader is willing to start at $t$ if and only if

$$\beta\lambda v - c \ge \beta\lambda\left[\frac{1}{4}\left[3 - 2\lambda(y^F - t)e^{-2\lambda(T-t)} + e^{-2\lambda(y^F-t)} - 4e^{-\lambda(T-2t+y^F)}\right] \cdot \left(v - \frac{c}{\lambda}\right)\right].$$

Reorganize, we have

$$\frac{1}{4}\left[3 - 2\lambda(y^F - t)e^{-2\lambda(T-t)} + e^{-2\lambda(y^F-t)} - 4e^{-\lambda(T-2t+y^F)}\right] \le \delta, \tag{14}$$

where $\delta = \frac{\beta\lambda v - c}{\beta\lambda v - \beta c}$ is denoted in Assumption 1. When $t = y^F$, the left-hand side of the above inequality is

$$1 - e^{-\lambda(T-y^F)} < 1 - e^{-\lambda(T-\hat{x}_0)} = \delta.$$

where the equality $1-e^{-\lambda(T-\hat{x}_0)}=\delta$ is given by Lemma 2. By the intermediate value theorem, the time threshold to ensure (14) holds as an equality, i.e.,

$$\frac{1}{4}\left[3-2\lambda(y^F-t)e^{-2\lambda(T-t)}+e^{-2\lambda(y^F-t)}-4e^{-\lambda(T-2t+y^F)}\right]=\delta, \tag{15}$$

is strictly positive, if

$$\frac{1}{4}\left[3-2\lambda y^F\cdot e^{-2\lambda T}+e^{-2\lambda y^F}-4e^{-\lambda(T+y^F)}\right]>\delta.$$

Substituting $y^F$ by the expression of $y_H^F$ of (9), the condition above is equivalent to

$$3-\left(4\sqrt{\frac{1}{1-2\varphi}}+2\lambda T-\frac{1}{1-2\varphi}-\ln\frac{1}{1-2\varphi}\right)e^{-2\lambda T}>4\delta.$$

Let $x_L^F$ be the solution of (15) with $t=x_L^F$ and $y^F=y_H^F$. By the monotonicity of (10) proved in Lemma 5, the solution is unique.

In the last step, we verify that $x_L^F\in[x_1^F,x_H^F]$. Observe that $x_H^F$, $x_L^F$ and $x_1^F$ solve $\pi_0'(t)=\delta$, $\pi_0(t)=\delta$ and $\pi_0''(t)=\delta$, respectively. Thus, $\pi_0'(x_L^F)\geq\pi_0(x_L^F)=\delta\geq\pi_0''(x_L^F)$, where the equality comes from (15). Consequently, since $\pi_0'$ is decreasing, $x_H^F\geq x_L^F$. In a similar vein, since $\pi_0''$ is decreasing, $x_L^F\geq x_1^F$.

## A.5 Proof of Corollary 1

**Part (i):** Note that we enter the one-person contest when $t>y^F$. Thus, the leader will start from $\hat{x}_0$, and it is clear that $BR_l^F(y)$ remains unchanged with respect to $y$ when $y<\hat{x}_0$.

When $y>\hat{x}_0$, note first that the partial derivative of $\Pi(x,y)$ in (12) with respect to $y$ is

$$\frac{\partial\Pi}{\partial y}=-\frac{1}{2}\lambda\left(e^{\lambda T}-e^{\lambda y}\right)^2e^{-2\lambda(T-x+y)}<0.$$

By the proof of Proposition 2, we know that $\pi_0''(x)\leq\Pi(x,y)\leq\pi_0'(x)$ for any $y\in[\hat{x}_0,T]$. Since $x_1^F$, $\hat{x}_0$ and $BR_l^F(y)$ are given by the solutions to equations $\pi_0''(x)=\delta$, $\pi_0'(x)=\delta$ and $\Pi(x,y)=\delta$, respectively, $BR_l^F(y)$ exists and $BR_l^F(y)\in[x_1^F,\hat{x}_0]$. Also, Lemma 5 has shown that $BR_l^F(y)$ is given by the intersection of $\Pi(BR_l^F(y),y)=\delta$, and at the intersection $\Pi(\cdot,y)$ must be decreasing. Applying implicit differentiation, we have

$$\frac{\partial\Pi(x,y)}{\partial y}+\frac{\partial\Pi(x,y)}{\partial x}\cdot\frac{dBR_l^F(y)}{dy}=0,$$

which indicates that $\frac{dBR_l^F(y)}{dy}<0$.

**Part (ii):** For the chaser, if she anticipates that her future selves will stop, she is willing to keep working if and only if

$$\lambda U_{ct}^{B} dt - c dt = \lambda \bar{U}_{ct} dt - c dt \geq 0.$$

Recall (7) of Lemma 1, it is clear that $BR_c^F(x_0, x_1)$ is irrelevant to $x_0$.

In terms of $x_1$, by (7) the chaser is unwilling to stop at time $x_1$ if and only if

$$\lambda \cdot \frac{1}{2}\left(1 - e^{-2\lambda(T-x_1)}\right) \cdot \left(v - \frac{c}{\lambda}\right) \geq c.$$

Solving for $x_1$, we have

$$x_1 \geq T - \frac{1}{2\lambda} \ln \frac{\lambda v - c}{\lambda v - 3c} = y_H^F.$$

By Lemma 2, we know that if $x_1 < y_H^F$, the chaser will stop at $y_H^F$. Thus, $BR_c^F(x_0, x_1)$ is constant with respect to any $x_1 \leq y_H^F$.

Suppose now $x_1 > y_H^F$, such that the chaser is unwilling to stop at $x_1$. Then by (7),

$$1 - e^{-\lambda(x_1 - BR_c^F(x_1))} + e^{-\lambda(x_1 - BR_c^F(x_1))} \cdot \frac{1}{2}\left(1 - e^{-2\lambda(T-x_1)}\right) = \varphi,$$

where we relabel $BR_c^F(x_0, x_1)$ as $BR_c^F(x_1)$. Since

$$\frac{\partial}{\partial x}\left[1 - e^{-\lambda(x-y)} + e^{-\lambda(x-y)} \cdot \frac{1}{2}\left(1 - e^{-2\lambda(T-x)}\right)\right] = \frac{\lambda}{2}\left(e^{2\lambda(T-x)} - 1\right) e^{\lambda(-2T+x+y)} > 0,$$

a larger $x_1$ makes the left-hand side of inequality (7) larger. Therefore, as a response, the chaser will have a later stopping time to balance the equation, since

$$\frac{\partial}{\partial y}\left[1 - e^{-\lambda(x-y)} + e^{-\lambda(x-y)} \cdot \frac{1}{2}\left(1 - e^{-2\lambda(T-x)}\right)\right] = -\frac{\lambda}{2}\left(1 - e^{-2\lambda(T-x)}\right) e^{\lambda(y-x)} < 0.$$

## A.6 Proof of Proposition 3

We first introduce the following lemma to specify that the chaser will exert non-stopping efforts once she has made one breakthrough in the hidden chasing contest.

**Lemma 6.** *In the hidden chasing contest, if at time $t$ the chaser has made the first breakthrough while the contest is not terminated, then $a_{ct} = 1$, $\forall t \in (0, T]$.*

*Proof.* The proof is similar to the proof of part (ii) of Proposition 1. We only need to replace $x_1^F$ by $x^N$ to complete the proof. □

Then we show that $(x^N, y^N) = (x_H^F, 0)$ is an equilibrium when $\varphi > 1 - \frac{1-\delta+0.5\delta^2}{1-\delta} e^{-\lambda T}$. First, note that if $y^N = 0$ is the chaser's stopping time, it indicates that the chaser will not work at all. The leader's starting time is $x_H^F = \hat{x}_0$, as this game is isomorphic to the one-person contest as illustrated by Lemma 2. On the other hand, given the leader starts at time $x^N = x_H^F$, if the chaser anticipates that her future selves will not work, she is willing to work at time $t$ if and only if

$$\lambda dt \cdot \bar{U}_{c,t+dt} + (1-\lambda dt) U^A_{c,t+dt} - c dt \geq U^A_{c,t+dt}.$$

Since $U^A_{c,t+dt} = 0$, the condition holds at time $t = 0$ if and only if $\lambda \cdot U^B_{c0} = \lambda \cdot \bar{U}_{c0} \geq c$. By (1) of Lemma A.2, $\bar{U}_{c0}$ is given by

$$\begin{aligned}\bar{U}_{c0} &= \left[1 - e^{-\lambda x_H^F} + e^{-\lambda x_H^F} \cdot \frac{1}{2}\left(1 - e^{-2\lambda(T - x_H^F)}\right)\right] \cdot \left(v - \frac{c}{\lambda}\right) \\ &= \left(1 - \frac{1-\delta+0.5\delta^2}{1-\delta} e^{-\lambda T}\right) \cdot \left(v - \frac{c}{\lambda}\right).\end{aligned}$$

Thus, shirking is optimal for the chaser if and only if

$$\varphi > 1 - \frac{1-\delta+0.5\delta^2}{1-\delta} e^{-\lambda T}.$$

Suppose $1 - \frac{1-\delta+0.5\delta^2}{1-\delta} e^{-\lambda T} \geq \varphi$. We characterize the equilibrium in four steps:

*Step 1: Characterize the chaser's best response given the leader's starting time.*

Suppose the leader starts at some $x^N$. By (1), since in the contest stage, the chaser will work continuously to the end, the chaser's continuation payoff at time $t < x^N$, if her first breakthrough has arrived before, is

$$U^B_{ct} = \bar{U}_{ct} = \left[1 - e^{-\lambda(x^N - t)} + e^{-\lambda(x^N - t)} \cdot \frac{1}{2}\left(1 - e^{-2\lambda(T - x^N)}\right)\right] \cdot \left(v - \frac{c}{\lambda}\right).$$

This function decreases with $t$. Thus, the chaser is willing to work at $t$ if she anticipates her future selves will not work for the first breakthrough if and only if $\lambda \bar{U}_{ct} \geq c$. Therefore, we can derive $y^N$ by

$$\begin{aligned}y^N &= \sup_{t \leq x^N} \left\{\lambda \cdot \left[1 - e^{-\lambda(x^N - t)} + e^{-\lambda(x^N - t)} \cdot \frac{1}{2}\left(1 - e^{-2\lambda(T - x^N)}\right)\right] \cdot \left(v - \frac{c}{\lambda}\right) \geq c\right\} \\ &= \sup_{t \leq x^N} \left\{1 - e^{-\lambda(x^N - t)} + e^{-\lambda(x^N - t)} \cdot \frac{1}{2}\left(1 - e^{-2\lambda(T - x^N)}\right) \geq \varphi\right\}.\end{aligned} \tag{16}$$

To guarantee $y^N \leq x^N$, the chaser must be willing to stop when the leader is ready to start; that is, $\frac{1}{2}\left(1 - e^{-2\lambda(T - x^N)}\right) < \varphi$, which implies

$$x^N > T - \frac{1}{2\lambda} \ln \frac{1}{1 - 2\varphi}. \tag{17}$$

If so, $y^N$ is determined by

$$e^{-\lambda y^N} = e^{-\lambda x^N} \cdot \frac{1 - \frac{1}{2}(1 - e^{-2\lambda(T - x^N)})}{1 - \varphi}. \tag{18}$$

*Step 2: Characterize the leader's best response given the chaser's stopping time.*

Suppose that the chaser stops at some $y^N > 0$, and then by the proofs of Lemma 2 and Lemma A.2, the leader's continuation payoff of exerting continuous effort from time $t \geq y^N$ until the contest terminates is

$$\bar{U}_{lt} = \left[ \left(1 - e^{-\lambda y^N}\right) \cdot \frac{1}{2} \left(1 - e^{-2\lambda(T-t)}\right) + e^{-\lambda y^N} \cdot \left(1 - e^{-\lambda(T-t)}\right) \right] \cdot \left(v - \frac{c}{\lambda}\right).$$

where the first item in the square bracket is the probability that the chaser has made the first breakthrough at time $t$, while the second item is the corresponding probability that the chaser has made no breakthrough at time $t$. Note that $\bar{U}_{lt} \geq 0$ for all $t < T$, thus, the leader is willing to keep working until the end once he has started, which gives us $\bar{U}_{lt} = U^B_{lt}$. Also, the leader is willing to start at time $t$ if and only if

$$\lambda dt v - c dt + (1 - \lambda dt) U^B_{l,t+dt} \geq U^B_{l,t+dt},$$

which yields

$$\left(1 - e^{-\lambda y^N}\right) \cdot \frac{1}{2} \left(1 - e^{-2\lambda(T-t)}\right) + e^{-\lambda y^N} \cdot \left(1 - e^{-\lambda(T-t)}\right) \leq \delta. \tag{19}$$

Thus, the leader's starting time is given by

$$x^N = \inf_{t \geq y^N} \left\{ \left(1 - e^{-\lambda y^N}\right) \cdot \frac{1}{2} \left(1 - e^{-2\lambda(T-t)}\right) + e^{-\lambda y^N} \cdot \left(1 - e^{-\lambda(T-t)}\right) \leq \delta \right\}.$$

Also, to guarantee that the leader is not willing to work at time $y^N$, we need

$$\left(1 - e^{-\lambda y^N}\right) \cdot \frac{1}{2} \left(1 - e^{-2\lambda(T-y^N)}\right) + e^{-\lambda y^N} \cdot \left(1 - e^{-\lambda(T-y^N)}\right) > \delta. \tag{20}$$

If (20) holds, the leader's starting time is given by the solution of (19) when it holds as equality. That is

$$\left(1 - e^{-\lambda y^N}\right) \cdot \frac{1}{2} \left(1 - e^{-2\lambda(T-t)}\right) + e^{-\lambda y^N} \cdot \left(1 - e^{-\lambda(T-t)}\right) = \delta, \tag{21}$$

which gives

$$e^{-\lambda y^N} = \frac{2\delta - 1 + e^{-2\lambda(T-x^N)}}{(1 - e^{-\lambda(T-x^N)})^2}. \tag{22}$$

Combine equation (18) and (22) we get

$$e^{\lambda x_H^N} = \frac{(1-e^{-\lambda(T-x_H^N)})^2}{2(1-\varphi)} \cdot \frac{1-\frac{1}{2}\left(1-e^{-2\lambda(T-x_H^N)}\right)}{\delta-\frac{1}{2}\left(1-e^{-2\lambda(T-x_H^N)}\right)} \tag{23}$$

of Proposition 3.

*Step 3: Show that the two best-response functions have only one intersection under the restriction that $x^N \geq y^N$, and thus the unique equilibrium exists.*

First, define

$$h_c(x) \equiv e^{-\lambda x} \cdot \frac{1-\frac{1}{2}(1-e^{-2\lambda(T-x)})}{1-\varphi} \text{ and } h_l(x) \equiv \frac{2\delta-1+e^{-2\lambda(T-x)}}{(1-e^{-\lambda(T-x)})^2}. \tag{24}$$

If the best responses of equations (18) and (22) form an equilibrium, the corresponding equilibrium strategy profile $(x^N, y^N)$ must be the solution of the equations $e^{-\lambda y^N} = h_c(x^N) = h_l(x^N)$. Function $h_c(x^N)$ decreases with $x^N$ since

$$h_c'(x^N) = -\frac{\lambda e^{-\lambda x^N}(1-e^{-2\lambda(T-x^N)})}{2(1-\varphi)} < 0; \tag{25}$$

also,

$$h_l'(x^N) = \frac{2\lambda e^{2\lambda(T-x^N)}}{(e^{\lambda(T-x^N)}-1)^3} \cdot \left(e^{-\lambda(T-x^N)} - (1-2\delta)\right), \tag{26}$$

which is positive if $x^N \geq T - \frac{1}{\lambda}\ln\frac{1}{1-2\delta}$. Also, note that the right-hand side of (18) is always non-negative. Therefore, the intersection of $h_c(x^N)$ and $h_l(x^N)$, if exists, must yield $y^N \geq 0$.

By (22), $y^N \geq 0$ implies $h_c(x^N) \geq 0$, which gives us $x^N \geq T - \frac{1}{2\lambda}\ln\frac{1}{1-2\delta}$.

We then derive the condition under which (17) holds. Observe that the right-hand side of (18) and (22) are decreasing and increasing, respectively, and $x^N$ is their intersection. Thus, $x^N > T - \frac{1}{2\lambda}\ln\frac{1}{1-2\varphi}$, if and only if the value of (18) is greater than that of (22). That is, it suffices to show that

$$\left.\frac{2\delta-1+e^{-2\lambda(T-x)}}{(1-e^{-\lambda(T-x)})^2}\right|_{x=T-\frac{1}{2\lambda}\ln\frac{1}{1-2\varphi}} < \left.\left[e^{-\lambda x^N} \cdot \frac{1-\frac{1}{2}(1-e^{-2\lambda(T-x^N)})}{1-\varphi}\right]\right|_{x=T-\frac{1}{2\lambda}\ln\frac{1}{1-2\varphi}}. \tag{27}$$

Reorganize the terms, we have

$$\frac{2(\delta-\varphi)}{(1-\sqrt{1-2\varphi})^2} < \frac{e^{-\lambda T}}{\sqrt{1-2\varphi}},$$

which implies $\delta < \Gamma(\varphi)$.

Next, we show that (20) holds when (17) holds. First, we show that (20) is equivalent to $y^N < x^N$. To see this, define

$$h(x,y) = (1-e^{-\lambda y}) \cdot \frac{1}{2}\left(1-e^{-2\lambda(T-x)}\right) + e^{-\lambda y} \cdot \left(1-e^{-\lambda(T-x)}\right), \tag{28}$$

which is decreasing with $x$. By (22), we can see that $h(x^N, y^N) = \delta$. Thus, statements $h(y^N, y^N) > \delta$ and $y^N < x^N$ are equivalent. Then it suffices to show $y^N < x^N$ when (17) holds, and we prove it by contradiction. Suppose instead that (17) holds but $y^N > x^N$, then by (18), we must have

$$\frac{1-\frac{1}{2}(1-e^{-2\lambda(T-x^N)})}{2(1-\varphi)} < 1,$$

which implies $x^N < T - \frac{1}{2\lambda}\ln\frac{1}{1-2\varphi}$. This contradicts with (17).

Last, the uniqueness is guaranteed by both $h_c(x)$ and $h_l(x)$ are strictly monotone.

## A.7 Proof of Proposition 4

Suppose the leader employs an $x^N$−start strategy, at any $t > x^N$, if the chaser anticipates that her future selves will not work for the first breakthrough (i.e., $a_{c\tau} = 0$ for all $\tau \in [t+dt, T]$ if $\tilde{t}_1 \notin [0,t]$), then she works in the time interval $[t, t+dt]$ if and only if

$$\lambda dt \cdot U^B_{c,t+dt} + (1-\lambda dt)U^A_{c,t+dt} - cdt = \lambda dt \cdot \bar{U}_{c,t+dt} + (1-\lambda dt)U^A_{c,t+dt} - cdt \geq U^A_{c,t+dt},$$

which gives us $\lambda \bar{U}_{ct} \geq c$.

Since $t > x^N$, the leader has already begun to work and will not stop. Thus, since the chaser will not stop either, the contest is equivalent to the contest stage in the public chasing contest. As a consequence, by Lemma A.2, $\bar{U}_{ct} = \frac{1}{2}(1-e^{-2\lambda(T-t)})(v-\frac{c}{\lambda})$, and thus $a_{ct} = 1$ if and only if $t \leq y^N_H = T - \frac{1}{2\lambda}\ln\frac{1}{1-2\varphi}$.

Conversely, suppose the chaser utilizes a $y^N$-stop strategy, for any $t < y^N$, by Lemma A.2 and Lemma 4,

$$\begin{aligned}\bar{U}_{lt} = \Bigg[&(1-e^{-\lambda t}) \cdot \frac{1}{2}\left(1-e^{-2\lambda(T-t)}\right) \\ &+ e^{-\lambda t}\left[\frac{3}{4}\left(1-e^{-2\lambda(y^N-t)}\right) - \frac{1}{2}\lambda(y^N-t)e^{-2\lambda(T-t)} + e^{-2\lambda(y^N-t)}\left(1-e^{-\lambda(T-y^N)}\right)\right]\Bigg] \cdot \left(v-\frac{c}{\lambda}\right).\end{aligned}$$

Since we still have $\bar{U}_{lt} \geq 0$ here, the $x$-start structure can be guaranteed by Lemma A.1. If the leader anticipates that his future selves will work from instant $t$ to the end of the contest, if

and only if

$$\lambda dt \cdot v - cdt + (1-\lambda dt)U^B_{l,t+dt} = \lambda dt \cdot v - cdt + (1-\lambda dt)\bar{U}_{l,t+dt} \geq U^B_{l,t+dt} = \bar{U}_{l,t+dt}.$$

Reorganizing the above expression gives us $U^B_{l,t+dt} \leq \delta$, where $\delta = \frac{\beta\lambda v - c}{\beta\lambda v - \beta c}$ is denoted in Assumption 1. Solving the inequality gives the equality

$$\begin{aligned}\delta = &(1-e^{-\lambda x^N_L})\cdot\frac{1}{2}\left(1-e^{-2\lambda(T-x^N_L)}\right)\\ &+e^{-\lambda x^N_L}\left[\frac{3}{4}\left(1-e^{-2\lambda(y^N_H-x^N_L)}\right)-\frac{1}{2}\lambda(y^N_H-x^N_L)e^{-2\lambda(T-x^N_L)}+e^{-2\lambda(y^N_H-x^N_L)}\left(1-e^{-\lambda(T-y^N_H)}\right)\right]\end{aligned} \tag{29}$$

of Proposition 4.

It remains to show that (29) is decreasing with $t$. Here, (29) is equivalent to

$$\left(1-e^{-\lambda(T-t)}\right)\cdot\frac{1}{2}\left(1-e^{-2\lambda(T-t)}\right)+e^{-\lambda(T-t)}\cdot\Pi(t,y).$$

The first order derivative with respect to $t$ is given by

$$\frac{\lambda e^{-(\lambda(3T-3t+y))}}{4}\cdot\left(\underbrace{-e^{\lambda(3T-4t+y)}\left(1-e^{-2(y-t)}\right)}_{<0}+\underbrace{4e^{\lambda(T-2t+y)}\left(1-e^{\lambda(T-y)}-e^{\lambda t}\right)}_{<0}+\underbrace{2\lambda(t-y)e^{\lambda(T-2t+y)}}_{<0}\right),$$

which is indeed negative.

Now it is sufficient to show that the equilibrium guarantees $y^N \geq x^N$. Since $y^N = T - \frac{1}{2\lambda}\ln\frac{1}{1-2\varphi}$ is independent of $x^N$, it is sufficient to show that the leader's best response to $y^N$ is not greater than $y^N$. We prove by contradiction. If not, by (22), there exists an instant $x \geq y^N$ such that for all $t \in (y^N, x)$,

$$(1-e^{-\lambda y^N})\cdot\frac{1}{2}\left(1-e^{-2\lambda(T-t)}\right)+e^{-\lambda y^N}\cdot\left(1-e^{-\lambda(T-t)}\right)\geq\delta. \tag{30}$$

Since the left-hand side of (30) is decreasing with $t$, we have

$$(1-e^{-\lambda y^N})\cdot\frac{1}{2}\left(1-e^{-2\lambda(T-y^N)}\right)+e^{-\lambda y^N}\cdot\left(1-e^{-\lambda(T-y^N)}\right)\geq\delta.$$

Substitute $y^N$ by $y^N_H$ and reorganize the above expression, we have $\delta \leq \Gamma(\varphi)$. Thus, when $\delta \geq \Gamma(\varphi)$, the best response to $y^N = T - \frac{1}{2\lambda}\ln\frac{1}{1-2\varphi}$ is less than $y^N$.

## A.8 Proof of Corollary 2

When $\delta < \Gamma(\varphi)$, by equation (16), $x^N$ is given by equation

$$(1-e^{-\lambda y^N})\cdot\frac{1}{2}\left(1-e^{-2\lambda(T-x^N)}\right)+e^{-\lambda y^N}\cdot\left(1-e^{-\lambda(T-x^N)}\right)=\delta.$$

Also, since $\Gamma(\varphi)<1-\sqrt{1-2\varphi}$ when $\delta<\Gamma(\varphi)$, we have $\delta<1-\sqrt{1-2\varphi}$. By Proposition 2, we know $x_0^F=x_H^F$; also, $x_1^F$ and $x_H^F$ are determined by the following two equations

$$\delta=\frac{1}{2}\left(1-e^{-2\lambda(T-x_1^F)}\right) \text{ and } \delta=1-e^{-2\lambda(T-x_H^F)}.$$

Thus, By

$$x_1^F\equiv\max\left\{0,T-\frac{1}{2\lambda}\ln\frac{1}{1-2\delta}\right\}$$

of Proposition 1 and (21),

$$\begin{aligned}\delta=\frac{1}{2}\left(1-e^{-\lambda(T-x_1^F)}\right)&=(1-e^{-\lambda y^N})\cdot\frac{1}{2}\left(1-e^{-2\lambda(T-x^N)}\right)+e^{-\lambda y^N}\cdot\left(1-e^{-\lambda(T-x^N)}\right)\\&\geq(1-e^{-\lambda y^N})\cdot\frac{1}{2}\left(1-e^{-2\lambda(T-x^N)}\right)+e^{-\lambda y^N}\cdot\frac{1}{2}\left(1-e^{-2\lambda(T-x^N)}\right)\\&=\frac{1}{2}\left(1-e^{-2\lambda(T-x^N)}\right),\end{aligned}$$

which implies $x^N\geq x_1^F$. Similarly, by

$$t\geq\hat{x}_0:=\max\left\{0,T-\frac{1}{\lambda}\ln\frac{1}{1-\delta}\right\}$$

of Lemma 2 and (21),

$$\begin{aligned}\delta=\left(1-e^{-\lambda(T-x_0^F)}\right)&=(1-e^{-\lambda y^N})\cdot\frac{1}{2}\left(1-e^{-2\lambda(T-x^N)}\right)+e^{-\lambda y^N}\cdot\left(1-e^{-\lambda(T-x^N)}\right)\\&\leq(1-e^{-\lambda y^N})\cdot\left(1-e^{-\lambda(T-x^N)}\right)+e^{-\lambda y^N}\cdot\left(1-e^{-\lambda(T-x^N)}\right)\\&=1-e^{-\lambda(T-x^N)},\end{aligned}$$

which implies that $x^N\leq x_0^F$.

When $\delta\geq\Gamma(\varphi)$, note that

- $x_1^F$ is the solution of $\frac{1}{2}(1-e^{-2\lambda(T-t)})=\delta$,
- $x_L^F$ is the solution of $\frac{1}{4}(3-2\lambda(y_H^F-t)e^{-2\lambda(T-t)})+e^{-2\lambda(y_H^F-t)}-4e^{-\lambda(T-2t+y^F)}=\delta$,
- $x_L^N$ is a linear combination of $x_1^F$ and $x_L^F$, when it also equals to $\delta$.

Since $x_1^F$ and $x_L^F$ decrease with $t$, the solution to the linear combination $x_L^N$ should also be located between the solutions of the first two equations.

## A.9 Proof of Corollary 3

**Part (i):** First, we consider $BR_l^N(y^N)$. By (20), the leader is willing to work at time $y^N$ if and only if

$$h(y^N, y^N) = (1 - e^{-\lambda y^N}) \cdot \frac{1}{2}\left(1 - e^{-2\lambda(T-y^N)}\right) + e^{-\lambda y^N} \cdot \left(1 - e^{-\lambda(T-y^N)}\right) \leq \delta, \tag{31}$$

where function $h(\cdot,\cdot)$ is defined by equation (28) in the Proof of Proposition 3. Define $H(y) = h(y, y)$, and then

$$H'(y) = -\frac{1}{2}\lambda \cdot e^{-\lambda(2T-y)}\left(e^{2\lambda(T-y)} + 2e^{\lambda y} - 1\right) < 0. \tag{32}$$

Let $\bar{y}_N$ be the unique solution to $H(y) = \delta$, thus, inequality (31) holds if and only if $y \geq \bar{y}^N$.

If (31) holds, define

$$\hat{\Pi}(x, y) = \left(1 - e^{-\lambda(T-x)}\right) \cdot \frac{1}{2}\left(1 - e^{-2\lambda(T-x)}\right) + e^{-\lambda(T-x)} \cdot \Pi(x, y),$$

and the proof of Proposition 4 has shown that $\hat{\Pi}(x, y)$ is decreasing with $x$. Also, by the proof of Corollary 1,

$$\frac{\partial\hat{\Pi}(x, y)}{\partial y} = e^{-\lambda(T-x)}\frac{\partial\Pi(x, y)}{\partial y} < 0.$$

Since $BR_l^N(y)$ is given by equation $\hat{\Pi}(x, y) = \delta$, applying implicit differentiation, we have

$$\frac{\partial\hat{\Pi}(x, y)}{\partial y} + \frac{\partial\hat{\Pi}(x, y)}{\partial x} \cdot \frac{dBR_l^N(y)}{dy} = 0,$$

which indicates that $\frac{dBR_l^N(y)}{dy} < 0$.

If (31) does not hold, the leader is not willing to work at $y^N$, and therefore the starting time is greater than $y^N$. By (22), the starting time is given by

$$e^{-\lambda y^N} = \frac{2\delta - 1 + e^{-2\lambda(T-BR_l^N(y^N))}}{(1 - e^{-\lambda(T-BR_l^N(y^N))})^2} = h_l(BR_l^N(y^N)). \tag{33}$$

By (26), we have $h_l'(x) > 0$ when $x > x_1^F$, and then it suffices to show that $BR_l^N(y^N) > x_1^F$

whenever (33) holds. Observe

$$
\begin{aligned}
h(x_1^F, x_1^F) =& (1-e^{-\lambda x_1^F}) \cdot \frac{1}{2}\left(1-e^{-2\lambda(T-x_1^F)}\right) + e^{-\lambda x_1^F} \cdot \left(1-e^{-\lambda(T-x_1^F)}\right) \\
=& (1-e^{-\lambda x_1^F})\delta + e^{-\lambda x_1^F} \cdot \left(1-e^{-\lambda(T-x_1^F)}\right) > \delta,
\end{aligned}
$$

and hence the leader is not willing to work at $x_1^F$, i.e., we have $BR_l^N(y^N) > x_1^F$. Since $h_l'(x) > 0$, while $e^{-\lambda y^N}$ decreases with $y^N$, to balance the equation (33), a greater $y^N$ yields a smaller $BR_l^N(y^N)$.

**Part (ii):** Next, consider $BR_c^N(x^N)$. If $x^N < y_H^N$, by the proof of Proposition 4, the chaser is unwilling to stop at $x^N$ and her stopping time is exactly $y_H^N$, which is constant with respect to $x^N$. If $x^N > y_H^N$, by (18), the chaser's stopping time is determined by

$$
e^{-\lambda BR_c^N(x^N)} = e^{-\lambda x^N} \cdot \frac{1-\frac{1}{2}(1-e^{-2\lambda(T-x^N)})}{1-\varphi} = h_c(x^N).
$$

Note that we have shown $h_c'(x^N) < 0$ in (25). Thus, to balance the equation above, $BR_c^N(x^N)$ must be increasing with respect to $x^N$ when $x^N > y_H^N$.

### A.10 Proof of Proposition 5

**Part (i):** Proposition 4 has shown that $x^N < y^N = y^F = \max\left\{0, T - \frac{1}{2\lambda}\ln\frac{1}{1-2\varphi}\right\}$ when $\delta \geq \Gamma(\varphi)$.

**Part (ii):** When $\delta < \Gamma(\varphi)$, recall Proposition 3 in which the equilibrium $(x^N, y^N)$ is given by the combination of (18) and (22), the left-hand side of which are both $e^{-\lambda y^N}$. The difference of the two right-hand sides is given by $h_c(x^N) - h_l(x^N)$ defined in (24). Taking their first-order derivatives, we have

$$
h_c'(x^N) - h_l'(x^N) = \frac{2\lambda e^{-\lambda(T-x^N)}}{(1-e^{-\lambda(T-x^N)})^3}\left((1-2\delta) - e^{-\lambda(T-x^N)}\right) - \frac{\lambda e^{-\lambda x^N}}{1-\varphi} \cdot \left(1 - \frac{1}{2}(1-e^{-2\lambda(T-x^N)})^2\right).
$$

Next, we show that $h_c'(x^N) - h_l'(x^N) < 0$, which can be guaranteed if $1-2\delta < e^{-\lambda(T-x^N)}$. We have shown $x^N > x_1^F$ in Corollary 2, and thus

$$
e^{-\lambda(T-x^N)} > e^{-2\lambda(T-x^N)} > e^{-2\lambda(T-x_1^F)} = 1-2\delta,
$$

which implies that $h_c'(x^N) - h_l'(x^N)$ is negative. Note that the difference equals 0 when $x^N = x_H^N$, and thus $h_c(x)$ and $h_l(x)$ are single-crossing, as illustrated in Figure 1.

Conversely, for any $y$, denote $h_c^{-1}(y)$ and $h_l^{-1}(y)$ the solutions to equations (18) and (22), respectively. Then $y > y^N$ if and only if $h_l^{-1}(y) < h_c^{-1}(y)$. Following the same arguments, we know that $y \leq y^N$ if and only if $h_l^{-1}(y) \geq h_c^{-1}(y)$. Or equivalently $y \geq h_l(h_c^{-1}(y))$.

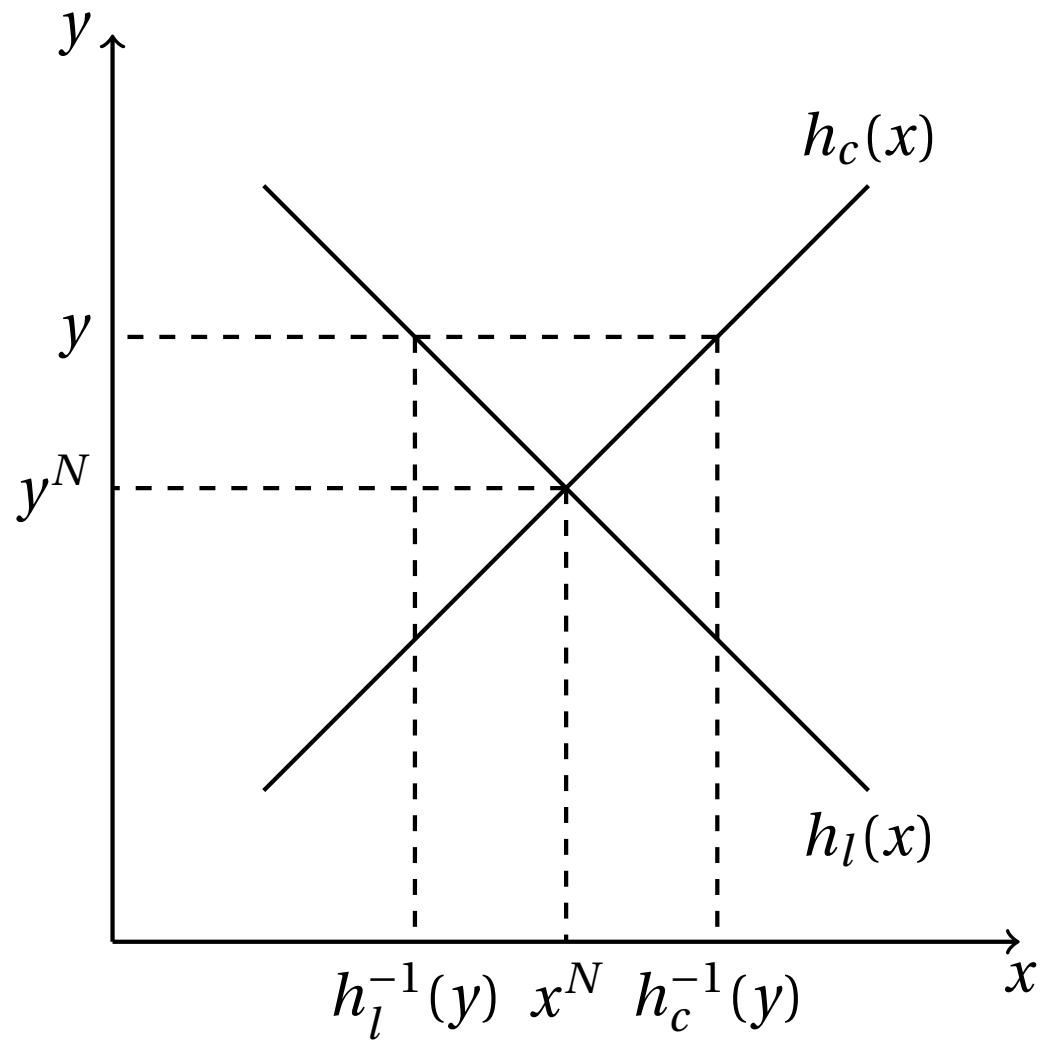

Figure 1

When $\delta < \varphi$, $y^F = y_L^F$. Substitute $y_L^F = T - \frac{1}{2\lambda}\ln\frac{1}{1-2\delta} - \frac{1}{\lambda}\ln\frac{1-\delta}{1-\varphi}$ to (18), and then we have $h_c^{-1}(y) = x_1^F$. However, by solving equation (22) when $x = x_1^F$, we have $h_l(x_1^F) = 0 < h_c(x_1^F)$. Thus, $e^{-\lambda y_L^F} > e^{-\lambda y^N}$, which implies $y_L^F < y^N$.

When $\delta \geq \varphi$, $y^F = y_H^F$. Substitute $y_H^F = T - \frac{1}{2\lambda}\ln\frac{1}{1-2\varphi}$ to (18) and we have $h_c^{-1}(y_H^F) = y_H^F$. However, by solving equation (22) when $x = y_H^F$, we have

$$h_l\left(h_c^{-1}(y_H^F)\right) = h_l(y_H^F) = \frac{1}{\lambda}\ln\left[\frac{(1-\sqrt{1-2\varphi})^2}{2(\delta-\varphi)}\right].$$

Therefore, $y_H^F > y^N$ if and only if $h_c(h_c^{-1}(y_H^F)) = h_c(y_H^F) > h_l(h_c^{-1}(y_H^F)) = h_l(y_H^F)$. Substitute the corresponding expressions for $h_c(y_H^F)$ and $h_l(y_H^F)$, and reorganize. Then $y_H^F \geq y^N$ if and only if

$$T - \frac{1}{2\lambda}\ln\frac{1}{1-2\varphi} > \frac{1}{\lambda}\ln\left[\frac{(1-\sqrt{1-2\varphi})^2}{2(\delta-\varphi)}\right]$$

which is equivalent to $\delta < \Gamma(\varphi)$. This is impossible since we have assumed the opposite.

## A.11 Proof of Proposition 6

It is straightforward that there is no competition effect when $\delta \geq \Gamma(\varphi)$ since Proposition 5 shows that in this case $y^N = y^F$.

When $\delta < \Gamma(\varphi)$, we have $y^F < y^N$ by Proposition 5. Recall function $H(y) = h(y, y)$ defined in the proof of Corollary 3, and we have shown that $H'(y) < 0$ by (32). Also, we know that $BR_l^N(y)$ decreases when $H(y) > \delta$. Observe that $H(T) = 0$ and $H(0) = 1 - e^{-\lambda T}$, and since $\hat{x}_0 > 0$ by assumption, we have $H(0) > \delta$. Therefore, there exists $\bar{y}^N \in (0, T)$ such that $h(\bar{y}^N) =$ $\delta$ by the intermediate value theorem. Thus, it suffices to show that $y^N < \bar{y}^N$, which, by Corollary 3, indicates that $y^F < y^N < \bar{y}^N$ and therefore $BR_l^N(y^F) > BR_l^N(y^N)$. In fact, $y^N < \bar{y}^N$ is guaranteed by the proof of Proposition 3, in which inequality (20) is verified.

## A.12 Proof of Proposition 7

**Part (i):** Recall Proposition 3, the chaser will work until time $y^N$when the inequality (16) is satisfied with an equal sign. Thus, for any $y < x_1^F$, in the hidden chasing contest, the leader's best response is given by

$$(1-e^{-\lambda y})\cdot\frac{1}{2}\left(1-e^{-2\lambda(T-t)}\right)+e^{-\lambda y}\cdot\left(1-e^{-\lambda(T-t)}\right)=\delta.$$

Solving the preceding quadratic function, we have

$$BR_l^N(y)=T-\frac{1}{\lambda}\ln\frac{1-e^{-\lambda y}}{\sqrt{1-2\delta(1-e^{-\lambda y})}-e^{-\lambda y}}. \tag{34}$$

Note that $BR_l^N(y)\to x_1^F$ when $y\to\infty$, and when $y\to 0$, the L'Hospital's rule indicates

$$BR_l^N(0)=T-\frac{1}{\lambda}\ln\left[\lim_{y\to 0}\frac{\lambda e^{-\lambda y}}{\lambda e^{-\lambda y}-\frac{\delta\lambda e^{-\lambda y}}{\sqrt{1-2\delta(1-e^{-\lambda y})}}}\right]=T-\frac{1}{\lambda}\ln\frac{1}{1-\delta}=x_H^F.$$

When $\delta<\varphi$, we have $x_0^F=x_H^F$ in Case (i). Meanwhile, $\bar{x}^F=e^{-\lambda y}x_H^F+(1-e^{-\lambda y})x_1^F$ is a linear function of $e^{-\lambda y}$, and coincides with $BR_l^N(y)$ when $e^{-\lambda y}=0$ and 1. Thus, to show $BR_l^N(y^F)>\bar{x}^F$, it is sufficient to show that the function

$$f(v)=\ln\frac{1-v}{\sqrt{1-2\delta(1-v)}-v}$$

is decreasing and convex. We first derive its first-order derivative

$$f'(v)=\frac{\delta(1-v)-(1-\sqrt{1-2\delta(1-v)})}{(1-v)\sqrt{1-2\delta(1-v)}(\sqrt{1-2\delta(1-v)}-v)},$$

which is negative if and only if $\delta(1-v)-(1-\sqrt{1-2\delta(1-v)})<0$. The partial derivative of the nominator of $f'(v)$ with respect to $\delta$ is

$$(1-v)\left(1-\frac{1}{\sqrt{1-2\delta(1-v)}}\right)<0,$$

which implies that $\delta(1-v)-(1-\sqrt{1-2\delta(1-v)})<0$. Since $f'(v)=0$ when $\delta=0$, $f'(v)<0$.

The second-order derivative of $f(v)$ is given by

$$f''(v) = -\frac{1}{(1-v)^2} + \frac{\left(\frac{\delta}{\sqrt{1-2\delta(1-v)}}\right)^2}{(\sqrt{1-2\delta(1-v)}-v)^2} + \frac{\delta^2}{\sqrt{1-2\delta(1-v)}^3(\sqrt{1-2\delta(1-v)}-v)}$$
$$= \frac{(2\delta+\zeta-1)^4}{8\delta\zeta^3(\zeta-v)^2(1+v-2\delta)^2}\cdot[4(1-\delta)\zeta+6\delta v-8\delta+4],$$

where $\zeta = \sqrt{1-2\delta(1-v)}$. The value is positive, if and only if $4(1-\delta)\zeta+6\delta v-8\delta+4>0$, which is guaranteed by the assumption that $\delta < \frac{1}{2}\left(1-e^{-2\lambda T}\right)$.

**Part (ii):** Note that there are three different combinations of the optimal starting and stopping time in the public and hidden chasing contests when $\delta > \varphi$, i.e.,

(a) when $\varphi < \delta \le \Gamma(\varphi)$, we have $(x^F, y^F) = (x^F_H, y^F_H)$, and $(x^N, y^N) = (x^N_H, 0)$ or $(x^N_H, y^N_L)$;

(b) when $\Gamma(\varphi) \le \delta \le 1-\sqrt{1-2\varphi}$, we have $(x^F, y^F) = (x^F_H, y^F_H)$, and $(x^N, y^N) = (x^N_L, y^N_H)$;

(c) when $\delta \ge 1-\sqrt{1-2\varphi}$, we have $(x^F, y^F) = (x^F_L, y^F_H)$, and $(x^N, y^N) = (x^N_L, y^N_H)$.

We employ the following two numerical examples at the boundary of Case (a) and (b), and Case (b) and (c), respectively, to show that there exists a vector $(\lambda, T, \delta, \varphi)$ and $(\lambda', T', \delta', \varphi')$, both of which guarantees $x^F_1 > 0$, such that the beep effect is positive given the realization of $(\lambda, T, \delta, \varphi)$ but is negative given $(\lambda', T', \delta', \varphi')$.

**Example 1.** Consider first the case that $\varphi \le \delta \le \Gamma(\varphi)$. When $\delta \ge \varphi$, we have $y^F = y^F_H = \max\left\{0, T - \frac{1}{2\lambda}\ln\frac{1}{1-2\varphi}\right\}$ by Proposition 2. Substituting this to the objective function

$$\bar{x}^F = \int_0^{x^F_1} \lambda e^{-\lambda t} x^F_1\, dt + \int_{x^F_1}^{y^F_H} \lambda e^{-\lambda t} t\, dt + e^{-\lambda y^F_H} x^F_H = x^F_1 + \frac{1}{\lambda}\left(e^{-\lambda x^F_1} - e^{-\lambda y^F_H}\right) + e^{-\lambda y^F}(x^F_H - y^F_H), \tag{35}$$

we have

$$T - \frac{1}{\lambda}\left[\frac{1}{2}\ln\frac{1}{1-2\delta} + \left(\sqrt{\frac{1}{1-2\varphi}} - \sqrt{\frac{1}{1-2\delta}}\right)e^{-\lambda T} + \frac{1}{1-2\varphi}\left(\ln\frac{1}{1-\delta} - \ln\sqrt{\frac{1}{1-2\varphi}}\right)\right].$$

Accordingly, substitute the expression of $y^F_H$ for the leader's best response in the hidden chasing contest of (34), we have

$$T - \frac{1}{\lambda}\ln\left[\frac{1-\sqrt{\frac{1}{1-2\varphi}}e^{-\lambda T}}{\sqrt{1-2\delta(1-\sqrt{\frac{1}{1-2\varphi}}e^{-\lambda T})} - \sqrt{\frac{1}{1-2\varphi}}e^{-\lambda T}}\right].$$

The difference is proportional to

$$\left[\ln\sqrt{\frac{1}{1-2\delta}}+\left(\sqrt{\frac{1}{1-2\varphi}}-\sqrt{\frac{1}{1-2\delta}}\right)e^{-\lambda T}+\frac{1}{1-2\varphi}\left(\ln\frac{1}{1-\delta}-\ln\sqrt{\frac{1}{1-2\varphi}}\right)\right]$$
$$-\ln\left[\frac{1-\sqrt{\frac{1}{1-2\varphi}}e^{-\lambda T}}{\sqrt{1-2\delta(1-\sqrt{\frac{1}{1-2\varphi}}e^{-\lambda T})}-\sqrt{\frac{1}{1-2\varphi}}e^{-\lambda T}}\right],$$

which is positive if and only if the beep effect is positive. Without loss of generality, we let $\lambda = 1$.[2]

We directly consider the case that $\delta = \Gamma(\varphi)$ at the boundary of Case (a) and (b). By Proposition 3 and Proposition 4, we have $y^N = x^N$. Then the difference is given by

$$-2\sqrt{\frac{1}{1-2\left(\left(\frac{1-\varphi}{\sqrt{1-2\varphi}}-1\right)e^{-T}+\varphi\right)}}+e^T\ln\left(\frac{1}{1-2\left(\left(\frac{1-\varphi}{\sqrt{1-2\varphi}}-1\right)e^{-T}+\varphi\right)}\right) \tag{36}$$
$$-2e^T\ln\left(\frac{\sqrt{\frac{1}{1-2\varphi}}-e^T}{\sqrt{\frac{1}{1-2\varphi}}-e^T\sqrt{2\sqrt{\frac{1}{1-2\varphi}}e^{-T}\left(\left(\frac{1-\varphi}{\sqrt{1-2\varphi}}-1\right)e^{-T}+\varphi\right)-2\left(\left(\frac{1-\varphi}{\sqrt{1-2\varphi}}-1\right)e^{-T}+\varphi\right)+1}}\right)$$
$$+2\sqrt{\frac{1}{1-2\varphi}}\ln\left(\frac{1}{1-\varphi\left(1-\frac{1-\varphi}{\sqrt{1-2\varphi}}\right)e^{-T}}\right)+2\sqrt{\frac{1}{1-2\varphi}}-\sqrt{\frac{1}{1-2\varphi}}\ln\left(\frac{1}{1-2\varphi}\right).$$

This is a function that only depends on the realization of $\varphi$ and $T$. Set $T = 0.5$. If $\varphi = 0.2$, the difference is 0.00469775, which indicates the beep effect is positive. Here, $x_1^F > 0$ because

$$\delta-\frac{1}{2}\left(1-e^{-2T}\right)=\varphi+\left(\frac{1-\varphi}{\sqrt{1-2\varphi}}-1\right)e^{-T}-\frac{1}{2}\left(1-e^{-2T}\right)=-0.0961688<0.$$

However, if $\varphi = 0.269$, the difference is $-0.00128828$ but $\delta-\frac{1}{2}\left(1-e^{-2T}\right)=-0.00128828$ is still negative. That is, the direction of the beep effect is reverted as $\varphi$ grows. Since $\delta = \Gamma(\varphi)$ is at the boundary of the two cases and the objective functions are continuous here, this creates a counter-example for both cases such that the beep effect can be either positive or negative.

**Example 2.** Now we turn to the case that $\delta \geq 1-\sqrt{1-2\varphi}$. In a similar vein, we consider directly the special case that $\delta = 1-\sqrt{1-2\varphi}$, which is at the boundary of Case (b) and (c). Substitute into the expression of $\hat{x}_0$ in Lemma 2, we have $\hat{x}_0 = y_H^F$. Also, since $x_L^F$ is the solution of (15), we have $x_L^F = \hat{x}_0 = T - \frac{1}{\lambda}\ln\frac{1}{1-\delta}$ when $\delta = 1-\sqrt{1-2\varphi}$. The objective function

$$\bar{x}^F = \int_0^{x_1^F} \lambda e^{-\lambda t} x_1^F dt + \int_{x_1^F}^{x_L^F} \lambda e^{-\lambda t} t dt + e^{-\lambda x_L^F} = x_1^F + \frac{1}{\lambda}\left(e^{-\lambda x_1^F} - e^{-\lambda x_L^F}\right)$$

can be written as

$$\begin{aligned}
\bar{x}^F =& x_1^F + \frac{1}{\lambda}\left(e^{-\lambda x_1^F} - e^{-\lambda \hat{x}_0}\right) \\
=& T - \frac{1}{\lambda}\left[\ln\sqrt{\frac{1}{2\sqrt{1-2\varphi}-1}} - e^{-\lambda T}\left(\sqrt{\frac{1}{2\sqrt{1-2\varphi}-1}} - \sqrt{\frac{1}{1-2\varphi}}\right)\right].
\end{aligned}$$

[2]This treatment is without loss of generality, because in the expression above every occurrence of $\lambda$ appears in the form of $\lambda T$.

Set $T = 0.5$ and $\lambda = 1$, and we will have $\varphi < 0.266113$ to guarantee $x_1^F > 0$. Therefore, when $\varphi = 0.2$, we have $\bar{x}^F = 0.235766$, while if $\varphi = 0.26$, $\bar{x}^F = 0.124824$.

For hidden chasing contest, when $y_H^N = T - \frac{1}{2\lambda}\ln\frac{1}{1-2\varphi}$ and $\delta = 1-\sqrt{1-2\varphi}$, equation (29) becomes

$$\begin{aligned}
&4(1-\sqrt{1-2\varphi}) = 2\left(1-e^{\lambda x^N}\right)\left(1-e^{-2\lambda(T-x^N)}\right) \\
&+ e^{\lambda x^N}\left(-4\sqrt{\frac{1}{1-2\varphi}}e^{-2\lambda(T-x^N)} + \frac{e^{-2\lambda(T-x^N)}}{1-2\varphi} + e^{-2\lambda(T-x^N)}\left(\ln\left(\frac{1}{1-2\varphi}\right) + -2\lambda(T-x^N)\right) + 3\right).
\end{aligned}$$

Set $T = 0.5$ and $\lambda = 1$, then the solution is $x^N = 0.237079$ when $\varphi = 0.2$, and is $x^N = 0.124168$ when $\varphi = 0.26$. Thus, $x^N > \bar{x}^F$ when $\varphi = 0.2$ and is $x^N < \bar{x}^F$ when $\varphi = 0.26$.

# Appendix B Additional Results

## B.1 Present Bias as Corporate Short-termism

Organizational short-termism (and *organizational myopia*) have been widely discussed in the literature on corporate behavior.[3] One key insight is that a short-term revenue-based evaluation (e.g., stock market pressure) could lead managers (agent) to overweight short-term earnings at the expense of investors (principal)'s long-term interests. Even if the *long-run self* of a manager is fully rational and time-consistent, she may still exhibit features of *short-run selves* who prioritize short-term interests. This "dual selves" conflict between short-run and long-run preferences allows the existing literature to view the individual as an organization (see, Thaler and Shefrin (1981), Fudenberg and Levine (2006), Brocas and Carrillo (2008), among others). Therefore, we can metaphorically view the two contestants in our chasing contest as two competing organizations, such as research labs or high-tech startups.

[3]See, Stein (1988), Stein (1989), Bolton et al. (2006), Aghion and Stein (2008), and Gryglewicz et al. (2020), for the discussion of organizational short-termism; and, Levinthal and March (1993), Miller (1994), and Sheppard and Chowdhury (2005), for the discussion of organizational myopia.

In the rest of this subsection, we present a mathematically equivalent interpretation between a firm's board of directors who focus on either long-term concerns (on R&D investment) or short-term revenue priorities, and the leader with *instantaneous-gratification* time preference in the chasing contest.

Suppose that the contestants are two firms. Each firm needs to allocate one unit of time resource between two tasks: one is on marketing strategies to boost short-term revenues, the other is to engage in the long-term R&D investments. Assume that attaining a technical breakthrough through R&D investment ahead of its competitor ensures a lasting competition advantage, thus eliminating the need for for future marketing efforts. Denote $a \in [0,1]$ the share of time resource allocated to the R&D contest, and normalize the instantaneous cost of exerting efforts to 0 for convenience. The breakthrough generates a prize $v$ for the winner, which is realized with probability $\lambda a dt$, while allocating time resource to marketing effort yields an instantaneous reward $z(1-a)dt$, where $z > 0$ represents the board member's marginal reward for short-term objectives. Suppose that the time allocation for each period is determined by two board members of each firm: one focuses on short-term sales performance and the other dedicates to the firm's long-term growth. Therefore, the payoff function for investing on the short-term marketing strategies is $z(1-a)dt$, and

$$\lambda a dt \cdot v + (1-a)zdt + (1-\lambda a dt)\cdot U_t,$$

for investing on the the R&D contest with its competing firm, where $U_t$ is the continuation payoff after time $t$. If the choice of $a$ is Paretian, we have

$$a \in \arg\max_{a' \in [0,1]} (1-\gamma)za'dt + \gamma\Big(\lambda a' dt \cdot v + (1-a')zdt + (1-\lambda a' dt)\cdot U_t\Big).$$

where $\gamma \in [0,1]$ represents the weight for the time allocation choice. The corresponding first order condition implies that $a = 1$, if and only if $\gamma\lambda v - z - \gamma\lambda U_t \geq 0$. Compared to the incentive compatibility condition of our chasing contest (expression (4) of Lemma 1), we can analogously interpret the effort cost $c$ as the board member's marginal reward from short-term objective (i.e., the opportunity cost for investing on long-term objective), and the present-biased leader's discounting factor for future rewards $\beta$ as the bargaining power of the board member with short-term concerns. This provides us a mathematical analogy between a procrastinated incumbent (a human) and a short-termistic firm (an organization).

## B.2 Numerical Results

### B.2.1 Beep Effect When $\delta \geq \varphi$

When $\delta \geq \varphi$, the beep effect is not unidirectional across the set of parameters with non-trivial values. We conduct two representative numerical examples to identify the potential pattern of the beep effect. In short, our numerical analyses reveal that the beep effect remains positive for any possible combinations of $\delta$ and $\varphi$, when $\lambda T$ is sufficiently large. Since the Poisson rate $\lambda$ is fixed, we treat $\lambda T$ as the approximate representation of the contest deadline. Without loss of generality, we normalize $\lambda = 1$ in the following two numerical examples.

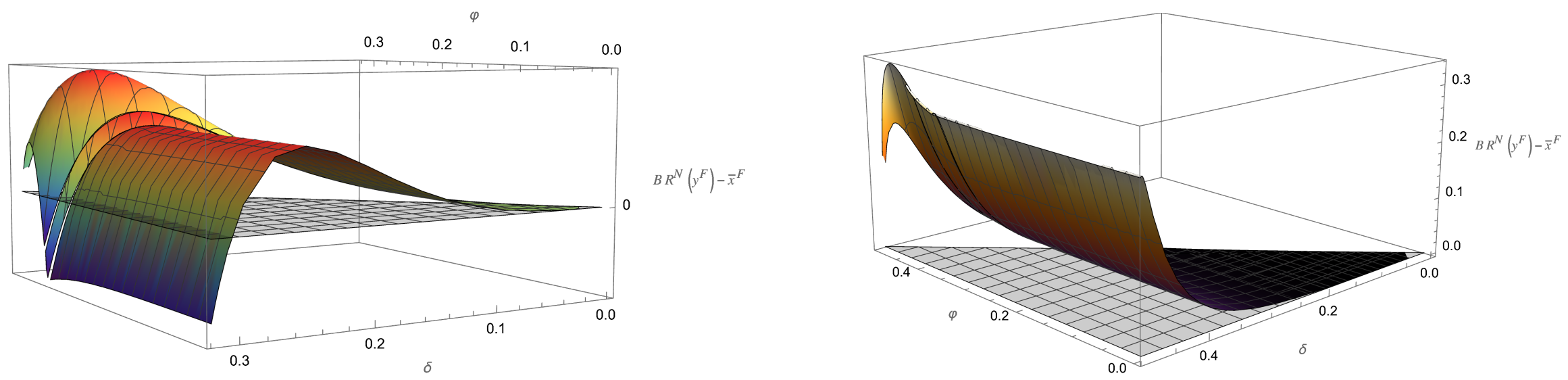

(a) $\lambda T = 0.5$ (b) $\lambda T = 2$

Figure 2: Beep effect when $\delta \geq \varphi$

Panel (a) of Figure 2 plots the beep effect when $\lambda T = 0.5$ and $\delta \geq \varphi$, which includes the three cases that: (i) $\varphi \leq \delta < \varphi + (\frac{1-\varphi}{\sqrt{1-2\varphi}} - 1)e^{-\lambda T}$, (ii) $\varphi + (\frac{1-\varphi}{\sqrt{1-2\varphi}} - 1)e^{-\lambda T} \leq \delta < 1 - \sqrt{1-2\varphi}$, and (iii) $\delta \geq 1 - \sqrt{1-2\varphi}$. Its vertical coordinate is the beep effect $BR_l^N(y^F) - \bar{x}^F$. Thus, given a fixed stopping time of the chaser, providing real-time feedback to the leader makes him more emergent to start working when the beep effect is positive with $BR_l^N(y^F) - \bar{x}^F > 0$. This implies that the beep effect is negative only when $\delta$ is sufficiently large, i.e., alarming a less procrastinating leader about his rival's progress is relatively ineffective.

Panel (b) of Figure 2 depicts the case when $\lambda T = 2$. It indicates that the beep effect becomes positive for all the possible combinations of $\delta$ and $\varphi$ when $\lambda T$ is greater than 0.7.

### B.2.2 Total Effect

We further employ numerical analysis to explore the possible patterns of the total effect, which is determined by the resultant force of the competition effect of Proposition 6 and the beep effect of Proposition 7.

Proposition 6 has demonstrated the absence of competition effect when $\delta \geq \Gamma(\varphi)$, the direction of the total effect is identical to that of the beep effect. We thus focus on the rest two cases when (i) $\delta < \varphi$ and (ii) $\varphi \leq \delta < \Gamma(\varphi)$. Without loss of generality, we still normalize $\lambda = 1$.

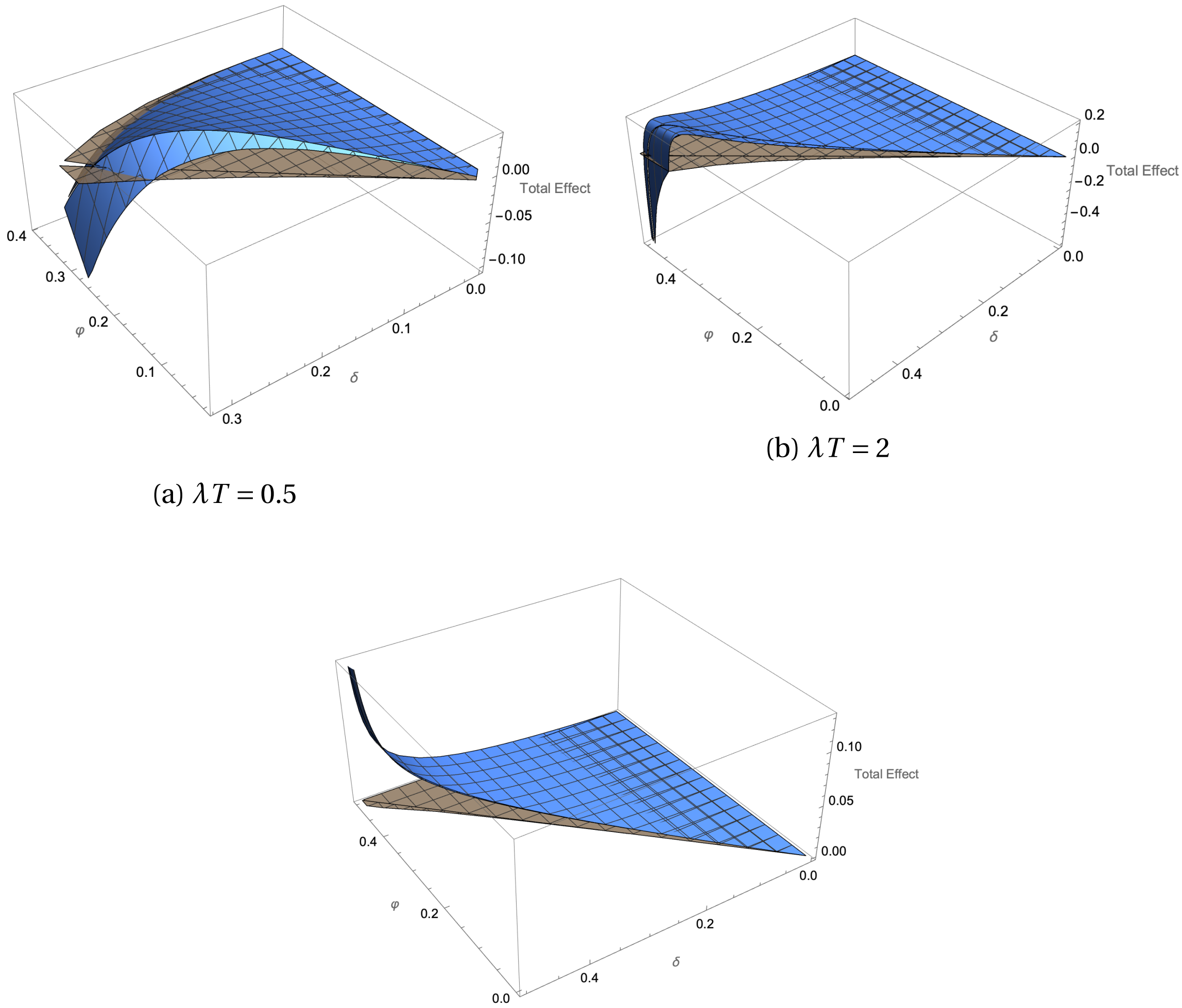

(a) $\lambda T = 0.5$

(b) $\lambda T = 2$

(c) $\lambda T = 4$

Figure 3: Total effect when $\delta < \Gamma(\varphi)$.

Note that when $\delta < \Gamma(\varphi)$, $x^N$ is determined by the unique solution of equation (23), which is derived by equation the right-hand sides of equations (18) and (22). By the proof of Proposition 3, the right-hand sides of equations (18) and (22) are decreasing and increasing, respectively. Therefore, by a similar logic of (27), $\bar{x}^F < x^N$ if and only if

$$\left.\frac{2\delta - 1 + e^{-2\lambda(T-x)}}{(1-e^{-\lambda(T-x)})^2}\right|_{x=\bar{x}^F} < \left.\left[ e^{-\lambda x^N} \cdot \frac{1-\frac{1}{2}(1-e^{-2\lambda(T-x^N)})}{1-\varphi} \right]\right|_{x=\bar{x}^F},$$

which is equivalent to

$$e^{-\lambda \bar{x}^F} > \frac{2(1-\varphi)}{(1-e^{-\lambda(T-\bar{x}^F)})^2} \cdot \frac{\delta - \frac{1}{2}\left(1-e^{-2\lambda(T-\bar{x}^F)}\right)}{1-\frac{1}{2}\left(1-e^{-2\lambda(T-\bar{x}^F)}\right)}.$$

Figure 3 demonstrates the total effect as a function of $\delta$ and $\varphi$, and the three subgraphs represent the case that $\lambda T = 0.5$, $\lambda T = 2$, and $\lambda T = 4$, respectively. Clearly, when $\lambda T$ becomes sufficiently large, the positive beep effect dominates the non-positive competition effect, which results in a positive total effect for all feasible combinations of $\delta$ and $\varphi$.

## B.3 Optimal Starting Strategy in Alternative Settings

### B.3.1 Symmetric Contest with One Breakthrough, Exponential Discounting, and Postponed Reward until Deadline

Consider a contest with $n \geq 2$ ex-ante homogeneous contestants, which is terminated either when one contestant achieves a breakthrough at time $t \in (0, T)$ or the predetermined deadline $T$ is reached. The contest game is symmetric in the sense that all searching contestants have an identical strategy space $a_{it} \in \{0, 1\}$ and instantaneous cost $c a_{it} dt$ for searching with the arrival rate $\lambda$. The prize for the breakthrough is still $v$. However, different from our chasing contest setting, here we assume that all agents discount exponentially with a common discount factor $\rho \in (0, 1)$. However, even if a contestant has completed the breakthrough, the prize $v$ is paid at the predetermined deadline $T$.

**Proposition B.1.** *Consider a continuous-time contest with a predetermined deadline and exponential discounting. When the prize $v$ is rewarded at the deadline $T$, the agents will work at time $t$ if and only if*

$$t \geq \max\left\{0, T - \frac{1}{n\lambda + \rho} \ln\left[\frac{\lambda((n\lambda + \rho)v - nc)}{\rho c}\right]\right\}. \tag{37}$$

*Proof.* Suppose the contest has not terminated at time $t < T$, in any symmetric Markov perfect equilibrium, if a contestant anticipates that her future selves and all the rest contestants will choose $a_{i\tau} = 1$ for any $\tau > t$. The present value of her expected reward is given by

$$e^{-\rho(T-t)} \int_t^T \lambda e^{-n\lambda(\tau - t)} d\tau \cdot v = \frac{1}{n} e^{-\rho(T-t)} \left(1 - e^{-n\lambda(T-t)}\right) \cdot v.$$

Next, if the contest is terminated at time $\tau > t$, the present value of the cost that the agent works continuously from time $t$ to $\tau$ is given by

$$\int_t^\tau e^{-\rho(s-t)} c ds = \frac{1}{\rho}\left(1 - e^{-\rho(\tau - t)}\right) \cdot c.$$

Hence, the present value of expected cost that the agent selects a $t$-start strategy is

$$\left[\left(\int_t^T n\lambda e^{-n\lambda(\tau-t)}\frac{1}{\rho}\left(1-e^{-\rho(\tau-t)}\right)d\tau\right)+e^{-n\lambda(T-t)}\cdot\frac{1}{\rho}\left(1-e^{-\rho(T-t)}\right)\right]\cdot c$$
$$=\frac{1}{n\lambda+\rho}\left(1-e^{-(n\lambda+\rho)(T-t)}\right)c.$$

Consequently, the continuation payoff at time $t$, if the contestants work from $t$ to the terminus, is given by

$$U_{it}^B=\bar{U}_{it}=\frac{1}{n}e^{-\rho(T-t)}\left(1-e^{-n\lambda(T-t)}\right)v-\frac{1}{n\lambda+\rho}\left(1-e^{-(n\lambda+\rho)(T-t)}\right)c.$$

Given the anticipation that her future selves as well as her rivals will work until the end of the contest, the agent is willing to work if and only if

$$\lambda dt\cdot e^{-\rho(T-t)}\cdot v-cdt+(1-n\lambda dt)\bar{U}_{it}\geq\bar{U}_{it},$$

which implies

$$\lambda e^{-\rho(T-t)}v-c-n\lambda\bar{U}_{it}=\lambda e^{-(n\lambda+\rho)(T-t)}v-\left[1-\frac{n\lambda}{n\lambda+\rho}\left(1-e^{-(\lambda n+\rho)(T-t)}\right)\right]c\geq 0. \tag{38}$$

Recollect terms we get

$$t\geq\max\left\{0,T-\frac{1}{n\lambda+\rho}\ln\left[\frac{\lambda((n\lambda+\rho)v-nc)}{\rho c}\right]\right\}.$$

Finally, we show that $\bar{U}_{it}\geq 0$ for all $t\geq T-\frac{1}{n\lambda+\rho}\ln\left[\frac{\lambda((n\lambda+\rho)v-nc)}{\rho c}\right]$. This implies that we can utilize the reasoning in Lemma A.1 to show that the contestants will not stop making effort once they have started. Observe

$$\frac{d\bar{U}_{it}}{dt}=-\frac{(\lambda n+\rho)v-nc}{n}e^{-(\lambda n+\rho)(T-t)}+\frac{\rho v}{n}.$$

This expression is decreasing with $t$ and it is negative $(-n(\lambda v-c))$ when $t=T$. Thus, $\bar{U}_{it}$ is either decreasing for all $t\in[0,T]$ or increasing for $t\in[0,\bar{t}]$ and decreasing otherwise for some $\bar{t}\in(0,T)$. Also, $\bar{U}_{iT}=0$, and hence it suffices to show that there exists $t^*<T-\frac{1}{n\lambda+\rho}\ln\left[\frac{\lambda((n\lambda+\rho)v-nc)}{\rho c}\right]$ such that $\bar{U}_{it^*}>0$.

Let $t^* = T - \frac{1}{\rho}\ln\frac{\lambda v}{c}$. We have

$$\bar{U}_{it^*} = \frac{\left(\frac{\lambda v}{c}\right)^{-\frac{\lambda n+\rho}{\rho}}\left(nc - v\left(-\rho\left(\frac{\lambda v}{c}\right)^{\frac{n\lambda}{\rho}} + n\lambda + \rho\right)\right)}{n(n\lambda+\rho)} \propto nc - v\left(-\rho\left(\frac{\lambda v}{c}\right)^{\frac{n\lambda}{\rho}} + n\lambda + \rho\right).$$

Since

$$\frac{d}{dc} nc - v\left(-\rho\left(\frac{\lambda v}{c}\right)^{\frac{n\lambda}{\rho}} + n\lambda + \rho\right) = -n\left(\left(\frac{\lambda v}{c}\right)^{\frac{\lambda n}{\rho}+1} - 1\right) < 0,$$

and $\bar{U}_{it^*} \to 0$ when $c \to \lambda v$, we have $\bar{U}_{it^*} \geq 0$ when $c \leq \lambda v$.

It remains to show that $t^* < T - \frac{1}{n\lambda+\rho}\ln\left[\frac{\lambda((n\lambda+\rho)v - nc)}{\rho c}\right]$ such that $\bar{U}_{it^*} > 0$. In fact,

$$t^* = T - \frac{1}{\rho}\ln\frac{\lambda v}{c} < T - \frac{1}{\rho}\ln\frac{\rho\lambda v + n\lambda(\lambda v - c)}{\rho c} < T - \frac{1}{n\lambda+\rho}\ln\frac{\rho\lambda v + n\lambda(\lambda v - c)}{\rho c},$$

which completes the proof. □

Note that when $\rho = 0$, the working condition of inequality (38) becomes $\lambda e^{-n\lambda(T-t)}(\lambda v - c) \geq 0$, i.e., each agent receives a nonnegative payoff from exerting effort from $t = 0$ until the terminus, if $\lambda v > c$ is satisfied. In other words, since each agent holds a constant instantaneous payoff at each time instant when there is no discounting, the choices of when the prize is awarded become irrelevant to an agent's instantaneous payoff of exerting effort at each time instant. A time-consistent agent will thus exert efforts from the beginning of the contest with a nonnegative payoff, which echoes the findings in existing literature on continuous-time dynamic contests.

### B.3.2 Infinite-horizon Chasing Contests

This subsection analyzes an alternative infinite-horizon version of our chasing contest and reveals that a predetermined deadline is necessary for the emergence of an optimal $x$-start strategy. Since both contestants face a stationary optimization problem over each time instant within the chasing and the contest stage until the end, we focus on analyzing the set of *stationary* MPE, in which both contestants will not alter their effort choices within each of the two stages.

**Proposition B.2.** *When $T \to \infty$, there exists no stationary MPE in the public chasing contest when $\lambda v/c \in (2/\beta - 1, 3)$ with $\beta < 1/2$, or $\lambda v/c \in (3, 2/\beta - 1)$ with $\beta > 1/2$. Otherwise, there exists a stationary MPE in which the chaser exerts non-stopping efforts throughout the contest, while the leader chooses to:*

*(i) exert non-stopping efforts throughout the contest, when $\lambda v/c \geq \max\{3, 4/\beta - 3\}$;*

*(ii) exert non-stopping efforts in the contest stage and shirks throughout the chasing stage, when* $\lambda v/c \in [\max\{3, 2/\beta - 1\}, \max\{3, 4/\beta - 3\}]$*;*

*(iii) shirks throughout the contest, when* $\lambda v/c < \min\{3, 2/\beta - 1\}$.

*Proof.* In this proof, we first demonstrate the non-existence of stationary MPE when the chaser shirks continuously in the contest in Step 1. Based on this observation, by assuming the chaser will exert non-stopping efforts throughout the contest, we characterize the leader's optimal effort choices throughout the contest in Step 2. Finally, given our characterization of the leader's best responses, we verify the conditions under which the chaser exert non-stopping efforts.

*Step 1:* First, observe that if the leader anticipates that his future selves will work after time $t$, his continuation payoff at time $t$ is given by

$$U_{lt}^{B} = \int_{t}^{\infty} \lambda e^{-\lambda(\tau - t)}(v - (\tau - t)c)dt = v - \frac{c}{\lambda}.$$

Therefore, she chooses to exert effort at time $t$ if and only if $\beta\lambda v - c \geq \beta\lambda(v - c/\lambda)$, which is impossible given that $\beta \in (0, 1)$.

On the other hand, if the leader anticipates that his future selves will shirk after time $t$, his continuation payoff at $t$ is 0, and he also shirks at time $t$ if and only if $\beta\lambda v - c < 0$, which is also impossible.

*Step 2:* Consider first the leader's choice in the contest stage. If he works continuously, his continuation payoff at time $t$ is

$$\bar{U}_{lt} = \int_{t}^{\infty} \lambda e^{-2\lambda(\tau - t)}(v - (\tau - t)c)dt = \frac{1}{2}\left(v - \frac{c}{\lambda}\right),$$

which is independent of $t$. Then $a_{lt} \equiv 1$ if and only if $\beta\lambda v - c \geq \beta\lambda \cdot \frac{1}{2}\left(v - \frac{c}{\lambda}\right)$, which implies $\lambda v \geq (2/\beta - 1)c$.

If not, the leader will not work in the contest stage. Given this, the continuation payoff at time $t$ in the chasing stage if the leader works in this stage is

$$U_{lt}^{A} = \int_{t}^{\infty} \lambda e^{-2\lambda(\tau - t)}(v - (\tau - t)c)dt = \frac{1}{2}\left(v - \frac{c}{\lambda}\right),$$

which equals to $\bar{U}_{lt}$. This is because the leader quits as long as the contest enters the contest stage. Thus, it is straightforward that the leader does not work in the chasing stage as well. If $\lambda v \geq (2/\beta - 1)c$, given the leader will work in the contest stage, the continuation payoff in the chasing stage is given by

$$U_{lt}^A = \int_t^\infty \lambda e^{-2\lambda(\tau - t)} \left[ v + \frac{1}{2}\left(v - \frac{c}{\lambda}\right) - 2(\tau - t)c \right] dt = \frac{3}{4}\left(v - \frac{c}{\lambda}\right).$$

Therefore, the leader also works in the chasing stage if and only if $\beta\lambda v - c \geq \beta\lambda \cdot \frac{3}{4}\left(v - \frac{c}{\lambda}\right)$, which implies $\lambda v \geq (4/\beta - 3)c$.

In short, given that the chaser will work continuously throughout the contest, the leader shirks when $\lambda v / c < 2/\beta - 1$ and works when $\lambda v/c > 4/\beta - 3$. When $2/\beta - 1 < \lambda v/c < 4/\beta - 3$, he works only in the contest stage.

*Step 3:* Denote $U_{ct}^B$ the chaser's continuation payoff at time $t$ in the contest stage. Since the time-consistent chaser faces a stationary optimization problem at each time instant throughout the contest, we can derive her optimal effort choices through the following Bellman equation

$$V_t = \max_{a_{ct} \in \{0,1\}} \left\{ \lambda a_{ct} dt \cdot U_{ct}^B - c a_{ct} dt + (1 - (a_{ct} + a_{lt})\lambda dt) V_{t+dt} \right\}.$$

Hence, $a_{ct} = 1$ is optimal if and only if (i) $\lambda U_{ct}^B - c \geq V_t$, and (ii) the transversality condition is satisfied such that $\lambda U_{ct}^B \geq c$. Note that requirement (i) holds automatically. This is because $V_{t+dt} = V_t \equiv V$ for some $V \geq 0$ by stationarity, and $\lambda U_{ct}^B - c \equiv (1 + a_{lt})V \geq V$ if $a_{ct} = 1$ is optimal.

When $\lambda v / c < 2/\beta - 1$, the leader shirks in the contest stage, and thus the chaser's continuation payoff at time $t$ in the contest stage is

$$U_{ct}^B = \int_t^\infty \lambda e^{-\lambda(\tau - t)} (v - (\tau - t)c) dt = v - \frac{c}{\lambda}.$$

Therefore, the transversality condition implies that $a_{ct} \equiv 1$ if and only if $\lambda v \geq 2c$. If the leader works in the contest stage ($\lambda v/c > 2/\beta - 1$), the chaser's continuation payoff in the contest stage is

$$U_{ct}^B = \int_t^\infty \lambda e^{-2\lambda(\tau - t)} (v - (\tau - t)c) dt = \frac{1}{2}\left(v - \frac{c}{\lambda}\right).$$

The transversality condition implies that $a_{ct} \equiv 1$ if and only if $\lambda v \geq 3c$.

We complete the proof by aligning all preceding piecewise thresholds of $\lambda v/c$ given that $\beta \in (0,1)$. □

### B.3.3 General Present-Future Preferences

In this subsection, we suppose that agents discount their future values in terms of the *present-future model* introduced by Harris and Laibson (2013). Although this different as-sumption about the leader's time preference will alter the specific time at which the chaser stops, it still permits the existence of the chaser's optimal $y$-stop strategy. To simplify computation, we consider a one-person contest in which a single agent with present-future preference needs to complete one breakthrough.

The following proposition reveals the absence of the optimal $x$-start strategy once we assume that the agent embodies a less drastic time discounting such as the IG model.

**Proposition B.3.** *Under the general present-future model, the agent works continuously for all* $t \in [0, \min\{\tilde{t}_1, T\}]$ *if and only if* $\lambda v \geq c$.

*Proof.* Note that at any time $t$, if the prize is paid at time $t + dt$, the instantaneous reward is given by $(1 - \eta dt + \eta dt \cdot \beta) \cdot v$. Thus, at the last infinitesimal of time, the agent works if and only if

$$(1 - \rho dt)\lambda dt \cdot (1 - \eta dt + \eta dt \cdot \beta) \cdot v - cdt \geq 0 \Leftrightarrow \lambda v > c.$$

In backward, if the agent anticipates that all her future selves will work, she also works at time $t$ if and only if

$$\lambda(1 - \rho dt)dt \cdot (1 - \eta dt + \eta dt \cdot \beta) \cdot v - cdt + (1 - \lambda dt)(1 - \rho dt)\bar{U}_{t+dt} \geq (1 - \rho dt)\lambda dt)\bar{U}_{t+dt}.$$

Let $dt \to 0$ and reorganize, we have $\lambda v - c \geq \lambda \bar{U}_t$.

It is enough to show that $\bar{U}_t = \alpha(v - c/\lambda)$ for some $\alpha \in [0, 1]$. First, if the agent starts at time $t$ and wins at date $\tau > t$, the present value of the prize is given by

$$\bar{v}(\tau) = e^{-\rho(\tau - t)} \left[ \left(1 - e^{-\eta(\tau - t)}\right) \cdot \beta + e^{-\eta(\tau - t)} \right] \cdot v = e^{-\rho(\tau - t)} \left(\beta + (1 - \beta)e^{-\eta(\tau - t)}\right) \cdot v.$$

Thus, the expected reward of winning the contest is given by

$$\int_t^T \lambda e^{-\lambda(\tau - t)} \bar{v}(\tau) d\tau = \left[ \beta \cdot \frac{1 - e^{-(\lambda + \rho)(T - t)}}{\lambda + \rho} + (1 - \beta) \cdot \frac{1 - e^{-(\lambda + \rho + \eta)(T - t)}}{\lambda + \rho + \eta} \right] \cdot \lambda v$$

Accordingly, the present value of the agent's cost when the contest terminates at date $t > \tau$ is given by

$$\begin{aligned}
\bar{c}(\tau) &= \int_t^\tau \eta e^{-\eta(s - t)} \left( \int_t^s e^{-\rho(z - t)} cdz + \beta \int_s^\tau e^{-\rho(z - t)} dz \right) ds + e^{-\eta(\tau - t)} \int_t^\tau e^{-\rho(\tau - t)} cd\tau \\
&= \left[ \frac{1 - e^{-\rho(\tau - t)}}{\rho} - (1 - \beta) \left( \frac{1 - e^{-\rho(\tau - t)}}{\rho} - \frac{1 - e^{-(\rho + \eta)(\tau - t)}}{\rho + \eta} \right) \right] \cdot c \\
&= \left[ \beta \frac{1 - e^{-\rho(\tau - t)}}{\rho} + (1 - \beta) \frac{1 - e^{-(\rho + \eta)(\tau - t)}}{\rho + \eta} \right] \cdot c.
\end{aligned}$$

Thus, his expected total cost is given by

$$\int_t^T \lambda e^{-\lambda(\tau-t)}\bar{c}(\tau)d\tau + e^{-\lambda(T-t)}\bar{c}(T)$$
$$= \left[\beta\cdot\frac{1-e^{-(\lambda+\rho)(T-t)}}{\lambda+\rho} + (1-\beta)\frac{1-e^{-(\lambda+\rho+\eta)(T-t)}}{\lambda+\rho+\eta}\right]\cdot c.$$

That is,

$$\bar{U}_t = \left[\beta\cdot\frac{\lambda}{\lambda+\rho}\left(1-e^{-(\lambda+\rho)(T-t)}\right) + (1-\beta)\frac{\lambda}{\lambda+\rho+\eta}\left(1-e^{-(\lambda+\rho+\eta)(T-t)}\right)\right]\cdot\left(v-\frac{c}{\lambda}\right).$$

Obviously, $\bar{U}_t \in [0, v-c/\lambda]$ and therefore the agent will work for all $t\in[0,\min\{\tilde{t}_1, T\}]$, and otherwise she quits immediately at date 0.[4] □

#### B.3.4 Experimentation

We deliberately exclude the learning process (see, e.g., Keller et al. (2005), Bonatti and Hörner (2011), and Murto and Välimäki (2011)) in our chasing contest model. This is not only for model simplicity, more importantly, an agent will not adopt the nonzero $x$-start strategy when she needs to experiment the feasibility of the project with no predetermined deadline. Again, with no loss of generality, we consider a one-person contest case with no (exponential) discounting as in Appendix B.3.3 to clarify the intuition.

Denote $p_0$ as the agent's prior belief at time 0 for the feasibility of the project, and $p_t$ as her posterior belief of feasibility at time $t$, with

$$p_t = \frac{p_0\, e^{-\int_0^t \lambda a_\tau d\tau}}{p_0\, e^{-\int_0^t \lambda a_\tau d\tau} + 1 - p_0},$$

where $a_t = \{0,1\}$. By Lemma 1, we can derive the corresponding incentive compatibility condition as

$$\beta p_t \lambda v \geq c + \beta\lambda U_t, \tag{39}$$

[4]Note that here since $\frac{1-e^{-(T-t)x}}{x}$ is decreasing of $x$ whenever $T-t>0$. We have

$$\frac{\lambda}{\lambda+\rho+\eta}\left(1-e^{-(\lambda+\rho+\eta)(T-t)}\right) < \frac{\lambda}{\lambda+\rho}\left(1-e^{-(\lambda+\rho)(T-t)}\right).$$

Thus,

$$\bar{U}_t < \frac{\lambda}{\lambda+\rho}\left(1-e^{-(\lambda+\rho)(T-t)}\right)\cdot\left(v-\frac{c}{\lambda}\right),$$

which is exactly the continuation payoff at time $t$ that the contestants work to the terminus when $\beta=1$ and $\eta=0$ (normal discounting); that is, the presence of the present bias in terms of general present-future preferences further cut down a time-consistent agent's continuation payoff at each instant of time.

where $\bar{U}_t$ is her continuation payoff at time $t$. Define

$$U_t = p_t\left[\int_t^y (\lambda v - c)a_\tau e^{-\int_t^\tau \lambda a_z dz} d\tau - e^{-\lambda(y-t)}(y-t)c\right] - (1-p_t)c\int_t^y a_\tau d\tau,$$

to represent the agent's continuation payoff at time $t \in [0, y]$, where $y$ is her optimal stopping time given the corresponding lowest posterior $p_{y|x}$ for exerting effort. Let $x \geq 0$ be the agent's optimal starting time. Note that given

$$p_{y|x} = \frac{p_0\, e^{-\lambda(y-x)}}{p_0\, e^{-\lambda(y-x)} + 1 - p_0} = \frac{c}{\beta\lambda v} \tag{40}$$

when he stops exerting effort, we have

$$y = \frac{1}{\lambda}\ln\left[\left(\frac{\beta\lambda v}{c} - 1\right)\frac{p_0}{1-p_0}\right] + x. \tag{41}$$

This implies that the agent will consistently adjust her optimal stopping time $y$ given her starting time $x > 0$ to update her belief until $p_y$. Therefore, if the agent shirks at the beginning of the one-person contest, we must have $\beta p_0 \lambda v < c + \beta\lambda U_0$. By (41), if the agent shirks for a period of time, her continuation payoff as well as her belief will remain unchanged during this shirking period, i.e., $U_{t'} = U_t$ and $p_t = p_{t'}$, $a_\tau = 0$, $\forall \tau \in [t, t']$.

The agent will thus never begin to exert effort, as the preceding incentive compatibility condition remains valid. In other words, the agent will choose to either work from the beginning or shirk throughout the infinite-horizon one-person contest.

Last, we briefly illustrate the results for the finite-horizon one-person contest case (i.e., with a predetermined deadline $T$): the agent will retrieve her $x$-start starting strategy when the contest duration is longer than the amount of time necessary to exhaust her incentives to exert positive efforts, and exert non-stopping effort throughout the contest otherwise. Even though the agent's behavior is different from the preceding infinite-horizon case, it essentially mirrors the economic rationale discussed in the preceding infinite-horizon case.

To see this, note that by (40), the agent's posterior belief will reach its lowest level for exerting positive effort only if she has consistently worked over a fixed duration. Denote $D \leq T$ a time duration during for which the posterior belief reaches its lowest level, that is, $D = t_2 - t_1$ if the agent chooses $a_t = 1$ for all $\tau \in [t_1, t_2]$ with $p_{t_2|t_1} = \frac{c}{\beta\lambda v}$. Given that the agent is present-biased (in term of instantaneous-gratification), it is optimal for the agent to set $t_2 = T$, and $t_1 \geq 0$ as her optimal starting time. However, when $D > T$, i.e., the deadline is too short to exhaust the agent's posterior belief to its minimum, it is clearly that the agent will choose to exert effort at time 0 when her incentive compatibility condition of (39) is satisfied.

## B.4 Present-Biased Chaser

This subsection examines the scenario in which the chaser is also affected by the present bias issue. Since the chaser is still required to complete two breakthroughs, she is motivated to stop exerting effort for the first breakthrough once she anticipates a low chance of completing both breakthroughs before the predetermined deadline. The novelty lies in the fact that the chaser, due to the presence of present bias, will also shirk initially if she feels the deadline is not sufficiently urgent; i.e., the present-biased chaser performs according to a *$x$-start-then-$y$-stop* strategy, in which she chooses both an optimal starting and an optimal stopping time.

Again, to avoid unnecessary technical complications, we consider the case in which the leader is absent. In such a one-person contest, we still term the agent as the *chaser* and the two stages of the contest the chasing stage and the contest stage, respectively. The chaser still has two breakthroughs that need to be accomplished before the deadline, and the final reward is $v$ if she succeeds. We assume that the present-biased chaser also behaves according to the MPE concept; that is, we implicitly assume that the chaser is *sophisticated*.

In the contest stage, since there is only one breakthrough left, her optimization problem is identical to the one-person chasing contest as illustrated by Lemma 2. Thus, the chaser selects an $x$-start strategy and her starting time is given by

$$\hat{x}_0 = \max\left\{0, T - \frac{1}{\lambda}\ln\frac{1}{1-\delta}\right\}.$$

The following result characterizes the present-biased chaser's equilibrium strategy in the chasing stage of the one-person contest.

**Proposition B.4.** *In the chasing stage, suppose the chaser is present-biased. Then if $\beta \geq 2c/(\lambda v)$, there exists an $x$-*start-then-*$y$*-stop equilibrium*, such that she starts at time*

$$x_1 = \max\left\{0, T - \frac{1}{\lambda}\ln\left[\frac{1}{(1-\beta)\varphi}\cdot\left(\frac{\beta(1-\delta)}{\beta-\varphi} - \ln\left[\frac{\beta(1-\delta)}{\beta-\varphi}\right]\right)\right]\right\},$$

*and if the first breakthrough does not come, she stops at*

$$y_1 = T - \frac{1}{\lambda}\ln\frac{\beta}{\beta-\varphi}.$$

*Otherwise, if $\beta < 2c/(\lambda v)$, the chaser will not work for any $t$.*

*Proof.* The chaser's stopping time is given by the latest time that the chaser is willing to work for the first breakthrough given her future selves will not do so; that is,

$$y_1 = \sup_{t\in[0,T]} \left\{\beta\lambda U_t^B - c \geq 0\right\}.$$

*Step 1: If $\beta < 2c/(\lambda v)$, the chaser will not work at all $t \in [0, T]$.*

For any $t \geq \hat{x}_0$, if the chaser achieves her first breakthrough at time $t$, her future selves will start working for the second breakthrough immediately. Thus, the chaser's continuation payoff at $t$ is given by

$$U_t^B = \bar{U}_t = \int_t^T \lambda e^{-\lambda(\tau-t)}(v-(\tau-t)c)dt - e^{-\lambda(T-t)}(T-t)c = \left(1-e^{-\lambda(T-t)}\right)\left(v-\frac{c}{\lambda}\right),$$

Thus, if the stopping time $y_1 \geq \hat{x}_0$, then $\beta\lambda\bar{U}_{y_1}^B - c = 0$, which implies $y_1 = T - \frac{1}{\lambda}\ln\frac{\beta}{\beta-\varphi}$. Here $y_1 \geq \hat{x}_0$, if and only if

$$\frac{1}{1-\delta} = \frac{\beta\lambda v - \beta c}{\beta\lambda v - \beta c - c} \leq \frac{\beta}{\beta-\varphi} = \frac{\beta\lambda v - \beta c}{(1-\beta)c},$$

which indicates $\beta \geq 2c/(\lambda v)$. That is, when $\beta \geq 2c/(\lambda v)$, the chaser stops at time $y_1 > \hat{x}_0$; otherwise, she shirks for all $t \geq \hat{x}_0$.

In addition, when $t < \hat{x}_0$, if the chaser achieves the breakthrough at time $t$, she will shirk until $\hat{x}_0$ and then work until the terminus. Therefore the continuation payoff at $t$ is given by

$$U_t^B = \bar{U}_{\hat{x}_0} = \left(1-e^{\lambda(T-\hat{x}_0)}\right)\left(v-\frac{c}{\lambda}\right) = v - \frac{c}{\beta\lambda}.$$

Thus, when $\beta < 2c/(\lambda v)$, given the future selves will work, the chaser works at $t < \hat{x}_0$ if and only if $\beta\lambda U_t^B - c = \beta\lambda v - 2c \geq 0$, which is impossible. That is, shirking for all $t \in [0, T]$ constitutes an intrapersonal subgame perfect equilibrium.

*Step 2: When $\beta \geq 2c/(\lambda v)$, the chaser will work for the first breakthrough from $[\hat{x}_0, y_1)$.*

Henceforth, we assume $\beta \geq 2c/(\lambda v)$, and we know that when $\beta \geq 2c/(\lambda v)$, for any $t > \hat{x}_0$, the chaser will work until time $y_1 > \hat{x}_0$. Thus, before $y_1$, when she anticipates that her future selves will work for the first breakthrough until $y_1$, she works at time $t$ if and only if

$$\beta\lambda U_t^B dt - cdt + \beta(1-\lambda)U_t^A \geq \beta U_t^A.$$

Reorganize, we have

$$\beta\lambda(U_t^B - U_t^A) \geq c. \tag{42}$$

Thus, the chaser's starting time $x_1$ is given by

$$x_1 = \inf_{t\in[0,y_1]} \left\{\beta\lambda(U_t^B - U_t^A) \geq c\right\}.$$

Consider first the case $t \geq \hat{x}_0$. Given the chaser will always work in the contest stage, if the chaser exerts effort from $t$ to $y_1$ in the chasing stage, her continuation payoff in the one-person contest is given by

$$\begin{aligned} U_t^A &= \int_t^{y_1} \lambda e^{-\lambda(\tau-t)} \left[\left(1-e^{-\lambda(y_1-t)}\right)\left(v-\frac{c}{\lambda}\right) - c(\tau-t)\right] d\tau - e^{-\lambda(y_1-t)}(y_1-t)c \\ &= \left(1-e^{-\lambda(y_1-t)} - \lambda(y_1-t)e^{-\lambda(T-t)}\right)\left(v-\frac{c}{\lambda}\right) - \left(1-e^{-\lambda(y_1-t)}\right)\cdot\frac{c}{\lambda}. \end{aligned} \tag{43}$$

Substituting the expressions of $U_t^A$ and $U_t^B$ when $t \geq \hat{x}_0$ into (42), the chaser works at time $t$ if and only if

$$\beta \cdot \frac{e^{-\lambda(y_1-t)} - (1-\lambda(y_1-t))e^{-\lambda(T-t)}}{1-\beta(1-e^{-\lambda(y_1-t)})} \geq \varphi.$$

Now we show that there exists $t' \in [\hat{x}_0, y_1)$, such that the chaser works at time $t \in (\hat{x}_0, y_1)$ if and only if $t \geq t'$. Define

$$G_1(t|y_1) = \frac{e^{-\lambda(y_1-t)} - (1-\lambda(y_1-t))e^{-\lambda(T-t)}}{1-\beta(1-e^{-\lambda(y_1-t)})},$$

where $t \in [\hat{x}_0, y_1)$. Its first-order derivative with respect to $t$ is

$$\begin{aligned} G_1'(t|y_1) &= \frac{\lambda e^{-\lambda(T-y_1+t)}}{\left(\beta+(1-\beta)e^{\lambda(y_1-t)}\right)^2}\cdot\left[(1-\beta)e^{\lambda T} - \beta e^{\lambda t} + (1-\beta)(\lambda y_1 - \lambda t - 2)e^{\lambda y_1}\right] \\ &\propto (1-\beta)e^{\lambda T} - \beta e^{\lambda t} + (1-\beta)(\lambda y_1 - \lambda t - 2)e^{\lambda y_1}. \end{aligned} \tag{44}$$

That is, $G_1'(t|y_1) > 0$ if and only if $(1-\beta)e^{\lambda T} - \beta e^{\lambda t} + (1-\beta)(\lambda y_1 - \lambda t - 2)e^{\lambda y_1} > 0$. Note that

$$\frac{d}{dt}\left[(1-\beta)e^{\lambda T} - \beta e^{\lambda t} + (1-\beta)(\lambda y_1 - \lambda t - 2)e^{\lambda y_1}\right] = -\beta e^{\lambda t} - (1-\beta)e^{\lambda y_1} < 0. \tag{45}$$

Therefore, $G_1(t|y_1)$ can be increasing, decreasing, or increasing at first and then decreasing, and there is no other possibilities. In addition,

$$G_1'(y_1|y_1) = \lambda(1-\beta-(2-\beta)e^{-\lambda(T-y_1)}) = -\frac{\beta-2\varphi+\beta\varphi}{\beta} = -\frac{\lambda(\beta\lambda v-2c)}{\beta(\lambda v-c)} < 0.$$

Thus, it is impossible that $G_1(t|y_1)$ is increasing in interval $(\hat{x}_0, y_1)$. Consequently, there exists $t' \in [\hat{x}_0, y_1)$, such that the chaser works at time $t \in (\hat{x}_0, y_1)$ if and only if $t \geq t'$.

Now we show that $t' = \hat{x}_0$, which indicates that the chaser would have started at time $\hat{x}_0$. It suffices to show that $\beta G(\hat{x}_0|y_1) \geq \varphi$. Substitute $\hat{x}_0$ and $y_1$ by their respective expressions,

$$\frac{\beta G(\hat{x}_0|y_1)}{\varphi} = \frac{c + (\beta\lambda v - (1+\beta)c)\ln\left[\frac{\beta\lambda v - (1+\beta)c}{(1-\beta)c}\right]}{\beta\lambda v - c},$$

which is greater than 1 if and only if

$$(\beta\lambda v - (1+\beta)c)\ln\left[\frac{\beta\lambda v - (1+\beta)c}{(1-\beta)c}\right] \geq \beta\lambda v - 2c.$$

Notice that

$$\frac{d}{dv}\left[(\beta\lambda v - (1+\beta)c)\ln\left[\frac{\beta\lambda v - (1+\beta)c}{(1-\beta)c}\right] \geq \beta\lambda v - 2c\right] = \beta\lambda\ln\left[\frac{\beta\lambda v - (1+\beta)c}{(1-\beta)c}\right],$$

which is positive if and only if $\beta \geq 2c/(\lambda v)$, which is satisfied by assumption. Thus, the chaser's starting time should be no later than $\hat{x}_0$ when $\beta \geq 2c/(\lambda v)$.

*Step 3: Identify the starting time when $\beta \geq 2c/(\lambda v)$.*

Consider the case that $t < \hat{x}_0$. By (43), since the chaser will work from $\hat{x}_0$ to $y_1$ in the chasing stage, the continuation payoff at time $\hat{x}_0$ in the chasing stage is

$$\begin{aligned} U^A_{\hat{x}_0} &= \left(1 - e^{-\lambda(y_1 - t)} - \lambda(y_1 - \hat{x}_0)e^{-\lambda(T-\hat{x}_0)}\right)\left(v - \frac{c}{\lambda}\right) - \left(1 - e^{-\lambda(y_1-\hat{x}_0)}\right)\cdot\frac{c}{\lambda} \\ &= \frac{(\lambda v - 2c)(\beta\lambda v - 2c)}{\lambda(\beta\lambda v - (1+\beta)c)} - \frac{(1-\beta)c}{\beta\lambda}\ln\left[\frac{\beta\lambda v - (1+\beta)c}{(1-\beta)c}\right]. \end{aligned}$$

Thus, for $t < \hat{x}_0$, if she works for the first breakthrough from $t < \hat{x}_0$ to $y_1 > \hat{x}_0$, her continuation payoff at time $t$ in the chasing stage is

$$\begin{aligned} U^A_t &= \int_t^{\hat{x}_0} \lambda e^{-\lambda(\tau - t)}\left[v - \frac{c}{\beta\lambda} - (\tau - t)c\right]dt + e^{-\lambda(\hat{x}_0 - t)}(U^A_{\hat{x}_0} - (\hat{x}_0 - t)c) \\ &= \left(1 - e^{-\lambda(\hat{x}_0 - t)}\right)\cdot\left(v - \frac{(1+\beta)c}{\beta\lambda}\right) + e^{-\lambda(\hat{x}_0-t)}U^A_{\hat{x}_0}. \end{aligned}$$

Thus, she is willing to work at time $t$ if and only if $\beta\lambda(U^B_t - U^A_t) \geq c$, which implies

$$e^{-\lambda(\hat{x}_0 - t)}\left(v - \frac{(1+\beta)c}{\beta\lambda} - U^A_{\hat{x}_0}\right) \geq \frac{1-\beta}{\beta}\frac{c}{\lambda}. \tag{46}$$

Here,

$$
\begin{aligned}
v-\frac{(1+\beta)c}{\beta\lambda}-U^A_{\hat{x}_0} &= \frac{(1-\beta)c}{\beta\lambda(\beta\lambda v-(1+\beta)c)}\cdot\left[(\beta\lambda v-(1+\beta)c)\cdot\ln\left[\frac{\beta\lambda v-(1+\beta)c}{(1-\beta)c}\right]\right]\\
&\geq \frac{(1-\beta)c}{\beta\lambda(\beta\lambda v-(1+\beta)c)}\cdot\left[(\beta\lambda v-(1+\beta)c)\cdot\ln\left[\frac{2c-(1+\beta)c}{(1-\beta)c}\right]\right]=0
\end{aligned}
$$

Thus, the left-hand side of (46) is increasing with $t$, which implies that once the chaser starts she will not stop until $\hat{x}_0$, and hence she will work until $y_1$.

We now identify the condition that the chaser is willing to start at $t=0$. Note that if $t=0$, (46) is revised to

$$
T\leq\frac{1}{\lambda}\ln\left[\frac{\lambda v-c}{](1-\beta)c}\cdot\left(\frac{(1-\beta)c}{\beta\lambda v-(1+\beta)c}+\ln\left[\frac{\beta\lambda v-(1+\beta)c}{(1-\beta)c}\right]\right)\right].
$$

If this inequality does not hold, the chaser is unwilling to start at time $t=0$, and her starting time $x_1$ is given by

$$
\begin{aligned}
x_1 &= T-\frac{1}{\lambda}\ln\left[\frac{\lambda v-c}{(1-\beta)c}\cdot\left(\frac{(1-\beta)c}{\beta\lambda v-(1+\beta)c}+\ln\left[\frac{\beta\lambda v-(1+\beta)c}{(1-\beta)c}\right]\right)\right]\\
&= T-\frac{1}{\lambda}\ln\left[\frac{1}{(1-\beta)\varphi}\cdot\left(\frac{\beta(1-\delta)}{\beta-\varphi}-\ln\left[\frac{\beta(1-\delta)}{\beta-\varphi}\right]\right)\right].
\end{aligned}
$$

□

Proposition B.4 implies that $y_1$ increases with $\beta$, and this indicates that the present bias makes the chaser stop earlier. Also, note that when $\beta=1$, i.e. there is no present bias, the stopping time is given by $y_0=T-\frac{1}{\lambda}\ln\frac{1}{1-\varphi}$. Similar to $\hat{x}_0$ in Lemma 2, $y_0$ represents the chaser's stopping time in the original chasing contest when the leader is absent.

## B.5 Naivety

### B.5.1 Naive Leader

Throughout this paper, we have assumed that the leader, the only party that suffers from present bias, is *sophisticated* about her present bias. Here, we consider an accommodated version of the chasing contest in which the leader is *naive* about his present bias. Given this alternative perception of the leader's behavior pattern, we need to redefine the equilibrium strategies of the two contestants. First, define

$$
\hat{u}_{lt}(h^T_{lt})=\tilde{u}_t(h^T_{lt})+\tilde{V}_t(h^T_{lt}), \tag{47}
$$

where $\tilde{u}$ and $\tilde{V}$ have been defined in Section 2 as the leader's payoff function of terminal history $h^T_{lt}$ when he has no present bias (with $\beta=1$). Also, for any strategy profile $\sigma$ and information set $I_{it}$, denote $\Phi_i(\sigma_c, \sigma_l, a|I_{it})$ an alternative strategy profile that substitutes $\sigma_{it}(I_{it})$ to $a$.

**Definition B.1.** *A pair* $(\sigma_c^*, \sigma_l^*) = \{(\sigma_{ct}^*, \sigma_{lt}^*)\}_{t\in[0,T]}$ *constitutes a* naivety-based equilibrium *if there exists a strategy profile* $\tilde{\sigma}_l^* = \{\tilde{\sigma}_{lt}^*\}_{t\in[0,T]}$*, such that for any information* $\{I_{it}\}_{t\in[0,T]}$*,*

$$\sigma_{ct}^*(I_{lt}) \in \arg\max_{a\in\{0,1\}} E_t\left[u_{ct}(\cdot)|\, I_{ct}, \Phi_c(\sigma_c, \sigma_l, a)\right],$$

$$\sigma_{lt}^*(I_{lt}) \in \arg\max_{a\in\{0,1\}} E_t\left[u_{lt}(\cdot)|\, I_{lt}, \Phi_l(\sigma_c, \tilde{\sigma}_l, a)\right],$$

*and*

$$\tilde{\sigma}_{lt}^*(I_{lt}) \in \arg\max_{a\in\{0,1\}} E_t\left[\tilde{u}_{lt}(\cdot)|\, I_{lt}, \Phi_l(\sigma_c, \tilde{\sigma}_l, a)\right].$$

That is, a naive-based equilibrium is the equilibrium of a fictitious game with three players, i.e. the chaser, the leader, and a *fictitious* leader with no present bias. Then the leader's action is the best response to the mistaken beliefs about her future *fictitious* selves and the chaser's actual strategy; while the chaser's action is the best reply to the leader's actual strategy.

**Proposition B.5.** *The equilibrium strategy profiles previously identified in the public and hidden chasing contest are all naivety-based equilibria.*

The proof is straightforward and is hence omitted. Proposition B.5 states that the equilibria we derived in Section 3 and 4 which are based on the presence of a sophisticated leader, are behaviorally indistinguishable from the naivety-based equilibrium. The intuition behind this comes from the identical optimal strategy of the leader in the two scenarios. Consider the case that the leader is ready to start working at moment $x$. Although a sophisticated and a naive leader may disagree with the preferences of their future selves, they both believe that their future selves will exert non-stopping efforts until the terminus; in other words, both the sophisticated and the naive leader will have an identical perception of $x$ given the same belief on the behavior of his future selves and his opponent. Therefore, although the naive leader will overestimate his future self-control, this overoptimism does not lead to an overoptimistic belief about his future behaviors because *"my future selves will work"* cannot be more optimistic at all.

However, we have concerns regarding the possible drawback behind the nature of the naivety-based equilibrium once we include epistemic reasoning. Note that Definition B.1 essentially formulates a specific epistemic state in which the two contestants *agree to disagree* (Aumann, 1976) on the leader's sophistication. That is, the naive leader believes that the preferences of his future selves are standard, and he also knows that the chaser believes that his future selves are present biased. In this state, the two contestants gamble on the leader's sophistication, but the leader never learns from his past naivety. An alternative epistemic equilibrium concept could be defined upon the assumption that the leader's anticipation about the chaser's behavior is consistent with the perception that his future selves are standard but may be mistaken in reality. However, when incorporating this epistemic reasoning requires us to also consider the possibility that the leader might "realize" that the chaser acts differently based on his previous incorrect belief. We are inclined to stop our analysis here, as such epistemic reasoning has strayed from the main focus of this paper.

### B.5.2 Naive Chaser

In Appendix B.4, we have considered a one-person contest in which a single *sophisticated* present-biased chaser needs to complete two breakthroughs. However, different from the indistinguishable behavior between a sophisticated leader and a naive leader as discussed in the preceding subsection, a naive chaser will behave differently compared to her sophisticated counterpart as long as there are two breakthroughs left.

First, note that when there is only one uncompleted breakthrough, the $x$-start strategy is robust even when the chaser is naive about her present bias. This is because the optimal starting time is given by the earliest time that the chaser is willing to work when she anticipates that her future selves will also work. When a sophisticated agent correctly anticipates her future selves will work, her naive counterpart, who is at least more optimistic about her present bias, will also make this anticipation. Thus, the starting time remains unchanged when naivety is introduced to the contest game.

However, when winning requires achieving more than one breakthrough, the chaser's naivety makes a difference. This is because when there are two breakthroughs left, a sophisticated chaser will stop at $y_1$ identified in Proposition B.4 when she believes that her chance to complete both breakthroughs is relatively low given the approaching deadline. By contrast, a naive chaser will mistakenly anticipate that her future selves have no present bias and will stop at time $y_0 > y_1$. Since this anticipation differs, the starting time of a naive chaser also varies.

**Proposition B.6.** *In the chasing stage, suppose the chaser is present-biased and is naive about it. Then the chaser works at time t if and only if*

$$G_1(t|y_0) = \frac{e^{-\lambda(y_0-t)} - (1-\lambda(y_0-t))e^{-\lambda(T-t)}}{1-\beta(1-e^{-\lambda(y_0-t)})} \geq \frac{\varphi}{\beta}.$$

*In particular, unless the chaser shirks for all $t \in [0,T]$, her strategy is $x$-start-then-$y$-stop for some $0 \leq x \leq y < y_1$.*

*Proof.* To a naive chaser, in the chasing stage, she possesses an overly optimistic belief that she will exert continuous effort from the time entering the contest stage to the terminus. Therefore, she mistakenly computes her continuation payoff in the contest stage at time $t$ as

$$\tilde{U}_t^B = \bar{U}_t = \left(1 - e^{-\lambda(T-t)}\right)\left(v - \frac{c}{\lambda}\right), \tag{48}$$

for any $t \in [0, T]$, even though her starting time in the contest stage will be $\hat{x}_0$ *de facto*. Also, she is under the mistaken belief in the chasing stage that she will exert non-stopping effort from the present until time $y_0$ instead of $y_1$. Therefore, by (43) of Proposition B.4, her "imaginary" continuation payoff in the chasing stage is given by

$$\tilde{U}_t^A = \left(1 - e^{-\lambda(y_0-t)} - \lambda(y_0 - t)e^{-\lambda(T-t)}\right)\left(v - \frac{c}{\lambda}\right) - \left(1 - e^{-\lambda(y_0-t)}\right)\frac{c}{\lambda}. \tag{49}$$

Since the chaser decides to work at time $t$ if and only if $\beta\lambda(\bar{U}_t - \tilde{U}_t^A) \geq c$. Using (48) and (49), it requires $\beta G_1(t|y_0) \geq \varphi$. Recall that (44) and (45)) have established that $G_1(\cdot|y_0)$ is either monotone or hump-shaped. Thus, the set of $t$ that guarantees $\beta G_1(t|y_0) \geq \varphi$ is either empty or an interval.

Finally, define

$$\mathscr{U}^A(y, t) = \left(1 - e^{-\lambda(y-t)} - \lambda(y - t)e^{-\lambda(T-t)}\right)\left(v - \frac{c}{\lambda}\right) - \left(1 - e^{-\lambda(y-t)}\right)\frac{c}{\lambda}.$$

The first derivative with respect to $y$ is given by

$$\frac{\partial \mathscr{U}^A(y, t)}{\partial y} = e^{-\lambda(y-t)}c - e^{-\lambda(T-t)}(1 + e^{\lambda(T+y-2t)}) \cdot (\lambda v - c),$$

which is positive if and only if

$$y \leq T - \frac{1}{\lambda}\ln\frac{1}{1-\varphi} = y_0.$$

Since $y_1 < y_0$, for any $t > 0$, we have $\tilde{U}_t^A = \mathscr{U}^A(y_0, t) > \mathscr{U}^A(y_1, t)$. Thus,

$$\beta\lambda(\bar{U}_t - \tilde{U}_t^A) - c < \beta\lambda(\bar{U}_t - \mathscr{U}^A(y_1, t)) - c = 0.$$

That is, the naive chaser would have stopped at time $y_1$, where the sophisticated still weakly prefers to work. □

The intuition that the naive chaser stops earlier is straightforward. At time $y_1$, the continuation payoffs of the naive and the sophisticated chaser in the contest stage are both $\bar{U}_{y_1}$, since they both work continuously until the terminus. However, since the naive chaser has a longer expected working length (i.e., $y_0 > y_1$), her continuation payoff in the chasing stage for working until $y_1$, $U^A_{y_1}$, is higher than her sophisticated self, which validates the equality $\beta\lambda(\bar{U}_{y_1} - U^A_{y_1}) - c = 0$. That is, the naive chaser's opportunity cost for exerting effort at time $t$ is higher compared to her sophisticated self, which consequently, induces the naive chaser to shirk when the sophisticated is still working.

However, we cannot conclude that the naive chaser will start later compared to her sophisticated counterpart, as the naif is motivated to start earlier as well. To see this, consider at an instant $t < \hat{x}_0$. Since the naive chaser anticipates that her future selves will exert non-stopping effort until the terminus once the first breakthrough arrives, her continuation payoff in the contest stage, $U^B_t$, is higher than her sophisticated self who is expected to start at $\hat{x}_0$. This motivates the naive chaser to start earlier. In fact, Proposition B.6 indicates that the naive chaser starts no later than her sophisticated self if and only if $\beta G_1(x_1|y_0) \geq \varphi$ holds, where $x_1$ is the starting time of her sophisticated self as illustrated by Proposition B.4.